\documentclass[10pt,journal,compsoc]{IEEEtran}

\ifCLASSINFOpdf
\else
\fi
\usepackage{verbatim}
\usepackage{amssymb}
\usepackage{amsmath}
\usepackage{makecell,multirow,diagbox,rotating}
\usepackage{longtable}
\usepackage{array}
\usepackage{supertabular}
\usepackage{booktabs}
\usepackage{tabu}
\usepackage{cite}
\usepackage[numbers,sort&compress]{natbib}
\usepackage{pifont}% http://ctan.org/pkg/pifont
\newcommand{\cmark}{\ding{51}}%
\newcommand{\xmark}{\ding{55}}%
\usepackage[dvipsnames]{xcolor}
\usepackage{color}
\usepackage{xcolor}
\usepackage{listings}
\definecolor{light-gray}{gray}{0.95}
\newcommand{\code}[1]{\colorbox{light-gray}{\texttt{#1}}}
\usepackage{url}
\usepackage{makecell,rotating}
\usepackage{colortbl}
\usepackage{float}
\usepackage{graphicx}
\usepackage{subfigure}
\graphicspath{{fig/}}
\usepackage{todonotes}
\usepackage[justification=centering]{caption}
\usepackage{enumitem}
\usepackage{tikz}
\usetikzlibrary{calendar,fpu}

\tikzset{
    moon colour/.style={
        moon fill/.style={
            fill=#1
        }
    },
    sky colour/.style={
        sky draw/.style={
            draw=#1
        },
        sky fill/.style={
            fill=#1
        }
    },
    southern hemisphere/.style={
        rotate=180
    }
}

\makeatletter
\pgfcalendardatetojulian{2010-01-15}{\c@pgf@counta} % 2010-01-15 07:11 UTC -- http://aa.usno.navy.mil/cgi-bin/aa_moonphases.pl?year=2010&ZZZ=END
\def\synodicmonth{29.530588853}
\newcommand{\moon}[2][]{%
    \edef\checkfordate{\noexpand\in@{-}{#2}}%
    \checkfordate%
    \ifin@%
        \pgfcalendardatetojulian{#2}{\c@pgf@countb}%
        \pgfkeys{/pgf/fpu=true,/pgf/fpu/output format=fixed}%
        \pgfmathsetmacro\dayssincenewmoon{\the\c@pgf@countb-\the\c@pgf@counta-(7/24+11/(24*60))}%
        \pgfmathsetmacro\lunarage{mod(\dayssincenewmoon,\synodicmonth)}
        \pgfkeys{/pgf/fpu=false}%%
    \else%
        \def\lunarage{#2}%
    \fi%
    \pgfmathsetmacro\leftside{ifthenelse(\lunarage<=\synodicmonth/2,cos(360*(\lunarage/\synodicmonth)),1)}%
    \pgfmathsetmacro\rightside{ifthenelse(\lunarage<=\synodicmonth/2,-1,-cos(360*(\lunarage/\synodicmonth))}%
    \tikz [moon colour=white,sky colour=black,#1]{
        \draw [moon fill, sky draw] (0,0) circle [radius=1ex];
        \draw [sky draw, sky fill] (0,1ex)
            arc (90:-90:\rightside ex and 1ex)
            arc (-90:90:\leftside ex and 1ex)
            -- cycle;
    }%
}

\usepackage{balance}
\setlist[itemize]{leftmargin=5mm}

\newcommand{\ling}[1]{{\textcolor{orange}{#1}}}

\newcommand{\revise}[1]{\textbf{\textcolor{red}{#1}}}

\begin{document}

%\title{Research on Third-Party Libraries in Android Apps: A Systematic Literature Review}
\title{Research on Third-Party Libraries in Android Apps: A Taxonomy and Systematic Literature Review}
%\title{Research on Third-Party Libraries in Android Apps: A Taxonomy and Comprehensive Survey}

\author{Xian Zhan, 
%~\IEEEmembership{Member,~IEEE,}
        Tianming Liu,
        Lingling Fan$^{*}$,
        Li Li, %~\IEEEmembership{Fellow,~OSA,}
        Sen Chen,
        Xiapu Luo$^{*}$, and
        Yang Liu% <-this % stops a space
\IEEEcompsocitemizethanks{
\IEEEcompsocthanksitem Xian Zhan and Xiapu Luo are with The Hong Kong Polytechnic University.
Lingling Fan is with College of Cyber Science, Nankai University, China.
Li Li and Tianming Liu are with Monash University, Australia.
Sen Chen is with College of Intelligence and Computing, Tianjin University, China. 
Yang Liu is with School of Computer Science and Engineering, Nanyang Technological University, Singapore. 
}% <-this % stops an unwanted space
\thanks{
$\bullet$ \ \ LingLing Fan (linglingfan@nankai.edu.cn) \&  \newline{}
Xiapu Luo (csxluo@comp.polyu.edu.hk)
 are the corresponding authors
%Manuscript received March 9 and revised August 15, October 15, 2019
}}

\begin{comment}

\markboth{Journal of \LaTeX\ Class Files,~Vol.~14, No.~8, January~2021}%

{Zhan \MakeLowercase{\textit{et al.}}: Research on Third-Party Libraries in Android Apps: A Systematic Literature Review}
\end{comment}

\IEEEtitleabstractindextext{%
	\begin{abstract}
    
Third-party libraries (TPLs) have been widely used in mobile apps, which play an essential part in the entire Android ecosystem. However, TPL is a double-edged sword. On the one hand, it can ease the development of mobile apps. On the other hand, it also brings security risks such as privacy leaks or increased attack surfaces (e.g., by introducing over-privileged permissions) to mobile apps. Although there are already many studies for characterizing third-party libraries, including automated detection, security and privacy analysis of TPLs, TPL attributes analysis, etc., what strikes us odd is that there is no systematic study to summarize those studies' endeavors. To this end, we conduct the first systematic literature review on Android TPL-related research. %, which is the first survey in this area. 
Following a well-defined systematic literature review protocol, we collected 74 primary research papers closely related to Android third-party library from 2012 to 2020.
After carefully examining these studies, we designed a taxonomy of TPL-related research studies and conducted a systematic study to summarize current solutions, limitations, challenges and possible implications of new research directions related to third-party library analysis.
We hope that these contributions can give readers a clear overview of existing TPL-related studies and inspire them to go beyond the current status quo by advancing the discipline with innovative approaches.

%Third-party library (TPL) is a double-edged sword. On the one hand, third-party libraries can facilitate the developing process of an app. On the other hand, they also bring various security risks, such as privacy leakage, over-privileged for TPLs. Nowadays, there are many research work on third-party library from different research perspectives, such as third-party detection techniques, security and privacy analysis of TPLs, empirical studies, library permission separation. What strikes us odd is that there is a lack of a systematic investigation for these state-of-the-art research. Therefore, it is hard to know the specific application scenarios of state-of-the-art research. Thus, in this paper, we aim at conducting a comprehensive and systematic investigation on current research of Android third-party library analysis. We first follow a well-defined systematic literature review methodology to collect 54 research papers regarding the third-party analysis. Then we design a taxonomy for these papers based on different research purposes. We review their research background, core techniques, research purposes, the pros and cons. We also conduct a comparative study on existing research and provide an overview to readers from different perspectives. We hope that our work could give readers a clear understanding of TPLs analysis like a road-map in this direction. We attempt to find their limitations and reveal the future research gap, which can inspire more researchers to conduct more valuable and attractive studies in this area.

	\end{abstract}
	\begin{IEEEkeywords}
		Third-party library, Android, Literature review, Applications
\end{IEEEkeywords}}

\maketitle

\IEEEraisesectionheading{\section{Introduction}
	\label{sec:introduction}}
%\label{sec:Introduction}

%\sen{We may highlight the findings/problems/challenges/opportunities, etc.} 
\IEEEPARstart{A}{ndroid} has gained tremendous popularity since it was published in 2007~\cite{MOBSCANNER2017ICSE-C}. Recent years have witnessed the booming market of Android apps. For example, more than 3 millions apps are available on the official Google Play Store~\cite{statista}.
%
%The statistic~\cite{statista} shows that the available apps have exceeded 3 million on the official Google Play Store. 
To facilitate the development of Android apps, lots of third-party libraries (TPLs) have been developed from various providers. These TPLs provide a wide variety of development kits, plentiful UIs, and various social media plugins, and have been adopted by many apps. For instance, ``Exodus Privacy'' provides the up-to-date statistics on third-party libraries in Google Play, and we can find that about 51\% of apps in Google Play include the analytic TPL, i.e., Google Firebase~\cite{exodus};
%more than 57\% of mobile apps include ad-libraries~\cite{PEDAL2015MobiSys}, 
and more than 60\% of the code in an Android app belongs to TPLs~\cite{Wang2017ICSE-C}. 
Unfortunately, TPL is a double-edged sword, which may bring unwanted security risks to mobile users. For example, malicious TPLs could be introduced into legitimate apps. Adversaries can repackage an app by adding malicious third-party libraries that can send premium SMS services and steal users' private information~\cite{li2017simidroid}. They also often modify ad libraries by changing the revenue destination of ad libraries to redirect the profits~\cite{li2017understanding}. 
Moreover, some TPLs have been reported with behaviors invading users' privacy, such as reading the contact list information or getting users' locations~\cite{privacyleak,privacyleak2014adhoc,pluto2016,YLDSN16,YLTSE19}.
The security issues caused by TPLs are mainly due to the permission mechanism of Android system, which creates a separate process and private storage for each app~\cite{DelDroid201983HAMMAD,hammad2017ICSA}. Specific permissions are required for apps to access the system services and resources. However, this permission model works at the app-level, meaning that TPLs and the host apps share the same privileges, which leads to over-privileged problems. To solve this problem, many researchers attempted to limit the privileges of TPLs~\cite{sanAdBox2013ICC,aframe2013ACSAC,PEDAL2015MobiSys,adsplit2012USENIX,AdDroid20120ASISCCS}.

Unscrupulous developers may resize the ad libraries to induce users to click the ads, which also may lead to revenue loss or affect the normal operation of apps.
We refer to all violations which affect the profits of developers and users' experience as ad fraud, which attracts many studies recently~\cite{FraudDroid2018FSE,Decaf2014NSDI,MadFraud2014mobisys,liu2020maddroid}.

Furthermore, 
vulnerabilities in third-party libraries can pollute the downstream clients, including apps or other TPLs that depend on these vulnerable TPLs. 
If a popular third-party library is compromised, %either by design or by malicious attacks,
the threats from this TPL could spread to a large number of apps and affect countless mobile devices~\cite{libscout2016ccs,Yasumatsu2019codaspy,OSSPOLICE2017CCS}. 
The vulnerability issues in TPLs will usually cause severe consequences because recent studies~\cite{libscout2016ccs,Yasumatsu2019codaspy,zhang2020empirical,ATVHunter2021ICSE} revealed that app developers seldom follow the fixed scheme of TPLs and tend to delay the replacement of outdated TPLs in apps, even if these TPLs include severe vulnerabilities and could pose serious threats to mobile devices or users.
%
%Furthermore, recent studies~\cite{libscout2016ccs,Yasumatsu2019codaspy} show that app developers seldom follow the fix scheme of TPLs and tend to delay the replacement of outdated TPLs in apps,
%even if these TPLs include severe vulnerabilities and could pose serious threats to mobile devices or users.
%
%For example, some third-party libraries have misused cryptographic APIs~\cite{libscout2016ccs,gao2019negative}. 
%
Without a doubt, without patching such vulnerabilities in TPLs, they will pose unexpected threats to the entire mobile ecosystem. Therefore, many studies~\cite{privacyleak,privacyleak2014adhoc,pluto2016,AdRisk2012Wisec} have been conducted to mitigate the vulnerability issues in TPLs.

TPLs may cause functionality issues in host apps because the direct or transitive dependency of TPLs can also bring dependencies conflicts, which may lead to app crash~\cite{DC2018FSE,RIDDLE2019ICSE}. 
An app may depend on multiple versions of the same TPL or class, but only one version can be loaded, which may lead to dependency conflict and bring some unexpected issues, such as runtime exceptions. Recent studies proposed several approaches to address this problem~\cite{DC2018FSE,huang2020interactive,Wang2020ICSME}.
Moreover, previous research~\cite{xian2019saner,Lili2016SANER,Wang2017ICSE-C} pointed out that TPLs can be noises that could affect the performance of other app analysis, such as the detection of repackaging and malicious apps.
%
%Based on aforementioned introduction, we can find that some studies (e.g., vulnerable TPL detection) need to depend on TPL identification.
%In addition, previous research~\cite{xian2019saner,Lili2016SANER,Wang2017ICSE-C} pointed out that TPLs as noises could affect the detection result of repackaging detection and malware detection. 
%
The earlier existing studies, such as the repackaged app detection~\cite{MassVet2015chen,DroidMOSS12CODASPY,YuruACSAC14,YLSANER16,YLTSE18} usually used the whitelist to exclude TPLs. However, the whitelist-based method has many limitations. For one thing, it is impossible for whitelist-based method to enumerate all TPLs.
%the whitelist cannot be exhaustive .
%For another, systems use whitelist to filter TPLs out based on their package name.
%TPLs in the whitelist are usually filtered out based on the package name. 
%Unfortunately, many obfuscated tools may encrypt the package name, leading to TPL packages overlooked. 
For another, whitelist-based method relies on package names to filter TPLs, while many apps can apply obfuscation techniques to encrypt the package name within the app, leading to the overlook of TPLs. 
As a result, many advanced tools~\cite{LibD2017ICSE,LibRadar2016ICSE,libscout2016ccs,LibID2019issta,ORLIS2018MOBILESoft} have been proposed to detect third-party libraries. 
%%%%%%%%%%%%%%%%%%%%%%%%%%%%%
%\revise{% can be delete this part
%According to previous studies~\cite{LibRoad2020TMC,LibExtractor2020wisec}, existing tools generally can be divided into two types: 1) clustering-based method, which does not require the prior knowledge to identify the in-app TPLs; 2) the similarity comparison-based method, which requires developers to collect the TPL files to build a predefined database to identify in-app TPLs.
%These methods extract the different code features and adopt the corresponding signature-based method to identify in-app TPLs, which can achieve better performance than the whitelist-based method.}

%\revise{highlight the contributions, give a brief conclusion about our work.}
{Based on the above description, we can find that TPLs play a significant role in the entire Android ecosystem. TPLs are essential participants in app development, maintenance, and subsequent detection. Besides, TPLs also can affect the quality, rating, and security of the host apps.
However, although there are many studies on TPLs, there is no systematic analysis of them. 
Due to the importance of TPLs,
researchers may be eager to know the current research status quo and the gap of state-of-the-art studies on TPLs. To fill this gap, in this paper, %we conduct a systematic review of existing research on {Android} TPLs and point out the research directions.
we present to the community the first systematic survey of studies on {Android} third-party libraries. Our contributions of this paper are as follows:

\begin{itemize}
    \item \textbf{A collection of Android TPL related publications.} Following a well-defined \textit{systematic literature review} (SLR) protocol~\cite{snowballing2014,SLR2007} and a thorough examination of the collected primary publications, we collect \textbf{74} primary papers closely related to the analysis of {Android} TPLs.
    
    \item \textbf{A comprehensive taxonomy.} We design a taxonomy of TPL-related research studies from different perspectives, including research objectives, targeted libraries, type of TPLs and type of program analysis. Based on the taxonomy, we conduct an in-depth comparative study on the existing research.
    
    \item \textbf{Useful insights.} We identify the advantages and disadvantages, trends, patterns, gaps via comparative analysis of the collected papers. We also concluded the essential findings of existing studies and proposed possible implications of new research directions.

\end{itemize}

 %based on their main research purposes to lead our whole work, including TPL detection, security issue identification, TPL privilege de-escalation, TPL updating and TPL attribute understanding.
%Following the taxonomy, we conduct an in-depth comparative study on the existing state-of-the-art techniques to summarize the challenges in third-party library-related research, current solutions, advantages and disadvantages, and possible implications of new research directions. 

We hope these contributions can give readers a clear overview of third-party library related research and inspire future researchers to go beyond the current status quo by presenting more useful work.
}
Besides, we also have the following essential findings:
\begin{itemize}

%\item \xian{We collect 64 primary literature on Android TPL-related research. We have mapped the existing research to the app industrial chain, i.e., the development and maintenance of apps. We hope that we can bridge the gap between academia and industry through this work. People in the industry can learn about state-of-the-art studies; meanwhile, the academia can know the limitations of current research and to implement more valuable work. }
%\item We give a comprehensive comparison of existing studies based on each category and highlight the challenges and solutions presented by our fellow researchers,
%allowing readers to have a deep understanding of the current landscape of Android TPL-related research.

\item Most existing tools usually focus on Java TPLs analyses, only a few studies focus on the native library analyses. Future research can pay attention to this direction (see Section~\ref{sec:Taxonomy}).

\item {For TPL identification, we still have a long way to go. Most TPL detection tools have high precision but low recall. Most tools cannot achieve a good resiliency to code obfuscation, especially dead code removal and package flattening. Even though there are many TPL identification, they still cannot achieve a good performance in version identification, partial imported TPL identification and optimization, and so on (see Section~\ref{sec:tplidentification}).}

\item We find existing studies present limited understanding about vulnerable TPLs. Only several types TPL vulnerabilities were studied. We suggest future researchers can pay more attention to how to reveal the entire landscape of TPL vulnerabilities (see Section 4.2).

\item Future research can focus on TPL recommendation, GUI-related TPL smell analysis, TPL updating system design, native libraries related research, library compatibility analysis, TPLs' dynamic features analysis, cross-language TPLs analysis (see Section 4.3 \& 4.4 \& 5.3).

\end{itemize}

%We discuss the limitations and possible implications of new research directions related to third-party library analyses. We aim to inspire our fellow researchers to come up with new ideas and solutions to make more contributions to the community.

%
The rest of the paper is organized as follows. Section~\ref{sec:relatedwork} introduces the related work.
Section~\ref{sec:search} introduce our literature search methodology.
%Section~\ref{sec:Background} introduces the basic concept of the third-party library and related research.
Section~\ref{sec:Taxonomy} provides the taxonomy on related research of TPLs.
Section~\ref{sec:review} introduces state-of-the-art research work based on our the research objectives.
Section~\ref{sec:implications} gives the implications for readers based on our investigation.
%Section~\ref{sec:threats} discusses the threats to the validity of our research.
Section~\ref{sec:conclusion} gives a conclusion of our work.

% Section~\ref{sec:Methodology} show some typical detection techniques.

% Section~\ref{sec:experiment} reports the evaluations of repackaging detection techniques.
% %Section~\ref{sec:discussion} elaborates some evasion techniques and points out some open problems on repackaging detection.

%\input{Background}
\section{Literature Search Methodology}
\label{sec:search}

Since we investigate the state-of-the-art research of third-party libraries in Android, we first introduce the methodology we use to find relevant literature and then present the basic information about our collected papers.

We follow the well-established guidelines~\cite{snowballing2014,SLR2007} to conduct our lightweight Systematic Literature Review (SLR). The overview of the SLR methodology we applied in this work is as below:

\begin{itemize}
	\item Define the research scope. This step is used to set up our research scope and clarify our research goal.
	\item Secondly, identify the keywords for searching.
	
	\item Conduct the search process. The search process consists of two parts: search on commonly-used publication repositories; and search on major venues, including both conferences and journals.
	
	\item Apply exclusion criteria on the search outcomes to enhance the analysis accuracy. In the process of keyword-based searching, it is inevitable to acquire some less related papers. To mitigate this, we evaluate the search results against the exclusive criterion defined in Section~\ref{sec:exclusion}.
	\item Conduct a backward-snowballing on the remaining papers in case of omission.
	\item Merge the results.
\end{itemize}

\subsection{Search Strategy}

%We now explain our search strategy in detail.

%1) The paper should be related to mobile third-party libraries. 2) We only focus on the Android platform.

\noindent \textbf{Search scope.}
We first define our research scope: {The papers should be related to third-party libraries in the Android ecosystem.}

\noindent \textbf{Search keywords.}
We define the search keywords which are applied to find %a large number of 
potentially related papers within the search scope. {The search keywords construction is an iterative process. Based on our research scope, Android TPL-related research, we first list some search keywords %by using brainstorm 
and find out their synonymous words. During the search process, we continuously refine and extend our search keywords.
Finally, we set up three groups of keywords shown in the Table~\ref{tbl:serchkeywords}. The first group of keywords limits our research scope. The second group of keywords consists of the modifiers of TPLs. The third group of keywords represents our research target; all of them are synonyms for third-party libraries. To ensure the collected papers are as complete as possible, we include ``-'' in the first group because we find some papers' titles do not include the platform information, but they are related to Android TPL-related research, such as OSSPoLICE~\cite{OSSPOLICE2017CCS}.
The search terms are composed of two groups of keywords arranged in sequence (i.e., $search\ term = g1 \bigcap g2 \bigcap g3  $, where $g1 \in group1$, $g2 \in group2$, $g3 \in group3$). The purpose of this step is to collect as many related papers as possible.
}

\begin{table}[t]
\centering
\caption{Keywords for paper repository search}
\vspace{-2mm}
\begin{tabular}{llll}
\toprule
   & \textbf{Group 1}    & \textbf{Group 2} & \textbf{Group 3} \\
\midrule
	 & android & third-party & lib* \\
     & mobile &  open-source     &   component \\
     & *phone   &   ad*  &     software \\
     &   APP    &   reuse  &  dependenc*\\
     & -       &      \\

\bottomrule

\end{tabular}
\vspace{-1mm}
\begin{center}
	``*'' refers to wildcard.
\end{center}
\label{tbl:serchkeywords}
\vspace{-2ex}
\end{table}

%%%%%%%%%%%%

\noindent \textbf{Search database.}
\textit{1) \underline{Repository}.} We first look for the potentially related papers in four well-known digital databases: {ACM Digital Library~\cite{acmdigital}, IEEE Xplore Digital Library~\cite{ieee}, SpringerLink~\cite{springer}, and ScienceDirect~\cite{sciencedirect}}. Besides, we employ Google Scholar as our auxiliary tool to complement our collection of papers. Since the search results of the online repositories usually include some irrelevant publications, we set some rules to delete the noises. The specific solutions will be introduced in Section~\ref{sec:exclusion} (exclusion criteria).

\begin{figure}[t]
  \centering
  \includegraphics[width=0.4\textwidth]{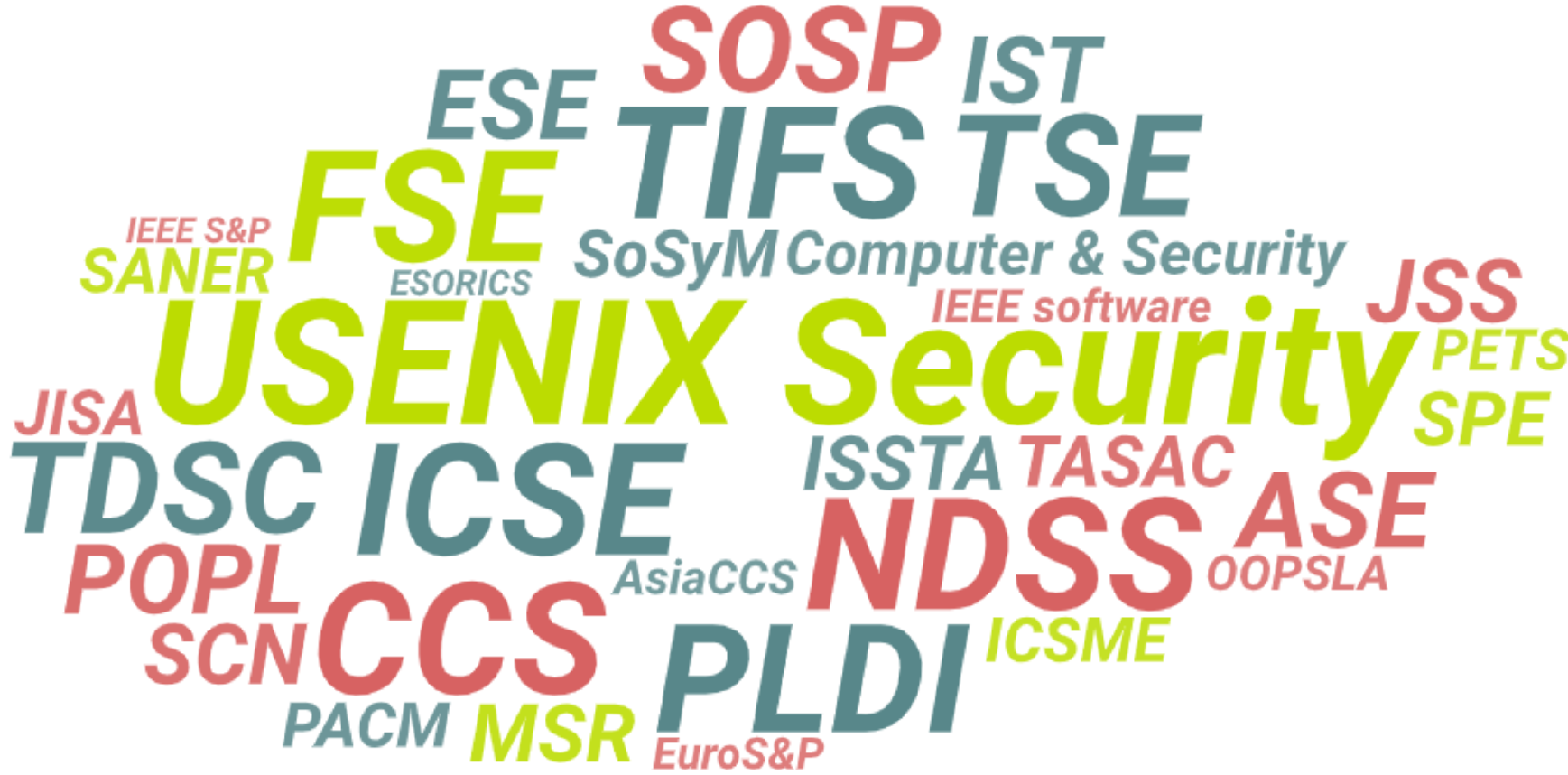}\\
  \caption{Word cloud of all major venue names of the examined publications }
  \label{fig:cloudword}
  \vspace{-3.3ex}
\end{figure}

\noindent\textit{2) \underline{Major Venue}.} 
%Unfortunately, some top-venue conferences (e.g., NDSS~\cite{NDSS}) are not included in the aforementioned digital databases, thus 
{ Some conferences and journals, such as NDSS~\cite{NDSS}, have policies (e.g., open proceedings) that cause their publications unavailable in the aforementioned digital repositories~\cite{lili2017static}.
To guarantee the completeness, we further supplement the well-known venues\footnote{\url{https://scholar.google.lu/citations?view_op=top_venues&hl=en&vq=eng}} in our search database to avoid the repository search do not miss any major publications. For this SLR, we select 20 top venues from software systems~\footnote{\url{https://scholar.google.lu/citations?view_op=top_venues&hl=en&vq=eng_softwaresystems}} that are from the field of software engineering and programming languages and other top 20 venues from security and privacy~\footnote{\url{https://scholar.google.lu/citations?view_op=top_venues&hl=en&vq=eng_computersecuritycryptography}} but we do not consider venues in the fundamental cryptography field (e.g., EUROCRYPT, CRYPTO, ASIACRYPT, FC, TCC). Fig.~\ref{fig:cloudword} provides a word cloud of the selected venues during the literature search process.  
}

\begin{comment}

We further supplement our paper dataset by adding the related literature of the top venues that are not included in the aforementioned digital databases such as NDSS~\cite{NDSS}, including both journals and conferences. 
Consequently, we choose the top 10 venues from the fields of software engineering and programming languages, and the top 10 venues from the area of security and privacy.

%For the primary venue search, we choose the top 10 venues from the field of software engineering and programming languages. Besides, we also select the top 10 venues from the direction of security and privacy. 
%Fig.~\ref{fig:cloudword} provides a word cloud of the major venues that we selected during the literature search process. 
Fig.~\ref{fig:cloudword} provides a word cloud of the major venues during the literature search process. 
%\revise{the description need to be revised by reality.}
\end{comment}

\noindent \textbf{Search Process.} {For the repository search, we use our search string that is formed as a conjunction of the three groups of the keywords to search the potential related work from the four well-known digital repositories one by one. To consolidate the collected list of relevant papers, we manually checked the searching results by going through the titles and abstracts of these publications to ensure that they are related to Android third-party libraries.
For the top venues search, we conduct our search process on DBLP~\footnote{\url{https://dblp.uni-trier.de/}}. Because the DBLP only provides the papers' titles, for some papers we cannot determine their content from the titles; we will exploit the Google Scholar to find these papers and check their abstracts.
}

\subsection{Exclusion Criteria}
\label{sec:exclusion}

In the process of keyword-based searching, it is inevitable to acquire some less related papers, e.g., repeated work, or even totally unrelated papers. {The coarse-granularity search process usually includes some irrelevant papers. For example, the state-of-the-art repository search engines (e.g., the one provided by Springer) may include many irrelevant papers~\cite{lili2017static,Li2018RebootingRO} because SpringerLink allows collecting information on the first 1,000 items from its search results.}
%e.g., non-English papers. 
To mitigate it and obtain more reliable results, 
%The matched findings are very likely contain some irrelevant or less relevant publications. In order to only focus on primary publications and obtain more reliable results, 
we define a set of exclusion criteria as follows to filter out the unrelated or less related publications. 
%The exclusion criteria can be categorized into three classes according to the exclusive targets, that are unrelated publications (1 - 2), less related publications (3 - 5). Details are as below:

\begin{enumerate}
  \item Non-English papers are filtered out. %We regard non-English papers unrelated since English is the only universal language for international academic mind exchange.
      
  \item {Papers that are irrelevant to Android platform are excluded. As shown in Table~\ref{tbl:serchkeywords}, in order to acquire as many related papers as possible, the search string may not include the term ``Android'' because we include the ``-'' in the group 1. Consequently, a large number of publications on other platforms, e.g., iOS and Windows, are included in the paper set, which should be filtered out. Besides, we also delete some papers from other ecosystems (e.g., npm) because the extracted features or identification methods are different from Android libraries.}
  \item Delete papers are not related to third-party libraries. Based on our search keywords, we may find some studies are related to third-party markets, third-party Android phones. Our research scope just focuses on Android TPLs; therefore, we also exclude these papers.
  \item This paper just focuses on extensive works. We have noticed that some papers, such as the short papers, posters, are for work of promising ideas in a preliminary stage. For this paper, if we have found the full formats in the later publications, we will delete the preliminary versions~\cite{lili2017static,kong2019Testing}. In accordance with international academic rules~\cite{snowballing2014,SLR2007}, publications that are less than five pages in IEEE/ACM-like double-column format or less than seven pages in LNCS-like single-column format should be treated as short papers. According to this item, we delete three papers~\cite{Wang2017ICSE-C, MAdLens2018infocom,LibD2017ICSE} from our dataset. For the first paper, we found the extensive version is in Chinese~\cite{haoyu2017automated}. Based on the first item, we also do not consider this paper. For the remaining two papers,  MadLens~\cite{MAdLens2018infocom} and LibD~\cite{LibD2017ICSE} that first was published in INFOCOM (2018) and ICSE (2017), respectively, then were extend to the journal papers and published in TMC (MAdLens, 2019)~\cite{jin2019madlens} and TSE (LibD, 2018)~\cite{LibD22018TSE}. For these papers, we keep the extended versions. We delete the posters if they just propose some basic ideas.

  \begin{comment}

  Immature work such as short papers and posters should be excluded. \textit{Short papers} are for work that is still in progress or of smaller scale, and \textit{Posters} are for work of promising ideas that are in early stages. In accordance with international academic rules~\cite{snowballing2014,SLR2007}, publications that are less than four pages (included) in IEEE/ACM-like double-column format or less than eight pages (included) in LNCS-like single-column format should be treated as short papers. %Considering this item, we delete these papers~\cite{Wang2017ICSE-C,MOBSCANNER2017ICSE-C} from our dataset.
  \end{comment}
  
  \item {Duplicated papers are removed.
  Some papers may have a preprint version online, the title or some contents of the published one and the preprint may have some differences. For these papers, if they have the same author list and similar or the same title, we consider they are suspicious pairs. Then we manually check whether their contents share a lot of content or not. If yes, we will keep the recent versions. If a paper published in a conference venue and was extended to a journal venue. We will remain the extensive publications.}
  
  \begin{comment}

  Firstly, we keep only one copy of the papers that are published in two or more resources, e.g., Google scholar and ACM.
  %some papers can be found in two or more resources, e.g. Google scholar and ACM, we should only keep one copy of the same papers. 
  Besides, we keep the conference version of the papers if they are extended to journal versions as they share similar core content and should also be treated as duplicated papers. 
  %some journal papers are the extension of the conference papers. These papers share the similar core content and should also be treated as duplicate papers. 
  %To recognize the papers, We extract the conference and journal papers sharing the same titles, abstracts and authors systematically, and check if the identified papers sharing the same core content. If this kind of duplication happens, we filter out these papers. For example, LibD2 is the journal version that expands the work of LibD~\cite{LibD2017ICSE}. Therefore, we just discuss LibD in this paper.
 To recognize the extended papers, we extract the conference and journal papers sharing similar titles, abstracts, or authors systematically and filter out the extension ones if any.
   \end{comment}
   
%\item Papers on a comparative evaluation of previous work should be filtered out, including but not limited to reviews of different approaches of testing Android apps.
\end{enumerate}

%After that, we quickly look through the collected papers, checking the title and abstract to ensure the paper set in the scope we define. Based on this rule, we delete three papers which are not on the Android platform from the collection. Following the SLR guidelines, we delete one poster~\cite{poster2014adhoney} and one short paper~\cite{Wang2017ICSE-C} from our collection of papers.

\subsection{Backward Snowballing}

To avoid the omission of related papers, we perform a backward-snowballing on the collected papers. Specifically, we manually check the references of each paper and find potentially related papers that are not in our paper repository or papers that cannot be found by using the defined keywords. %Finally, we manually check the paper repository by using the method we mentioned in the previous steps. 
We finally add two research papers~\cite{FLEXDROID201NDSS,OSSPOLICE2017CCS} into our paper repository, whose titles do not contain our predefined keywords.
%We can find that the titles of the two papers do not include the keywords in our predefined searching items. 

\subsection{Statistics of Selected Publications}

\begin{table}[t]
  \centering
    \caption{Summary of the selection of primary publications}
    \vspace{-2mm}
  \begin{tabular}{rc}
    \toprule[1.3pt]
    \textbf{Search Process} & \textbf{\#Related paper} \\
    \midrule[0.8pt]
    Repository and Major Venus Search & 5, 953 \\
    After reading the titles/abstracts & 95 \\
    After filtering out irrelevant topics   &  82 \\
     After selecting extensive papers & 79 \\
  %  After selecting extensive papers & 72 \\
    After deleting duplicate papers & 76 \\
    Final selection & \textbf{\underline{74}} \\
    \bottomrule[1.2pt]
  \end{tabular}
	\vspace{-2mm}
  \label{tbl:sum_pubs}
\end{table}

Table~\ref{tbl:sum_pubs} concludes the statistics of literature during the collection process. We obtained 5,953 papers in total by searching from the aforementioned database and venues.
After reading the titles and abstracts, we delete the irrelevant papers, and the number of papers meeting the requirement decreases to 95. {The discard rate is very high here primarily due to two reasons: the first one is that SpingerLink may include many false positives by using our search string~\cite{lili2017static}; the second reasons is due to the predefined search keywords. To ensure the completeness, our search strings contain ``lib*'', ``ad*'', ``open-source software'' that may include some papers from other platforms except Android. Thus,
we get more than 5900 irrelevant papers during the search process.
The whole selection process is performed by the first author and other co-authors also help to conduct the cross-validation.} We then go through the remaining papers by reading their abstract, introduction, and conclusion and finally select \textbf{74} research papers, which are the main research subjects for the following analysis.

%\vspace{-2pt}
 \begin{table}[thpb]
\vspace{-2pt}
\centering
\caption{Primary information of our paper repository}
\vspace{-2mm}
\begin{tabular}{lcr}
\toprule[1.3pt]
\textbf{Tool/System Ref.}  & \textbf{Year}   & \textbf{Venue}  \\
\midrule
\rowcolor{gray!20}
Wang et al.~\cite{Wang2020ICSME} & 2020 & ICSME \\
LibDetect Analysis~\cite{libdetect2020ASE} & 2020 & ASE \\
\rowcolor{gray!20}
LibHarmo~\cite{huang2020interactive}   &   2020  & ESEC/FSE \\
LibDX~\cite{LibDX2020SANER}  & 2020  & SANER  \\
\rowcolor{gray!20}
LibExtractor~\cite{LibExtractor2020wisec} & 2020 & WiSec   \\
LibRoad~\cite{LibRoad2020TMC} & 2020 & Mobile Computing (J) \\
\rowcolor{gray!20}
Ahasanuzzaman et al.~\cite{md2020studying} & 2020 & TSE(J)  \\
MadDroid~\cite{liu2020maddroid}  & 2020 & WWW \\

\rowcolor{gray!20}
Ahasanuzzaman et al.~\cite{ahasanuzzaman2020longitudinal} & 2020 & EMSE \\
MAdLens~\cite{jin2019madlens}  & 2019 & TMC (J) \\
\rowcolor{gray!20}
Yasumatsu et al.~\cite{Yasumatsu2019codaspy}  &  2019  & CODASPY \\ 
LibID~\cite{LibID2019issta}  & 2019 & ISSTA  \\
\rowcolor{gray!20}
RIDDLE~\cite{RIDDLE2019ICSE} & 2019 & ICSE \\
MadLife~\cite{chen2019revisiting} & 2019 & WWW \\
\rowcolor{gray!20}
Salza et al.~\cite{TPLsurvey2019} &  2019 & Spring Science \\
APPCOMMUNE~\cite{APPCOMMUNE2019SANER} & 2019 & SANER \\
\rowcolor{gray!20}
DECCA~\cite{DC2018FSE} & 2018  & ESEC/FSE \\
FraudDroid~\cite{FraudDroid2018FSE} & 2018 & ESEC/FSE \\
\rowcolor{gray!20}
LibPecker~\cite{libpecker2018}    & 2018 & SANER \\

% MadLens~\cite{MAdLens2018infocom}   & 2018 & INFOCOM \\
ORLIS~\cite{ORLIS2018MOBILESoft} & 2018 & MOBILESoft \\
\rowcolor{gray!20}
Salza et al.~\cite{Salza2018ICPC} & 2018 & ICPC \\
LibD2~\cite{LibD22018TSE}        & 2018 & TSE(J)  \\
\rowcolor{gray!20}
Dong et al.~\cite{Dong2018HotMibile}  & 2018 & HotMobile \\
Han et al.~\cite{identifyads2018WPC} & 2018 & WPC(J) \\
 %Wireless Personal Communications(J) \\
 \rowcolor{gray!20}
Ogawa et al.~\cite{CANDARW2018}   & 2018 & CANDARW \\
FLEXDROID~\cite{FLEXDROID201NDSS} & 2017 & NDSS \\
%LibD~\cite{LibD2017ICSE} & 2017 & ICSE \\
\rowcolor{gray!20}
Droid-V~\cite{Watanabe2017MSR}  & 2017 & MSR\\
OSSPOLICE~\cite{OSSPOLICE2017CCS}  & 2017  & CCS  \\
\rowcolor{gray!20}
AppLibRec~\cite{AppLibRec2017Internetware}  & 2017 & Internetware \\
Derr et al.~\cite{Derr2017ccs} & 2017 & CCS \\
\rowcolor{gray!20}
Zhan et al.~\cite{splitads2017ACISP} & 2017 & ACISP \\
Gui et al.~\cite{Gui2017WhatAO}   & 2017 & CoRR  \\
%MOBSCANNER~\cite{MOBSCANNER2017ICSE-C} & 2017 &  ICSE-C \\
\rowcolor{gray!20}
Son et al.~\cite{son2016mobile} & 2016 & NDSS \\
LibCage~\cite{Libcage2016ESORICS} & 2016 & ESORICS \\
\rowcolor{gray!20}
LibFinder~\cite{LibFInder2016sp} & 2016 & S\&P \\
LibRadar~\cite{LibRadar2016ICSE} & 2016 & ICSE-C \\
\rowcolor{gray!20}
LibScout~\cite{libscout2016ccs} & 2016 & CCS \\
LibSift~\cite{LibSift2016soh}  & 2016 & APSEC  \\
\rowcolor{gray!20}
Pluto~\cite{pluto2016} & 2016 & NDSS \\
Li et al.~\cite{Lili2016SANER} & 2016 & SANER \\
\rowcolor{gray!20}
Ruiz et al.~\cite{Israel2016software} & 2016 & IEEE Software(J) \\
Rastogi et al.~\cite{ad2016NDSS} & 2016 & NDSS \\
\rowcolor{gray!20}
Wei et al.~\cite{price2016NDSS} & 2016 & NDSS \\
Madscope~\cite{Madscope2015Mobisys} & 2015 & MobiSys \\
\rowcolor{gray!20}
PEDAL~\cite{PEDAL2015MobiSys} & 2015 & MobiSys \\
Book et al.~\cite{adempirical2015compscience} & 2015 & Computer Science(J) \\
\rowcolor{gray!20}
Paturi et al.~\cite{paturi2015NDSS} & 2015  & NDSS \\
Gui et al.~\cite{ICSE2015jiapping} & 2015 & ICSE \\
\rowcolor{gray!20}
ClickDroid~\cite{cho2015empirical} & 2015 & ARES \\
K\"{u}hnel et al.~\cite{kuhnel2015fast} & 2015 &  Trustcom \\
\rowcolor{gray!20}
AdDetect~\cite{AdDetect2014ISSNIP} &  2014 & ISSNIP \\
APKLancet~\cite{Apklancet2014ASIACCS} & 2014 & ASIACCS \\
\rowcolor{gray!20}
COMPAC~\cite{COMPAC2014Wang} & 2014 & CODASPY \\
DECAF~\cite{Decaf2014NSDI} & 2014 & NSDI \\
\rowcolor{gray!20}
Duet~\cite{Duet2014wisec} & 2014 & WiSec \\
Madfraud~\cite{MadFraud2014mobisys} & 2014 & MobiSys \\
\rowcolor{gray!20}
NativeGuard~\cite{NativeGuard2014Wisec} & 2014 & Wisec \\
Moonsamy et al.~\cite{privacyleak} & 2014 & ISITA \\
\rowcolor{gray!20}
Short et al.~\cite{privacyleak2014adhoc} & 2014 & MASS \\
Ullah et al.~\cite{wkshps2014} & 2014 & INFOCOM WKSHPS \\
\rowcolor{gray!20}
Ruiz et al.~\cite{Mojia2014software} & 2014 & IEEE Software(J) \\
Brahmastra~\cite{bhoraskar2014brahmastra}  & 2014 & USENIX Security  \\
\rowcolor{gray!20}
AFrame~\cite{aframe2013ACSAC} & 2013 & ACSAC \\
SanAdBox~\cite{sanAdBox2013ICC} & 2013 & ICC \\
\rowcolor{gray!20}
Book et al.~\cite{book2013spsm} & 2013 & SPSM \\
Book et al.~\cite{MOST2013} & 2013 & MoST \\
\rowcolor{gray!20}
Tongaonkar et al.~\cite{PAM2013} & 2013 & PAM \\
AdDroid~\cite{AdDroid20120ASISCCS} & 2012 & ASIACCS  \\
\rowcolor{gray!20}
AdRisk~\cite{AdRisk2012Wisec} & 2012 & WiSec \\
AdSplit~\cite{adsplit2012USENIX} & 2012 & USENIX Security \\
\rowcolor{gray!20}
Bauer et al.~\cite{structure2012ICSM}  & 2012 & ICSM \\
Leontiadis et al.~\cite{Leontiadis12HotMobile} & 2012 & HotMobile \\
\rowcolor{gray!20}
Stevens et al.~\cite{Most2012} & 2012 & MoST \\
Vallina-Rodriguez et al.~\cite{vallina2012breaking}  & 2012 & IMC \\
\bottomrule
\end{tabular}
\vspace{-1mm}
	\begin{center}	
\textit{Note that: if a tool is not given the name, we use the \\ first author's name to represent it.}
	\end{center}
	\vspace{-3ex}
\label{tbl:listinfo}
\vspace{-3mm}
\end{table}

\begin{figure}[t]
%\vspace{-3ex}
\centering
 \includegraphics[scale=0.51]{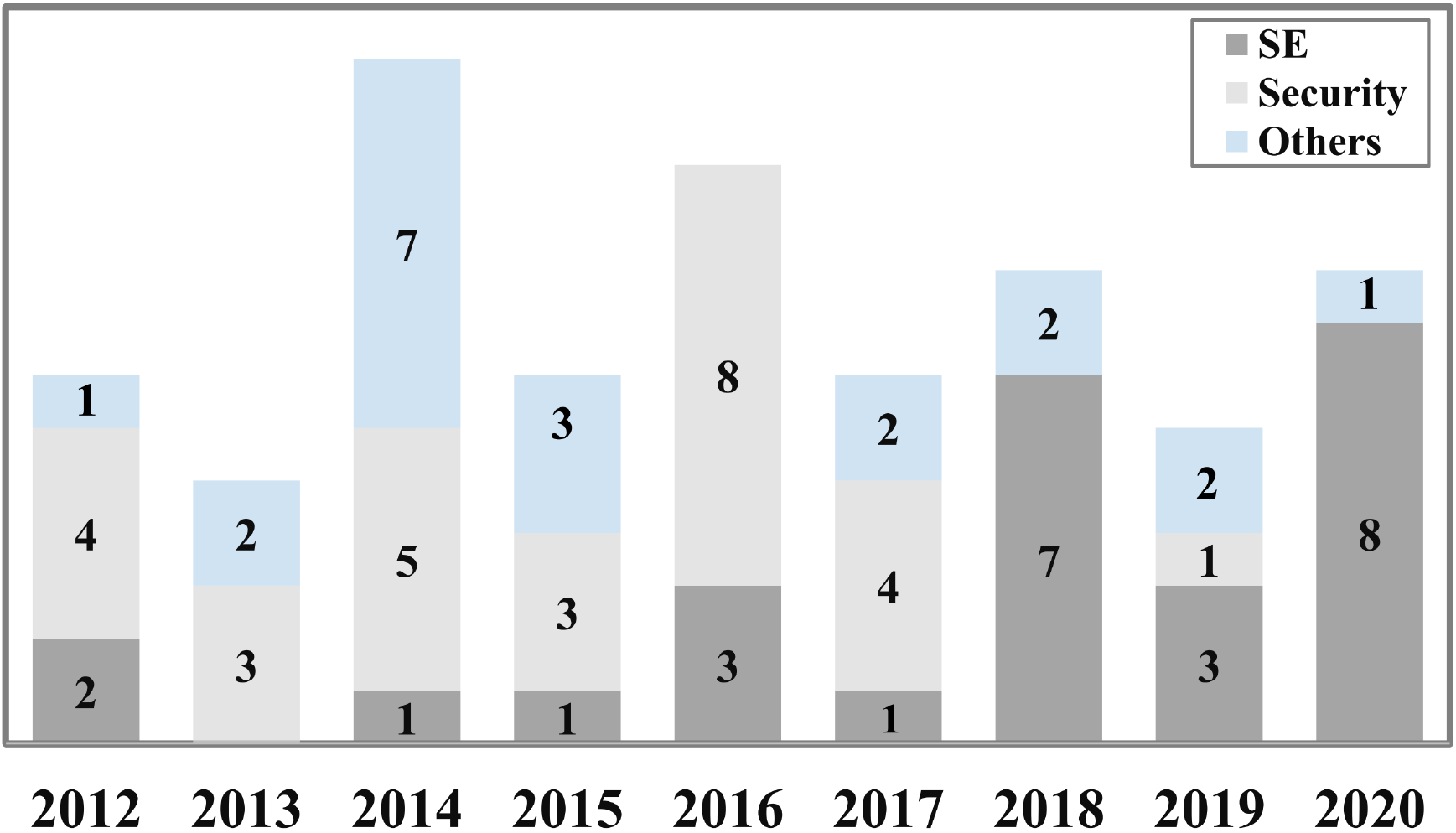}
 \caption{Distribution of the Repository of literature from 2012 to 2020 based on Venue Differences (SE: Software Engineering; Others: mainly include fields from Networking and Programming Language)
 }
  \label{fig:yeardis}
  \vspace{-2ex}
\end{figure}

{Fig.~\ref{fig:yeardis} shows the distribution of the collected literature through the published year (2012-2020). The period from 2012 to 2016 witnessed a fluctuation in the number of publications about Android third-party libraries. The number of publications reached the peak in 2014 (11). % and also kept a consistently high research trend. 
The papers in security and software engineering reached their peak in 2016 and 2018, respectively.
According to Fig.~\ref{fig:yeardis}, we can find the early work most published in networking and reached the peak in 2014. The research on the security-related issues of Android third-party libraries basically remained stable from 2012 to 2015 and increased abruptly in 2016, reached eight papers. And then the following years from 2017 to 2020 witnessed its gradual decline.
Since 2018, an increasing number of researchers have focused on software engineering. The recent three years have witnessed a sharp increase.
Table~\ref{tbl:listinfo} enumerates all the 74 papers and the corresponding publication year, venues, and their tool name if any. 
As for some work without a tool name, we use the first author name to denote them.
}
%Based on our paper repository, we find that over 84.9\% of papers are conference papers while 5.6\% publications of our collection are from workshops, and only about 9.5\% of the work comes from journals. 

%\subsection{Statistics on Paper Repository}

%\begin{figure}[htpb]
%\vspace{-3ex}
%\centering
%\subfigure[Distribution of the Repository of literature from 2012 to 2018 based on the Venue Differences]{%
% \includegraphics[scale=0.53]{yearbar.pdf}
 %\label{fig:subfigure1}
% }
%\quad

%\subfigure[Distribution of Repository of literature based on the %Different Research Direction]{%
%	\includegraphics[scale=0.53]{secandLA.png}
	%\label{fig:subfigure2}
%}
%\caption{Distribution of the Repository of literature from 2012 to 2018 based on the Venue Differences}
%\label{fig:yeardis}
%\vspace{-5ex}
%\end{figure}

%Based on the aforementioned SLR methodology, we finally obtain our paper repository which includes 53 papers on mobile third-party library analysis. In this part, we list some statistic information about the paper repository.

%Table~\ref{tbl:listinfo} enumerates all the papers and corresponding publication year, venues, and their tool name. As for some work without a tool name, we use the first author name to indicate them. Based on our paper repository, we can find that over 81\% papers come from conference papers while 5.6\% publications of our collection are from workshop and only about 13\% of the state-of-the-art work comes from journals. 

\section{Taxonomy of Examined Publications}
\label{sec:Taxonomy}

\begin{figure*}[t]
	\centering
	\includegraphics[scale=0.95]{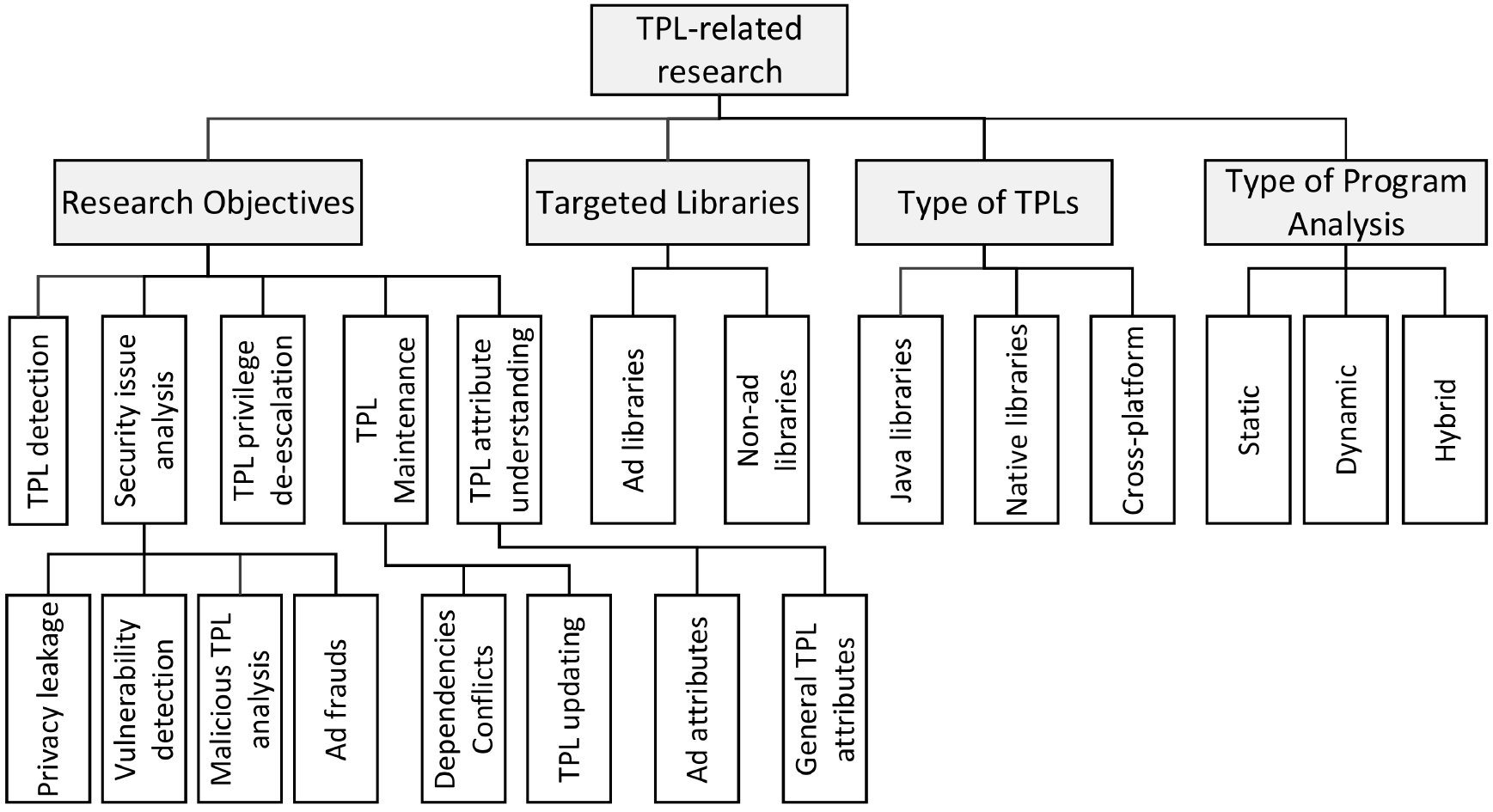}
	\caption{Taxonomy of TPL-related research}
	\label{fig:taxonomy}
	\vspace{-2ex}
\end{figure*}

To define a taxonomy for existing Android TPL-related research, we first tried to choose appropriate dimensions and properties in existing surveys~\cite{assessAndroidsecurity2017TSE,kong2019Testing}. We hope the taxonomy can help characterize existing work and gain insights into the state-of-the-art research as well as assess different techniques. Fig.~\ref{fig:taxonomy} shows a high-level view of the taxonomy diagram unfolding in four dimensions (i.e., \textbf{Research Objectives, Targeted Libraries, Type of TPLs} and \textbf{Type of Program Analysis}). We give a detailed introduction about each dimension and sub-dimension in the following sub-sections.

\subsection{Research Objectives}
{This dimension categorizes existing research with respect to the purposes of their analysis. Different studies have different problems they want to solve, we enumerate five sub-dimensions in this category:  1) TPL detection, 2) TPL security-related issue analysis, 3) TPL privilege de-escalation, 4) TPL maintenance, and 5) TPL attribute understanding. We now explain each of them as follows.
}

%\xian{Fig.~\ref{fig:taxonomy} shows the taxonomy of TPL-related research in our collected paper repository, which is grouped into five categories according to the research topics: 1) TPL detection, 2) TPL security-related issue analysis, 3) TPL privilege de-escalation, 4) TPL maintenance, and 5) TPL attribute understanding. We now explain each of them as follows.}

\textbf{TPL detection} aims to find the TPLs used in Android apps. TPL dependency information is not transparent, not to mention there are many direct and transitive dependencies. On the other hand, TPL detection has many significant application prospects in assisting the downstream tasks, such as malicious app detection, repackaged Android app detection, vulnerable in-app TPL detection, software composition analysis. We also find that TPL detection is becoming a hot topic in recent years; therefore, it is necessary to understand existing TPL detection techniques.
%\tm{(TPL detection has significant application prospects in assisting downstream tasks such as malware detection, repackaged app detection, vulnerable TPL identification, and TPL isolation.)}
%TPL detection aims to find the TPLs used in Android apps, including the TPL names and TPL versions, if possible. We find that library detection is a hot topic in recent years, and our repository contains ten related papers.
%

\textbf{TPL security issue analysis}
%Considering the second category, TPL security-related issues identification,
 accounts for the {largest proportion (31\%, 23/74) in collected papers. On the one hand, we can see that many researchers were committed to the research on TPL security; on the other hand, TPLs do exist many security issues. Understanding current research status and their risk to users and devices is utterly necessary.}
For this dimension, we mainly discuss the following five parts: 1) privacy leakage detection and analysis, 2) vulnerability identification, 3) malicious TPL detection and analysis, 4) ad frauds.
{Based on existing research, we can find privacy leakage is not uncommon in TPLs~\cite{tang:ase19,tangtse:21a,privacyleak,privacyleak2014adhoc,price2016NDSS}. For example, some ad libraries can collect user's demographics information, but these TPLs may leak users' privacy without notice.
Aiming at this phenomenon, existing research investigated how many TPLs can cause privacy leakage\cite{price2016NDSS}, how was the data leaked~\cite{pluto2016} so on and so forth.
Vulnerability identification involves in-app TPL vulnerability detection. Malicious TPL detection involves finding TPLs with malicious behaviors, such as dynamically loading malicious payloads, leading to revenue loss. Note that malicious TPL detection does not necessarily need to identify the specific TPLs. Most research just needs to find the parts that belong to a TPL and identify the malicious behaviors. That is the difference between TPL identification and malicious TPL detection.
Ad frauds mean that unscrupulous developers violate the ad developing rules by placing ads that close to or cover the UIs, which may affect the user experience and induce extra clicks or impressions~\cite{FraudDroid2018FSE}.
}
%Among our collected papers, 

\textbf{TPL privilege de-escalation} is responsible for separating the privilege of TPLs from the host app. Therefore, we also can call it TPL isolation. 
Android system allows apps to access the system resources with corresponding privileges, but the permission mechanism is working at the app-level, which means the in-app TPLs share the same permissions with host app~\cite{DelDroid201983HAMMAD,hammad2017ICSA}. We can find this permission mechanism can bring potential risks because it causes the TPL over-privileged.
TPL isolation usually aims to separate TPLs from host apps by allocating them different storage space, permissions, and process IDs to ensure the TPLs cannot use the permissions of the host app to conduct sensitive behaviors.

\textbf{TPL maintenance} plays an essential role in app development, which can help keep the quality and health of apps.
{As for the TPL maintenance, we mainly introduce the research on dependency conflicts and TPL updating. Dependency conflicts are mainly due to the considerable direct and transitive dependencies within TPLs. Dependency conflict occurs when the loaded version cannot cover the features required by the app, leading to runtime exceptions. TPL updating usually involves vulnerabilities and compatibility issues. For instance, the old version of TPL may be detected within a known vulnerability and the new version has fixed the vulnerability. We can investigate app developers' responses to the updating TPLs.
Existing TPL updating research primarily can be divided into two parts, namely the TPL automated updating tool implementation and TPL updating analysis.
}
%Researchers also conduct a series of research on TPL updating. In sum, TPL updating primarily can be divided into two parts, namely the TPL automated updating tool implementation, and TPL updating analysis. %The delay of TPL update may bring some security risks, understanding the related issues are really indeed \tm{this sentence need modify}. 

{\textbf{TPL attribute understanding} includes miscellaneous aspects of TPLs, such as the library recommendation, the impact of TPLs' new features on apps, rating analysis, permission analysis and so on.
For this dimension, we mainly describe these studies from two aspects: general TPL-related attributes, and attributes related to the special TPLs, ad libraries. Besides, we find most of research in this dimension usually adopts empirical study, case study and user study as the evaluation methods. }
%we mainly describe these studies from two aspects: TPL-related attributes and the special one, i.e., ad libraries. \tm{(TPL-related attributes, and attributes related to the special TPLs, ad libraries.)}
%and other TPL related attribute understanding, which can help readers understand TPLs better.

%TPL isolation usually aims to separate TPLs from host apps by allocating them different storage space, permissions, and process IDs to ensure the TPLs cannot use the permissions of the host app to conduct sensitive behaviors. Considering the security and privacy analysis of TPLs, we divide them into four parts: 1) violation detection, 2) vulnerability analysis, 3) ad fraud analysis, and 4) privacy leakage. 
%In the early days, most of the research was aimed at the privacy leaks of third-party libraries or vulnerable libraries. After that, recent research focused on ad frauds. 
% Researchers also conducted a series of empirical studies relating to library updating, the relation between the rating of apps and TPLs, as well as Ad libraries analysis in Android apps. 
%As for the remaining research, it includes library exterior analysis and library recommendation. There is only one research paper focusing on library recommendation.
%
%
In particular, some articles may be involved in multiple categories, and each category is not entirely independent for this dimension. Therefore, there are intersections among different categories. 
For example, PEDAL~\cite{PEDAL2015MobiSys} attempts to implement privilege de-escalation for ad libraries (TPL isolation). However, it also needs to identify TPLs first. Thus, PEDAL also implements a tool named \textit{Separator} to identify ad libraries (TPL detection).

\subsection{Targeted Libraries}

{This dimension classifies collected papers based on the ad and non-ad libraries. The non-ad libraries mean not only for ad libraries but also for general TPLs.
TABLE~\ref{tbl:ad&nonad} characterize the publications selected from our SLR in terms of the research objectives based on the ad and non-ad libraries.
Among our collected papers, we find {about a half (35/74)} of the collected papers focus on ad libraries. One of the main reasons is that developers can make profits by embedding ad libraries in their apps. The ad library is an essential type of TPLs, which bridges the advertisers, developers, and customers. Besides, ad libraries usually need to get users' information and push customized content to target users. The ad libraries are more likely to collect users' privacy information without users' attention and lead to privacy leakage~\cite{Most2012}. Such a category can help us understand existing research on ad and non-ad libraries and figure out the current research gap.
}

\begin{table*}[t]
  \centering
  \caption{The categorization of collected papers based on the ad and non-ad libraries}
  \vspace{-1ex}
    \begin{tabular}{lcc||lcc||lcc}
    \toprule
    \textbf{Tool} & \multicolumn{1}{l}{\textbf{Ad}} & \multicolumn{1}{l}{\textbf{All TPLs}} & \textbf{Tool} & \multicolumn{1}{l}{\textbf{Ad}} & \multicolumn{1}{l}{\textbf{All TPLs}} & \multicolumn{1}{l}{\textbf{Tool}} & \multicolumn{1}{l}{\textbf{Ad}} & \multicolumn{1}{l}{\textbf{All TPLs}} \\
    \midrule
    \rowcolor[rgb]{ .929,  .929,  .929} LibDX~\cite{LibDX2020SANER} &       & \color{Green}{\cmark} & \cellcolor[rgb]{ 1,  1,  1}AdRisk~\cite{AdRisk2012Wisec} & \multicolumn{1}{l}{\cellcolor[rgb]{ 1,  1,  1}\color{Green}{\cmark}} & \cellcolor[rgb]{ 1,  1,  1} & \multicolumn{1}{l}{LibHarmo~\cite{huang2020interactive}} &       & \color{Green}{\cmark} \\
    LibExtractor~\cite{LibExtractor2020wisec} &       & \color{Green}{\cmark} & \cellcolor[rgb]{ .929,  .929,  .929}MadDroid~\cite{liu2020maddroid} & \multicolumn{1}{l}{\cellcolor[rgb]{ .929,  .929,  .929}\color{Green}{\cmark}} & \cellcolor[rgb]{ .929,  .929,  .929} & \multicolumn{1}{l}{DECCA~\cite{DC2018FSE}} &       & \color{Green}{\cmark} \\
    \rowcolor[rgb]{ .929,  .929,  .929} LibRoad~\cite{LibRoad2020TMC} &       & \color{Green}{\cmark} & \cellcolor[rgb]{ 1,  1,  1}MadLife~\cite{chen2019revisiting} & \multicolumn{1}{l}{\cellcolor[rgb]{ 1,  1,  1}\color{Green}{\cmark}} & \cellcolor[rgb]{ 1,  1,  1} & \multicolumn{1}{l}{RIDDLE~\cite{RIDDLE2019ICSE}} &       & \color{Green}{\cmark} \\
    LibID~\cite{LibID2019issta}  &       & \color{Green}{\cmark} & \cellcolor[rgb]{ .929,  .929,  .929}Rastogi et al.~\cite{ad2016NDSS} & \cellcolor[rgb]{ .929,  .929,  .929} \color{Green}{\cmark} & \cellcolor[rgb]{ .929,  .929,  .929} & \multicolumn{1}{l}{Wang et al.~\cite{Wang2020ICSME} } &       & \color{Green}{\cmark} \\
    \rowcolor[rgb]{ .929,  .929,  .929} LibPecker~\cite{libpecker2018} &       & \color{Green}{\cmark} & \cellcolor[rgb]{ 1,  1,  1}LibFinder~\cite{LibFInder2016sp} & \cellcolor[rgb]{ 1,  1,  1} & \cellcolor[rgb]{ 1,  1,  1}\color{Green}{\cmark} & \multicolumn{1}{l}{Yasumatsu et al.~\cite{Yasumatsu2019codaspy}} &       & \color{Green}{\cmark} \\
    ORLIS~\cite{ORLIS2018MOBILESoft} &       & \color{Green}{\cmark} & \cellcolor[rgb]{ .929,  .929,  .929}K\"{u}hnel et al.~\cite{kuhnel2015fast} & \multicolumn{1}{l}{\cellcolor[rgb]{ .929,  .929,  .929}\color{Green}{\cmark}} & \cellcolor[rgb]{ .929,  .929,  .929} & \multicolumn{1}{l}{APPCOMMUNE~\cite{APPCOMMUNE2019SANER}} &       & \color{Green}{\cmark} \\
    \rowcolor[rgb]{ .929,  .929,  .929} Han et al.~\cite{identifyads2018WPC} &       & \color{Green}{\cmark} & \cellcolor[rgb]{ 1,  1,  1}APKLancet~\cite{Apklancet2014ASIACCS} & \cellcolor[rgb]{ 1,  1,  1} & \cellcolor[rgb]{ 1,  1,  1}\color{Green}{\cmark} & \multicolumn{1}{l}{Salza et al~\cite{TPLsurvey2019}} &       & \color{Green}{\cmark} \\
    OSSPoLICE~\cite{OSSPOLICE2017CCS} &       & \color{Green}{\cmark} & \cellcolor[rgb]{ .929,  .929,  .929}Duet~\cite{Duet2014wisec} & \multicolumn{1}{l}{\cellcolor[rgb]{ .929,  .929,  .929}} & \cellcolor[rgb]{ .929,  .929,  .929}\color{Green}{\cmark} & \multicolumn{1}{l}{Salza et al.~\cite{Salza2018ICPC} } &       & \color{Green}{\cmark} \\
    \rowcolor[rgb]{ .929,  .929,  .929} LibD~\cite{LibD22018TSE} &       & \color{Green}{\cmark} & \cellcolor[rgb]{ 1,  1,  1}Madfraud~\cite{MadFraud2014mobisys} & \multicolumn{1}{l}{\cellcolor[rgb]{ 1,  1,  1}\color{Green}{\cmark}} & \cellcolor[rgb]{ 1,  1,  1} & \multicolumn{1}{l}{Ogawa et al.~\cite{CANDARW2018} } &       & \color{Green}{\cmark} \\
    LibScout~\cite{libscout2016ccs} &       & \color{Green}{\cmark} & \cellcolor[rgb]{ .929,  .929,  .929}DECAF~\cite{Decaf2014NSDI}  & \multicolumn{1}{l}{\cellcolor[rgb]{ .929,  .929,  .929}\color{Green}{\cmark}} & \cellcolor[rgb]{ .929,  .929,  .929} & \multicolumn{1}{l}{Derr et al.~\cite{Derr2017ccs}} &       & \color{Green}{\cmark} \\
    \rowcolor[rgb]{ .929,  .929,  .929} LibRadar~\cite{LibRadar2016ICSE} &       & \color{Green}{\cmark} & \cellcolor[rgb]{ 1,  1,  1}Dong et al.~\cite{Dong2018HotMibile} & \multicolumn{1}{l}{\cellcolor[rgb]{ 1,  1,  1}\color{Green}{\cmark}} & \cellcolor[rgb]{ 1,  1,  1} & \multicolumn{1}{l}{Ahasanuzzaman et al.~\cite{md2020studying}} & \color{Green}{\cmark} &  \\
    LibSift~\cite{LibSift2016soh} &       & \color{Green}{\cmark} & \cellcolor[rgb]{ .929,  .929,  .929}FraudDroid~\cite{FraudDroid2018FSE} & \multicolumn{1}{l}{\cellcolor[rgb]{ .929,  .929,  .929}\color{Green}{\cmark}} & \cellcolor[rgb]{ .929,  .929,  .929} & \multicolumn{1}{l}{Ahasanuzzaman et al.~\cite{ahasanuzzaman2020longitudinal}} & \color{Green}{\cmark} &  \\
    \rowcolor[rgb]{ .929,  .929,  .929} PEDAL~\cite{PEDAL2015MobiSys} & \multicolumn{1}{l}{\color{Green}{\cmark}} &       & \cellcolor[rgb]{ 1,  1,  1}Zhan et al.~\cite{splitads2017ACISP} & \cellcolor[rgb]{ 1,  1,  1} & \cellcolor[rgb]{ 1,  1,  1}\color{Green}{\cmark} & \multicolumn{1}{l}{MAdLens~\cite{jin2019madlens}} & \color{Green}{\cmark} &  \\
    AdDetect~\cite{AdDetect2014ISSNIP} & \multicolumn{1}{l}{\color{Green}{\cmark}} &       & \cellcolor[rgb]{ .929,  .929,  .929}FLEXDROID~\cite{FLEXDROID201NDSS} & \cellcolor[rgb]{ .929,  .929,  .929} & \cellcolor[rgb]{ .929,  .929,  .929}\color{Green}{\cmark} & \multicolumn{1}{l}{Gui et al.~\cite{Gui2017WhatAO}} & \color{Green}{\cmark} &  \\
    \rowcolor[rgb]{ .929,  .929,  .929} Wei et al.~\cite{price2016NDSS} & \multicolumn{1}{l}{\color{Green}{\cmark}}    &  & \cellcolor[rgb]{ 1,  1,  1}LibCage~\cite{Libcage2016ESORICS} & \cellcolor[rgb]{ 1,  1,  1} & \cellcolor[rgb]{ 1,  1,  1}\color{Green}{\cmark} & \multicolumn{1}{l}{Ullah et al.~\cite{wkshps2014}} & \color{Green}{\cmark} &  \\
    Son et al~\cite{son2016mobile}  & \multicolumn{1}{l}{\color{Green}{\cmark}} &       & \cellcolor[rgb]{ .929,  .929,  .929}ClickDroid~\cite{PEDAL2015MobiSys} & {\cellcolor[rgb]{ .929,  .929,  .929}\color{Green}{\cmark}} & \cellcolor[rgb]{ .929,  .929,  .929} & \multicolumn{1}{l}{Madscope~\cite{Madscope2015Mobisys}} & \color{Green}{\cmark} &  \\
    \rowcolor[rgb]{ .929,  .929,  .929} Pluto~\cite{pluto2016} &   \multicolumn{1}{l}{ \color{Green}{\cmark}}   &  & \cellcolor[rgb]{ 1,  1,  1}NativeGuard~\cite{NativeGuard2014Wisec} & \cellcolor[rgb]{ 1,  1,  1} & \cellcolor[rgb]{ 1,  1,  1} & \multicolumn{1}{l}{Book et al.~\cite{book2013spsm}} & \color{Green}{\cmark} &  \\
    Paturi et al.~\cite{paturi2015NDSS} &       & \color{Green}{\cmark} & \cellcolor[rgb]{ .929,  .929,  .929}COMPAC~\cite{COMPAC2014Wang} & \cellcolor[rgb]{ .929,  .929,  .929} & \cellcolor[rgb]{ .929,  .929,  .929}\color{Green}{\cmark} & \multicolumn{1}{l}{Tongaonkar et al.~\cite{PAM2013}} & \color{Green}{\cmark} &  \\
    \rowcolor[rgb]{ .929,  .929,  .929} Moonsamy et al.~\cite{privacyleak} &   \color{Green}{\cmark}    &  & \cellcolor[rgb]{ 1,  1,  1}AFrame~\cite{aframe2013ACSAC} & \cellcolor[rgb]{ 1,  1,  1} & \cellcolor[rgb]{ 1,  1,  1}\color{Green}{\cmark} & \multicolumn{1}{l}{Book et al.~\cite{MOST2013}} & \color{Green}{\cmark} &  \\
    Short et al.~\cite{privacyleak2014adhoc} &       & \color{Green}{\cmark} & \cellcolor[rgb]{ .929,  .929,  .929}SanAdBox~\cite{sanAdBox2013ICC} & \multicolumn{1}{l}{\cellcolor[rgb]{ .929,  .929,  .929}\color{Green}{\cmark}} & \cellcolor[rgb]{ .929,  .929,  .929} & \multicolumn{1}{l}{Vallina-Rodriguez et al.~\cite{vallina2012breaking}} & \color{Green}{\cmark} &  \\
    \rowcolor[rgb]{ .929,  .929,  .929} Leontiadis et al.~\cite{Leontiadis12HotMobile} &  \color{Green}{\cmark}     &  & \cellcolor[rgb]{ 1,  1,  1}AdDroid~\cite{AdDroid20120ASISCCS} & \multicolumn{1}{l}{\cellcolor[rgb]{ 1,  1,  1}\color{Green}{\cmark}} & \cellcolor[rgb]{ 1,  1,  1} & \multicolumn{1}{l}{Gui et al.~\cite{ICSE2015jiapping}} &   \color{Green}{\cmark}    &  \\
    Stevens et al.~\cite{Most2012} &   \color{Green}{\cmark}    &  & \cellcolor[rgb]{ .929,  .929,  .929}AdSplit~\cite{adsplit2012USENIX} & \multicolumn{1}{l}{\cellcolor[rgb]{ .929,  .929,  .929}\color{Green}{\cmark}} & \cellcolor[rgb]{ .929,  .929,  .929} & \multicolumn{1}{l}{Ruiz et al.~\cite{Mojia2014software}} &   \color{Green}{\cmark}    &  \\
    \rowcolor[rgb]{ .929,  .929,  .929} Droid-V~\cite{Watanabe2017MSR} &       & \color{Green}{\cmark} & \cellcolor[rgb]{ 1,  1,  1}Ruiz et al.~\cite{Israel2016software}  & \multicolumn{1}{l}{\cellcolor[rgb]{ 1,  1,  1}\color{Green}{\cmark}} & \cellcolor[rgb]{ 1,  1,  1} & \multicolumn{1}{l}{Li et al.~\cite{Lili2016SANER}} &       & \color{Green}{\cmark} \\
    Bauer et al.~\cite{structure2012ICSM}  &       & \color{Green}{\cmark} & \cellcolor[rgb]{ .929,  .929,  .929}Zhan et al.~\cite{libdetect2020ASE}  & \cellcolor[rgb]{ .929,  .929,  .929} & \cellcolor[rgb]{ .929,  .929,  .929} \color{Green}{\cmark} &  Book et al.~\cite{adempirical2015compscience}     &   \color{Green}{\cmark}    &  \\

    \rowcolor[rgb]{ .929,  .929,  .929} Salza et al.~\cite{TPLsurvey2019} &       & \color{Green}{\cmark} & \cellcolor[rgb]{ 1,  1,  1}Brahmastra~\cite{bhoraskar2014brahmastra}  & \cellcolor[rgb]{ 1,  1,  1} & \cellcolor[rgb]{ 1,  1,  1} \color{Green}{\cmark} &       &       &  \\

    \bottomrule
    \end{tabular}%
  \label{tbl:ad&nonad}%
  \vspace{-2ex}
\end{table*}%

\begin{figure}[t]
	\centering
\includegraphics[scale=0.52]{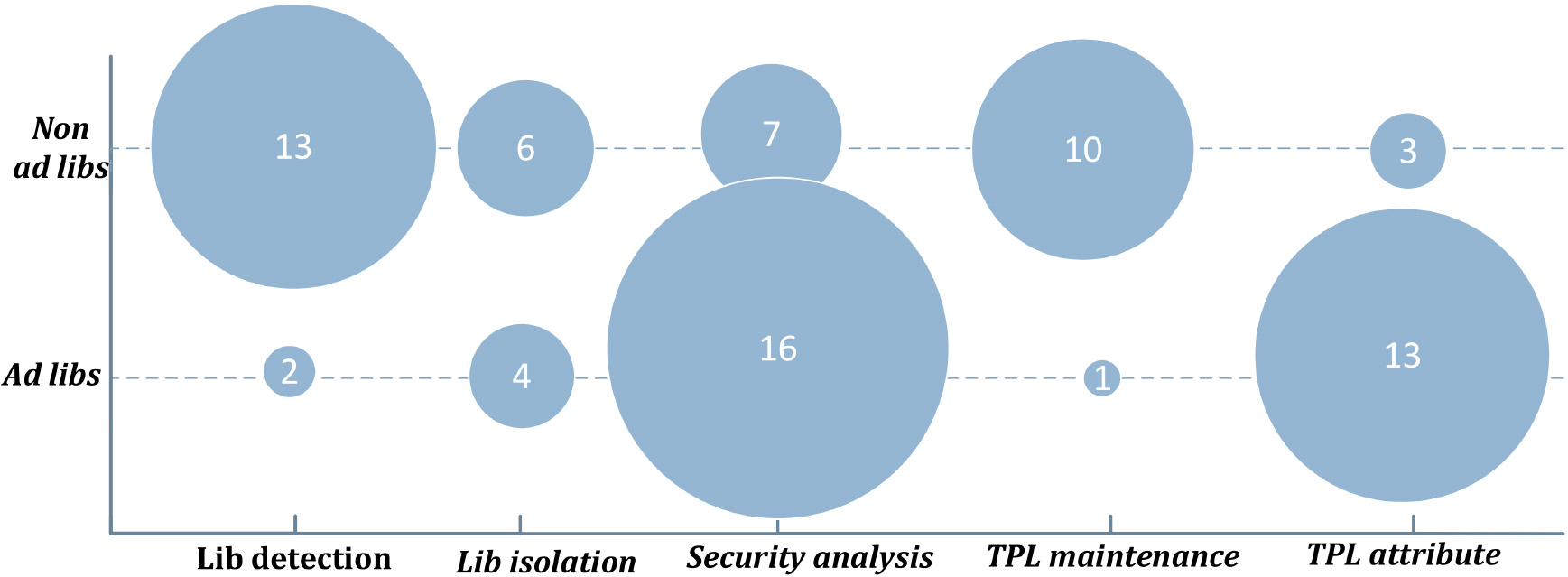}
	\caption{Distribution of ad/non-ad related-papers from our collected paper repository (Note that the total number is 75 because PEDAL belongs to both TPL detection and Lib isolation)}
	\label{fig:bulb}
	\vspace{-3ex}
\end{figure}

Fig.~\ref{fig:bulb} is a bulb graph that illustrates the number of existing research publications related to ad/non-ad libraries.
As can be seen from Fig.~\ref{fig:bulb}, ad libraries are the main target for both security-related research and attribute analysis.
For security and privacy analysis (third column), we find that ad libraries became the primary targets for many adversaries, which account for 70\% (16/23) of the whole research on security. This is because malicious developers can redirect the ad account information to gain illegal revenues, and thus many adversaries conduct various attacks on ad libraries.
Besides, ad libraries usually can collect targeted information from users and push the corresponding content to users base on the collected information~\cite{privacyleak,MadFraud2014mobisys,MAdLens2018infocom,privacyleak2014adhoc}. Therefore, that also attracts many researchers to focus on these problems. 
%the UI design of ad libraries can also more or less affect users' experience~\cite{jin2019madlens}, which can directly affect the rating of an app. This situation has led to a large number of 
%studies on these issues. Apart from above, we also can find many research focuses on ad attribute analysis. 
As for the TPL attribute analysis, ad libraries have some typical characteristics, such as the GUI and pushed content. Many interesting points still deserve to dig deep. Compared with other types of TPLs, UI design of ad libraries can affect users' experience~\cite{jin2019madlens}, which can directly affect the rating of an app. Existing research analyzed the traffic consumption~\cite{PAM2013}, display frequency, timing and location~\cite{Gui2017WhatAO}, the effects on apps, display contents~\cite{liu2020maddroid}.
In contrast, we can find that research w.r.t library detection and maintenance usually tend to not distinguish the types of TPLs; that can make sense by considering their research goals. It is more valuable to analyze all types of TPLs for these studies.

\subsection{Type of TPLs}
{This dimension classifies existing research based on different type of TPLs.
Android TPLs can be realized by different languages, such as the most widely used third-party libraries implemented by Java or Kotlin~\cite{Kotlin}, native libraries implemented with C/C++, some libraries with respect to GUI plugins implemented by multiple languages (e.g., Java and JavaScript). A specific languages often have their own unique characteristics. For instance, the reflection mechanism is not available in C language. The extracted CFGs from TPLs developed by C language before and after optimization are quite different. Java has package managers, it can easily import the third-party dependencies but C/C++ has not such a package manager.
For different languages, 
detection techniques and analysis approaches are usually different. 
For native libraries, researchers usually adopt the binary code to conduct the analysis and usually choose the string constants as the feature. For java library analysis, researchers usually change the Java bytecode into a appropriate intermediate representation(IR) to conduct the subsequent analysis. 
Thus, some approaches designed for one type of TPLs could not support other type of libraries.
Based on this dimension, we can understand current research trend and think about the future research direction.
}

\subsection{Type of Program Analysis}

{This dimension classifies existing TPL-related studies based on the type of program analysis. The type of program analysis employed in TPL-related research could be static and dynamic.
Static analysis can help us understand some TPL structure and code features, which is usually used in TPL identification and security analysis.
Dynamic analysis can detect the runtime behaviors, which can capture some features that static analysis cannot, such as the dynamic loaded malicious code, managing some runtime privileges.
The two analysis approaches are complementary to each other. Static analysis is usually more effective, and dynamic can capture the dynamic interaction behaviors and more accurate. For this categorization, we analyze existing research from three sub-dimensions: i.e., static analysis, dynamic analysis, and hybrid analysis.}

% Table generated by Excel2LaTeX from sheet 'SDH'
\begin{table*}[t]
  \centering
  \caption{The categorization of collected papers based on the type of program analysis}
  \vspace{-1ex}
    \begin{tabular}{llcllrrlrrrr}
    \toprule
    \textbf{Tool} & \multicolumn{1}{c}{\textbf{S}} & \textbf{D} & \multicolumn{1}{c}{\textbf{H}} & \textbf{Tool} & \multicolumn{1}{c}{\textbf{S}} & \multicolumn{1}{c}{\textbf{D}} & \multicolumn{1}{c}{\textbf{H}} & \multicolumn{1}{l}{\textbf{Tool}} & \multicolumn{1}{c}{\textbf{S}} & \multicolumn{1}{c}{\textbf{D}} & \multicolumn{1}{c}{\textbf{H}} \\
    \midrule
    \rowcolor[rgb]{ .929,  .929,  .929} LibDX~\cite{LibDX2020SANER} & \color{Green}{\cmark} &       &       & AdRisk~\cite{AdRisk2012Wisec} & \multicolumn{1}{l}{\color{Green}{\cmark}} &       &       & \multicolumn{1}{l}{LibHarmo~\cite{huang2020interactive}} & \multicolumn{1}{l}{\color{Green}{\cmark}} &       &  \\
    LibExtractor~\cite{LibExtractor2020wisec} & \color{Green}{\cmark} &       &       & MadDroid~\cite{liu2020maddroid} &       & \multicolumn{1}{l}{\color{Green}{\cmark}} &       & \multicolumn{1}{l}{DECCA~\cite{DC2018FSE}} & \multicolumn{1}{l}{\color{Green}{\cmark}} &       &  \\
    \rowcolor[rgb]{ .929,  .929,  .929} LibRoad~\cite{LibRoad2020TMC} & \color{Green}{\cmark} &       &       & MadLife~\cite{chen2019revisiting} &       & \multicolumn{1}{l}{\color{Green}{\cmark}} &       & \multicolumn{1}{l}{RIDDLE~\cite{RIDDLE2019ICSE}} & \multicolumn{1}{l}{\color{Green}{\cmark}} &       &  \\
    LibID~\cite{LibID2019issta}  & \color{Green}{\cmark} &       &       & Rastogi et al.~\cite{ad2016NDSS} &       &       & \color{Green}{\cmark} & \multicolumn{1}{l}{Wang et al.~\cite{Wang2020ICSME} } & \multicolumn{1}{l}{-} & \multicolumn{1}{l}{-} & \multicolumn{1}{l}{-} \\
    \rowcolor[rgb]{ .929,  .929,  .929} LibPecker~\cite{libpecker2018} & \color{Green}{\cmark} &       &       & LibFinder~\cite{LibFInder2016sp} & \multicolumn{1}{l}{\color{Green}{\cmark}} &       &       & \multicolumn{1}{l}{Yasumatsu et al.~\cite{Yasumatsu2019codaspy}} & \multicolumn{1}{l}{\color{Green}{\cmark}} &       &  \\
    ORLIS~\cite{ORLIS2018MOBILESoft} & \color{Green}{\cmark} &       &       & K\"{u}hnel et al.~\cite{kuhnel2015fast} & \multicolumn{1}{l}{\color{Green}{\cmark}} &       &       & \multicolumn{1}{l}{APPCOMMUNE~\cite{APPCOMMUNE2019SANER}} &       & \multicolumn{1}{l}{\color{Green}{\cmark}} &  \\
    \rowcolor[rgb]{ .929,  .929,  .929} Han et al.~\cite{identifyads2018WPC} & \color{Green}{\cmark} &       &       & APKLancet~\cite{Apklancet2014ASIACCS} & \multicolumn{1}{l}{\color{Green}{\cmark}} &       &       & \multicolumn{1}{l}{Salza et al~\cite{TPLsurvey2019}} & \multicolumn{1}{l}{-} & \multicolumn{1}{l}{-} & \multicolumn{1}{l}{-} \\
    OSSPoLICE~\cite{OSSPOLICE2017CCS} & \color{Green}{\cmark} &       &       & Duet~\cite{Duet2014wisec} & \multicolumn{1}{l}{\color{Green}{\cmark}} &       &       & \multicolumn{1}{l}{Salza et al.~\cite{Salza2018ICPC} } & \multicolumn{1}{l}{\color{Green}{\cmark}} &       &  \\
    \rowcolor[rgb]{ .929,  .929,  .929} LibD~\cite{LibD22018TSE} & \color{Green}{\cmark} &       &       & Madfraud~\cite{MadFraud2014mobisys} &       & \multicolumn{1}{l}{\color{Green}{\cmark}} &       & \multicolumn{1}{l}{Ogawa et al.~\cite{CANDARW2018} } &       & \multicolumn{1}{l}{\color{Green}{\cmark}} &  \\
    LibScout~\cite{libscout2016ccs} & \color{Green}{\cmark} &       &       & DECAF~\cite{Decaf2014NSDI}  &       & \multicolumn{1}{l}{\color{Green}{\cmark}} &       & \multicolumn{1}{l}{Derr et al.~\cite{Derr2017ccs}} & \multicolumn{1}{l}{-} & \multicolumn{1}{l}{-} & \multicolumn{1}{l}{-} \\
    \rowcolor[rgb]{ .929,  .929,  .929} LibRadar~\cite{LibRadar2016ICSE} & \color{Green}{\cmark} &       &       & Dong et al.~\cite{Dong2018HotMibile} &       & \multicolumn{1}{l}{\color{Green}{\cmark}} &       & \multicolumn{1}{l}{Ahasanuzzaman et al.~\cite{md2020studying}} & \multicolumn{1}{l}{\color{Green}{\cmark}} &       &  \\
    LibSift~\cite{LibSift2016soh} & \color{Green}{\cmark} &       &       & FraudDroid~\cite{FraudDroid2018FSE} &       & \multicolumn{1}{l}{\color{Green}{\cmark}} &       & \multicolumn{1}{l}{Ahasanuzzaman et al.~\cite{ahasanuzzaman2020longitudinal}} & \multicolumn{1}{l}{\color{Green}{\cmark}} &       &  \\
    \rowcolor[rgb]{ .929,  .929,  .929} PEDAL~\cite{PEDAL2015MobiSys} &       &       & \color{Green}{\cmark} & Zhan et al.~\cite{splitads2017ACISP} & \multicolumn{1}{l}{\color{Green}{\cmark}} &       &       & \multicolumn{1}{l}{MAdLens~\cite{jin2019madlens}} & \multicolumn{1}{l}{\color{Green}{\cmark}} &       &  \\
    AdDetect~\cite{AdDetect2014ISSNIP} & \color{Green}{\cmark} &       &       & FLEXDROID~\cite{FLEXDROID201NDSS} &       &       &       & \multicolumn{1}{l}{Gui et al.~\cite{Gui2017WhatAO}} & - & - & - \\
    \rowcolor[rgb]{ .929,  .929,  .929} Wei et al.~\cite{price2016NDSS} & -    & -     & -    & LibCage~\cite{Libcage2016ESORICS} &       & \multicolumn{1}{l}{\color{Green}{\cmark}} &       & \multicolumn{1}{l}{Ullah et al.~\cite{wkshps2014}} &       & \multicolumn{1}{l}{\color{Green}{\cmark}} &  \\
    Son et al~\cite{son2016mobile}  &       & \color{Green}{\cmark} &       & ClickDroid~\cite{cho2015empirical} &       & \multicolumn{1}{l}{\color{Green}{\cmark}} &       & \multicolumn{1}{l}{Madscope~\cite{Madscope2015Mobisys}} &       & \multicolumn{1}{l}{\color{Green}{\cmark}} &  \\
    \rowcolor[rgb]{ .929,  .929,  .929} Pluto~\cite{pluto2016} &       &       & \color{Green}{\cmark} & NativeGuard~\cite{NativeGuard2014Wisec} &       &       & \color{Green}{\cmark} & \multicolumn{1}{l}{Book et al.~\cite{book2013spsm}} & \multicolumn{1}{l}{\color{Green}{\cmark}} &       &  \\
    Paturi et al.~\cite{paturi2015NDSS} &       &       & \color{Green}{\cmark} & COMPAC~\cite{COMPAC2014Wang} & \multicolumn{1}{l}{\color{Green}{\cmark}} &       &       & \multicolumn{1}{l}{Tongaonkar et al.~\cite{PAM2013}} &       & \multicolumn{1}{l}{\color{Green}{\cmark}} &  \\
    \rowcolor[rgb]{ .929,  .929,  .929} Moonsamy et al.~\cite{privacyleak} &       &       & \color{Green}{\cmark} & AFrame~\cite{aframe2013ACSAC} & \multicolumn{1}{l}{\color{Green}{\cmark}} &       &       & \multicolumn{1}{l}{Book et al.~\cite{MOST2013}} & \multicolumn{1}{l}{\color{Green}{\cmark}} &       &  \\
    Short et al.~\cite{privacyleak2014adhoc} &       & \color{Green}{\cmark} &       & SanAdBox~\cite{sanAdBox2013ICC} &       & \multicolumn{1}{l}{\color{Green}{\cmark}} &       & \multicolumn{1}{l}{Vallina-Rodriguez et al.~\cite{vallina2012breaking}} & \multicolumn{1}{l}{\color{Green}{\cmark}} &       &  \\
    \rowcolor[rgb]{ .929,  .929,  .929} Leontiadis et al.~\cite{Leontiadis12HotMobile} &       & \color{Green}{\cmark} &       & AdDroid~\cite{AdDroid20120ASISCCS} &       &       & \color{Green}{\cmark} & \multicolumn{1}{l}{Gui et al.~\cite{ICSE2015jiapping}} & \multicolumn{1}{l}{-} & \multicolumn{1}{l}{-} & \multicolumn{1}{l}{-} \\
    Stevens et al.~\cite{Most2012} &       &       & \color{Green}{\cmark} & AdSplit~\cite{adsplit2012USENIX} &       &       & \color{Green}{\cmark} & \multicolumn{1}{l}{Ruiz et al.~\cite{Mojia2014software}} & \multicolumn{1}{l}{-} & \multicolumn{1}{l}{-} & \multicolumn{1}{l}{-} \\
    \rowcolor[rgb]{ .929,  .929,  .929} Droid-V~\cite{Watanabe2017MSR} & \color{Green}{\cmark} &       &       & Ruiz et al.~\cite{Israel2016software}  & \multicolumn{1}{l}{\color{Green}{\cmark}} &       &       & \multicolumn{1}{l}{Li et al.~\cite{Lili2016SANER}} & \multicolumn{1}{l}{\color{Green}{\cmark}} &       &  \\
    Bauer et al.~\cite{structure2012ICSM}  & \color{Green}{\cmark} &       &       & Zhan et al.~\cite{libdetect2020ASE}  & \multicolumn{1}{l}{-} & - & -     & \multicolumn{1}{l}{Book et al.~\cite{adempirical2015compscience}} & \multicolumn{1}{l}{-} & \multicolumn{1}{l}{-} & \multicolumn{1}{l}{-} \\
   \rowcolor[rgb]{ .929,  .929,  .929} Salza et al.~\cite{TPLsurvey2019} & -    & -    & -    & Brahmastra~\cite{bhoraskar2014brahmastra} &       &       & \color{Green}{\cmark} &       &       &       &  \\
 
    \bottomrule
    \end{tabular}%
    	\begin{center}	
\textit{S: static analysis, D: dynamic analysis, H: hybrid analysis (static analysis \& dynamic analysis), - "not applicable"}
	\end{center}
  \label{tab:sdh}%
  \vspace{-1ex}
\end{table*}%

{TABLE~\ref{tab:sdh} presents the categorization of collected papers based on the type of program analysis.
Based on the TABLE~\ref{tab:sdh}, we can find some patterns. For example, 1) all TPL detection tools adopt static analysis methods to identify the in-app TPLs. %Based on our conjecture, we consider that 
There are two main reasons: the first one is that
it is difficult to capture some dynamic features of some TPLs; the code coverage rate of the dynamic analysis is also limited, which may lead to many false negatives. The second reason is that it is also difficult to find the exact version of TPLs by using dynamic analysis. 2) All ad fraud detections use dynamic analysis. 3) For TPL attributes analysis, researchers usually employ empirical study, case study, and user study. The research approaches also adopt interview and statistical analysis. Thus, the program analysis method is not suitable for most of these studies. 4) When it comes to extracting graphical interfaces, UI states, traffic features, it usually needs to adopt the dynamic analysis, such as MadDroid~\cite{liu2020maddroid}, Ullah et al.~\cite{wkshps2014}, and MadLife~\cite{chen2019revisiting}. 5) Most privacy leakage detections and malicious TPL detections usually use dynamic analysis or hybrid analysis.
}

\subsection{Summary}

{Based on the taxonomy, we calculate the distribution of collected papers in each dimension. The specific results can be seen from TABLE~\ref{tbl:dis_tax}. For research objectives, most of the research focused on security issues analysis on TPLs, which accounted for 31\% of the total collected papers. The proportion of TPL maintenance is the least. TPL detection plays an essential role for downstream tasks; many research such as the vulnerable TPL identification, TPL isolation may involve in TPL detection. Thus, the total number of papers in each category is greater than the total number of collected papers.
For dimension with respect to targeted libraries, we can find that nearly half of the research is focused on the ad libraries. For the dimension regarding the type of TPLs, we can find that existing studies usually concentrate on Java libraries; the native libraries and other libraries that are written by other languages are seldom thoroughly explored. We only find two papers can handle native libraries, i.e., OSSPOLICE~\cite{OSSPOLICE2017CCS} and NativeGuard~\cite{NativeGuard2014Wisec}. OSSPOLICE is a TPL detection tool that can identify Java and native libraries. NativeGuard is the first work that isolates the native libraries from the host app.
Only one paper can solve the cross-language TPLs, i.e., LibDX~\cite{LibDX2020SANER}.
We encourage future researchers to fill this gap by exploring more of these TPLs instead of only focus on Java libraries. As for the method of program analysis, most of the studies adopt static analysis. Many open challenges such as dynamic loading and reflection still have not well-studied.}

% \usepackage{multirow}
\begin{table}[ht]
  \centering
  \caption{The distribution of collected papers in different dimensions}
  \vspace{-1ex}
\begin{tabular}{|c||c|c|c|}
\hline
\textbf{Dimension} & \textbf{Sub-dimension} & \textbf{\#} & \textbf{\%} \\ \hline
\hline
\multirow{5}{*}{\begin{tabular}[c]{@{}c@{}}Research \\ \\ Objectives\end{tabular}} & TPL detection & 15 & 20\% \\ \cline{2-4} 
 & \begin{tabular}[c]{@{}c@{}}Security issue \\ analysis\end{tabular} & 23 & 31\% \\ \cline{2-4} 
 & \begin{tabular}[c]{@{}c@{}}TPL privilege\\ de-escalation\end{tabular} & 10 & 14\% \\ \cline{2-4} 
 & TPL maintenance & 11 & 15\% \\ \cline{2-4} 
 & \begin{tabular}[c]{@{}c@{}}TPL attribute\\ understanding\end{tabular} & 16 & 22\% \\ \hline
\multirow{2}{*}{\begin{tabular}[c]{@{}c@{}}Targeted \\ Libraries\end{tabular}} & Ad libraries & 35 & 47\% \\ \cline{2-4} 
 & No-ad libraries & 39 & 53\% \\ \hline
\multirow{3}{*}{\begin{tabular}[c]{@{}c@{}}Type \\ of TPLs\end{tabular}} & Java libraries & 71 & 96\% \\
 & Native libraries & 2 & 3\% \\
 & \begin{tabular}[c]{@{}c@{}}cross-paltform\\ languages\end{tabular} & 1 & 1\% \\ \hline
\multirow{3}{*}{\begin{tabular}[c]{@{}c@{}}Type of \\ program\\ analysis\end{tabular}} & Static analysis & 36 & 49\% \\
 & Dynamic analysis & 17 & 23\% \\
 & Hybrid analysis & 10 & 13\% \\ \hline
\end{tabular}
    	\begin{center}	
\textit{ \#: the number of corresponding papers, \%: the percentage of the selected papers }
	\end{center}
\label{tbl:dis_tax}
\vspace{-3ex}
\end{table}

\section{Review of TPL Research}
\label{sec:review}

In this part, we discuss different research on Android TPLs in detail based on the research objectives. {We use this categorization to organize the following content due to two reasons: 1) this categorization can completely cover our collected papers, 2) we can compare the advantages and disadvantages of existing methods because they target the same objectives.}
%aforementioned taxonomy in Fig.~\ref{fig:taxonomy}. %Section~\ref{sec:Taxonomy}. 

%###################################################################

\subsection{TPL Detection}
\label{sec:tplidentification}

In this section, we introduce the research background of TPL detection and {provide a brief description of current research. In particular, we}
%then give a brief statement of current research. We also 
provide a taxonomy of these state-of-the-art techniques from three different perspectives. We also summarize obfuscation-resilient capability and discuss the defects of existing tools.

	%we split six parts to introduce TPL detection. Section~\ref{sec:tpldetection:researchbackground} introduces current research status and brief background, Section~\ref{sec:tpldetection:existing problems} reports the limitations in previous methods. Section~\ref{sec:tpl_detection:current_research} presents an overview of state-of-art detection tools. Section~\ref{sec:tpldetection:methods} elaborates the methodology of these tools in depth. Section~\ref{sec:tpldetection:codeobfuscation} compares the capability of obfuscation-resilient. Section~\ref{sec:tpldetection:systemconnections} describes the relations between the state-of-the-art systems.}

\subsubsection{Research Background}
\label{sec:tpldetection:researchbackground}

Research such as repackaged app detection and mobile malware detection needs to first identify third-party libraries as these TPLs could be the noises of the host code during the detection process, which would decrease the accuracy of the results. License violation and TPL vulnerability detection require the specific in-app versions.
Prior research~\cite{PEDAL2015MobiSys} has shown that about 57\% of apps contain ad libraries.
Wang et al.~\cite{Wang2017ICSE-C} also pointed out that on average, more than 60\% of the code in an Android app belongs to TPLs.
CLANDroid~\cite{CLANDroid2016ICPC} showed that TPLs could affect the detection accuracy. 
Li et al.~\cite{Lili2016SANER} explained why TPLs could affect the detection results and give motivating examples.
As we can see, TPL detection has essential functions for downstream tasks.
Zhan et al.~\cite{xian2019saner} summarized the method of different repackaging systems on how to filter out TPLs. 
% They found that most repackaging detection techniques exploit whitelist-based method to filter out TPLs. 
% Wukong~\cite{Wukong2015issta} and PiggyApp~\cite{PiggyApp13CODASPY} are two repackaging detection systems and use the clustering-based method to remove TPLs.
They found that most repackaging detection techniques~\cite{MassVet2015chen,DroidEagle2015,DroidMOSS12CODASPY,Juxtapp13DIMVA,ViewDroid14,chen14ICSE,FSquaDRA2014IFIP,Andarwin2013ESORICS} exploit whitelist-based method to filter out TPLs, while Wukong~\cite{Wukong2015issta} and PiggyApp~\cite{PiggyApp13CODASPY} use clustering-based method.
Besides, other research such as TPL isolation also needs to identify in-app TPLs first~\cite{PEDAL2015MobiSys}.
Detecting TPLs in Android apps is evidently an essential task for many downstream research tasks.

In the beginning, %most of the TPL detection techniques are used as a branch of repackaging detection, malware detection to filter TPLs out.
most research uses the whitelist-based method to remove TPLs because it is relatively easy to conduct. However, the whitelist-based method is not very reliable, which has many inevitable shortcomings: 
\textit{1) It is hard to maintain a complete list of libraries.} 
% Since the whitelist-based method just collects the package names of TPLs to identify whether an app contains the listed TPLs or not, it is impossible to maintain a complete TPL list. 
Existing studies, such as ViewDroid~\cite{ViewDroid14}, MassVet~\cite{MassVet2015chen} only choose commonly-used TPLs as their whitelists to identify TPLs. Obviously, it is inevitable for this method to miss some TPLs.
\textit{2) Such a method cannot discover new TPLs.} Relying only on the collected whitelist to identify TPLs, this method fails to identify newly-emerged TPLs that are not included in the list. 
\textit{3) Such a method depends on the package name of TPLs, which is not resilient to code obfuscation such as package renaming and package flattening}. Given the limitations of the whitelist-based method, many researchers start to explore more effective methods to detect TPLs, leading to the emerging of independent TPL detection studies. These approaches pay more attention to the features of TPL themselves and attempt to extract unique features from TPLs,
% These methods usually extract unique birthmarks from libraries, 
such as the code semantic features, UI features, or string features, to identify TPLs. TPLs can be identified with the help of the machine learning-based methods or similarity comparison algorithms. 
% especially it cannot handle the code obfuscation.
%Given that TPLs are widely used in mobile apps and have a serious impact on the many detection results, many researchers start to work on library detection and take it as an independent research area.

\subsubsection{Existing Research}
\label{sec:tpl_detection:current_research}
We collected 14 research work on TPL detection, as shown in TABLE~\ref{tbl:libdetection}. 
%provides a list of the related work from our paper repository.
%Note that LibD2~\cite{LibD22018TSE} is an extension work of LibD~\cite{LibD2017ICSE}, so we discuss them together.
In previous research~\cite{libdetect2020ASE}, we have conducted a comprehensive comparison on 11 of the 14 studies from the perspective of practical usage and implementation performance. We also published a benchmark that can be used to evaluate TPL detection tools of Android. 
%\ling{In previous research~\cite{libdetect2020ASE}, we have conducted a comprehensive comparison on Java library detection tools of Android from the perspective of practical usage and implementation performance, including 11 of the 14 studies that are publicly available. }
%We also published a benchmark that can be used to evaluate TPL detection tools. 
%We mainly compare these publicly available tools from the practical usage and implementation performance in that paper. 
%For more details, please refer to our paper~\cite{libdetect2020ASE}. 
In this paper, we conduct a systematic survey on different TPL detection tools and also make a supplement to previous work.

\begin{table}[t]
	\vspace{-1pt}
	\centering
	
	\caption{A summary of lib detection techniques}
	\vspace{-2mm}
	\scalebox{0.91}{
	\begin{tabular}{llc}
		\toprule
		\textbf{Function}  &\textbf{Tool/First Author} & \textbf{Year} \\
		\midrule
		
		\multirow{11}{*}{\textbf{Lib Detection}}  
	              & 	LibDX~\cite{LibDX2020SANER} & 2020 \\
		                 &  LibExtractor~\cite{LibExtractor2020wisec}  & 2020 \\
		                 &  LibRoad~\cite{LibRoad2020TMC}  & 2020 \\
		
		                  & LibID~\cite{LibID2019issta}  & 2019 \\
		
	                  	& LibPecker~\cite{libpecker2018} & \multicolumn{1}{c}{2018} \\

		%                                & \multicolumn{1}{>{\columncolor{gray!20}}c}{$LibD^{2}$~\cite{LibD22018TSE}} & \multicolumn{1}{>{\columncolor{gray!20}}c}{2018} \\
		& \multicolumn{1}{l}{ORLIS~\cite{ORLIS2018MOBILESoft}} & \multicolumn{1}{c}{2018} \\
		& Han et al.~\cite{identifyads2018WPC} & 2018 \\
		%                                & \multicolumn{1}{>{\columncolor{gray!20}}c}{MOBSCANNER~\cite{MOBSCANNER2017ICSE-C}} & \multicolumn{1}{>{\columncolor{gray!20}}c}{2017} \\
		& OSSPoLICE~\cite{OSSPOLICE2017CCS} & 2017  \\
		& LibD~\cite{LibD2017ICSE,LibD22018TSE} & 2017  \\
		& \multicolumn{1}{l}{LibScout~\cite{libscout2016ccs}} & \multicolumn{1}{c}{2016} \\
		& \multicolumn{1}{l}{LibRadar~\cite{LibRadar2016ICSE}} & \multicolumn{1}{c}{2016} \\
		& LibSift~\cite{LibSift2016soh} & 2016 \\
		\midrule
		\textbf{Ad Detection + Isolation} & PEDAL~\cite{PEDAL2015MobiSys} & \multicolumn{1}{c}{2015} \\
		\midrule
		\textbf{Ad Detection}                     & AdDetect~\cite{AdDetect2014ISSNIP} & 2014 \\
		%\midrule
		%Repackaging Detection    &\multicolumn{1}{>{\columncolor{gray!20}}c}{ Wukong~\cite{Wukong2015issta}} & \multicolumn{1}{>{\columncolor{gray!20}}c}{2015} \\
		%+ Lib Detection             & PiggyAPP~\cite{PiggyApp13CODASPY} & 2013 \\

		\bottomrule
	\end{tabular}
  }  
	\label{tbl:libdetection}

	\vspace{-1ex}
\end{table}

As shown in TABLE~\ref{tbl:libdetection}, the research on TPL detection originated in 2014.
Both \textbf{PEDAL} and \textbf{AdDetect} are ad library identification tools that cannot specify other types of TPLs. Both of them first extract ad library-related features (e.g., APIs, permissions, UIs, traffic features) to construct the feature vectors and then use a {binary} classifier to distinguish ad libraries and non-ad libraries. %Another particular tool is LibSift that just separate different TPL candidates from the host apps but it cannot identify the specific TPLs as well. 
{\textbf{LibSift} is another special tool that can only separate different TPL candidates from host apps but cannot identify specific TPLs either.}
Hence, the real specific TPL identification {research} started from LibRadar (2016). 

\textbf{LibRadar} is a component of Wukong~\cite{Wukong2015issta}, a repackaged app detection {tool.}
Wukong {leverages} LibRadar to find in-app TPLs and filter them out because these in-app TPLs could be noises that affect the detection accuracy. The intuition behind LibRadar is that TPLs are widely-used by many apps. Thus, LibRadar can find these TPLs by clustering them without prior knowledge.
LibRadar constructs the TPL candidate instances based on the package hierarchy structures. And then, it extracts the Android APIs of each instance as the code features. These TPLs would be clustered into big groups because the same libraries are used by many different apps~\cite{Wukong2015issta}. 
Similar to LibRadar, \textbf{LibD} and \textbf{LibExtractor} also adopt the clustering-based method to construct the TPL feature database.
LibD pointed out that the method of LibRadar in library instance construction could be problematic. Researchers of LibD find that the package structure of different versions of the same TPLs could be different; using the package structure to construct the TPL candidates could lead to mis-identification. 
Therefore, LibD proposes a new method to adopt the package inclusion and inheritance relationship (named ``homogeny graph'') as the module decoupling features to improve the construction of in-app TPL instances.
LibExtractor uses six class dependency relations to construct the in-app TPL instances and encode the class dependency into the code features. LibExtractors also adopts a clustering algorithm to identify TPL components of large-scale input apps and then identifies malicious libraries.
Both \textbf{Han et al.~\cite{identifyads2018WPC}} and LibD proposed to adopt the opcode of basic block in control flow graph (CFG) as the code feature of TPLs, but unlike LibD, Han et al.~\cite{identifyads2018WPC} is a similarity comparison-based tool that first collects the TPL files in advance and then adopts the similarity comparison method to identify in-app TPLs.

\textbf{LibScout} is the first tool that claims it can pinpoint the precise library versions used in apps. 
By leveraging LibScout, researchers can detect known vulnerabilities in TPLs that are still used by apps. \textbf{ORLIS} and \textbf{LibScout} use the same code feature (fuzzy method signature, the specific definition can refer to the Section~\ref{sec:TPL_detect:tool cmp} ``\textbf{Extracted features}'') but different identification methods to identify in-app TPLs.
\textbf{OSSPoLICE} can detect the potential software license violations and the known vulnerabilities of in-app TPL versions. {OSSPoLICE first uses the string constants and fuzzy method signature as the first stage code feature to identify the potential TPLs. Among these potential TPLs, it uses the function centroid as the fine-grained feature to identify the specific TPL version.}

Researchers of \textbf{LibPecker} found that tools such as LibScout and ORLIS use a relaxed TPL profile (i.e., the code feature granularity is too coarse), leading to more false negatives. Hence, LibPecker attempts to improve the performance of TPL identification by adopting the internal class dependencies inside a TPL to generate more strict TPL signatures. Meanwhile, it introduces the adaptive class similarity threshold and weight class similarity score when calculating the TPL similarity to ensure better precision and recall. 
\textbf{LibID} is another library version identification tool by formulating the TPL identification problem into a binary integer programming models. LibID uses more semantic information that includes CFG and class dependency in feature extraction to improve the resiliency to code obfuscation.

To improve the detection efficiency, \textbf{LibRoad} adopts a combo strategy to identify in-app TPLs. For non-obfuscated parts of TPLs, LibRoad adopts the package name-based matching policy to identify the in-app TPLs; while for obfuscated parts of TPLs, LibRoad adopts the signature-based matching policy by comparing the features of in-app TPLs with the features in the TPL database. 
\textbf{LibDX} is a cross-platform open-source software detection tool. Unlike other tools, LibDx directly extracts code features from the binaries.
As can be seen that even if there are many TPL detection tools, they have different concerns and application scenarios and adopt different techniques in TPL detection. 
More details about each tool will be introduced in the following sections.

\subsubsection{Taxonomy}
\label{Sec:TPL_detect:taxonomy}

\begin{figure} [t]
	\centering
%	\includegraphics[scale=0.5]{fig/TPL_det_tax.pdf}\\
    \includegraphics[scale=0.5]{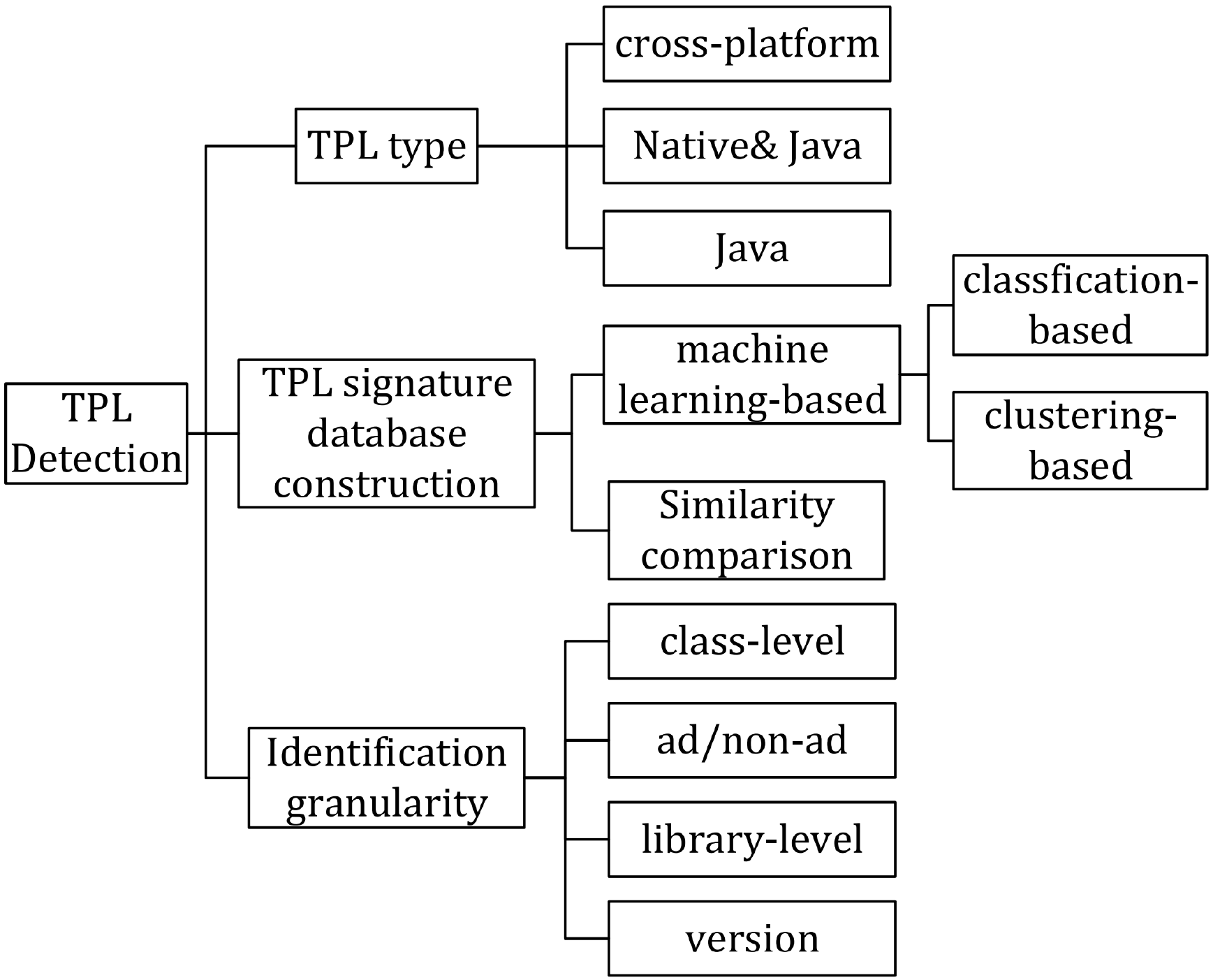}\\
	\vspace{-2mm}
	\caption{The taxonomy of TPL detection tools}
	\label{fig:tpl_det_tax}
	\vspace{-3ex}
\end{figure}

{We can classify these TPL detection tools from different perspectives. In this paper, we propose {three} classification schemes, as shown in Fig.~\ref{fig:tpl_det_tax}.
Based on the \textbf{detected type of TPLs}, we can further divide existing tools into three types: 1) cross-platform TPL detection tools, 2) native and Java library detection tools, and 3) Java library detection tools.
LibDX is the only cross-platform TPL detection tool that can detect various types of TPLs and is not just limited to Android. 
The remaining tools are Android TPL detection tools and most of them can only identify Java TPLs. 
OSSPoLICE can detect both native code libraries (C/C++) and Java Libraries. Other tools only can detect Java libraries.
Besides, three tools are special among these Java library detection tools, i.e., LibSift, AdDetect, and PEDAL. 
AdDetect and PEDAL only can distinguish the ad/non-ad libraries. 
LibSift only can identify %which parts belong to TPL modules 
{the parts belonging to TPL modules}, without reporting specific TPLs.
}

{Based on the \textbf{method of TPL feature database construction}, we can classify existing tools into two categories: 1) machine learning-based method and 2) similarity comparison-based method. 
The machine learning-based method can be sub-divided into two categories, the classification-based method, and the clustering-based method. 
AdDetect and PEDAL adopt the classification-based method to distinguish the ad/non-ad libraries.
%They first %separate the TPLs from the host app and then extract ad related features, such as API, UI, permissions and then 
%use classification algorithm to distinguish the ad/non-ad libraries. 
LibExtractor, LibD, and LibRadar are clustering-based tools. The intuition is that TPLs are used by many apps. Taking a considerable number of apps as input, the same TPLs will be clustered together. 
Therefore, clustering-based tools usually require millions of apps as input to ensure generating enough TPL signatures. 
Most existing tools are similarity comparison-based tools, which do not require a substantial number of apps as input but require developers to collect TPL files to construct the feature database. Developers use the same algorithm to generate the profile for in-app TPLs and feature database. By comparing the similarity value with the TPLs in the database, these tools can identify the in-app TPLs.
}

%\vspace{1mm}
%\noindent $\bullet$ \textbf{Identification Granularity}

%In fact, different TPL detection tools have different identification granularities. 
Based on the \textbf{identification granularity}, we divide current tools into four types, i.e., the class-level, ad/non-ad level, version-level, and library-level. ORLIS can report the classes that belong to TPLs. PEDAL and AdDetect only can distinguish the ad and non-ad libraries. The library-level identification means that the tool just can identify which TPL is used by apps but cannot determine the specific version. Version-level identification means that the tool not only can identify this TPL and also can specify the version. We give this classification based on whether the literature states it can detect version-level TPL or not. LibID, LibScout, and OSSPoLICE claim that they can identify the TPL at version-level. 
LibRoad, LibPecker, and \cite{identifyads2018WPC} mainly report the TPLs are presented in apps.
We can find that existing clustering-based tools only can identify TPL at the library-level.
In fact, it is challenging for clustering-based tools to identify the TPL at version-level. The code similarity of different versions of the same TPLs is various. 
%It is difficult for clustering algorithms to find perfect parameters to divide different TPLs into different clusters. Besides, the labeling process is also time-consuming and labor-intensive. The most essential problem is that this verification process cannot guarantee error-free. The clusters may include a single version or multiple versions. 
{It is difficult for clustering algorithms to find perfect parameters to divide different TPLs into different clusters. Each cluster may include a single version or multiple versions. Besides, the labeling process is also time-consuming, labor-intensive, and most importantly, error-prone.}

\subsubsection{State-of-the-art Techniques}
\label{sec:TPL_detect:tool cmp}

Basically, the TPL detection process can be summarized into four steps: 1) pre-processing, 2) library instance construction, 3) feature extraction, 4) library instance identification~\cite{libdetect2020ASE}.
The pre-processing mainly decompiles the input app and transforms the bytecode into an appropriate intermediate representation to facilitate subsequent processing. 
The library instance construction mainly implements the module decoupling algorithm to find the boundaries of different TPLs and constructs the TPL instance candidates. 
Then TPL detection tools can 
%adopt different program analysis techniques to extract the code features 
{extract different code features} for TPL instance candidates to represent them,
%and convert them into a kind of feature representation, 
such as the graph, hash values, or feature vectors, etc. 
The last step is library identification by using the similarity comparison techniques to compare the in-app TPL features with the features in the database to identify the specific TPLs/versions. %\ling{Comment: do all of them use similarity to identify TPLs in the last step?}

In Section~\ref{sec:tpl_detection:current_research}, we give a brief introduction about each TPL detection tool. To allow readers to better understand the existing TPL detection tools, we give a more detailed introduction about each tool by comparing their commonalities and differences based on the TPL detection process in this section, as shown in TABLE~\ref{tbl:tool_cmp}. 
LibDX is the only cross-platform TPL detection tool. LibDX extracts the read-only DATA segment (composed of string constants) of binaries and fuzzy filename as the code features. Then LibDX adopts a gene map method to implement the binary-to-binary match comparison.  
Because its approach is far from other TPL detection tools that focus on Java TPLs, we do not compare LibDX with other tools here.

\vspace{1mm}
\noindent$\bullet$ \textbf{ Pre-processing}

\noindent In TABLE~\ref{tbl:tool_cmp}, we also can see the connection of current TPL detection tools. 
%As we mentioned before, according to the difference of the TPL signature database construction method, TPL identification tools can be divided into two categories: the clustering-based method and the similarity comparison method. The classification-based method cannot identify the specific TPLs, and existing classification-based tools only can identify the ad and non-ad libraries. \tm{(what's the use of the frist half of this paragraph, seems unrelated to preprocessing)}
We can find that most tools choose Apktool~\cite{apktool} as the reverse-engineering tool that can keep the complete package structure of the decompiled code. Baksmali~\cite{baksmali} is another reverse-engineering tool, but it just work on the dex file directly. Researchers of LibRadar develop a decompilation tool (i.e., LIBDEX) for quickly handling the dex file, while its functions are similar to baksmali.
Compared with Apktool, baksmali is more efficient because it does not need to handle resource files. 
Androguard~\cite{Androguard} is used to extract the class dependency, and Soot~\cite{soot} is used to extract the CFG (control flow graph). %\tm{(consider deleting two 'usually' in this sentence)}

% \usepackage{multirow}
\settowidth\rotheadsize{\theadfont{LibExtractorrr}}
\begin{table*}[]
\centering
\caption{The comparison of existing state-of-the-art TPL detection tools}
\vspace{-1ex}
\begin{tabular}{ccccccccccccccc}
\toprule[1.1pt]
\multicolumn{1}{l}{} &  & \rothead{\textbf{LibExtractor}} & \rothead{\textbf{LibRoad}} & \rothead{\textbf{LibID}} & \rothead{\textbf{LibPecker}} & \rothead{\textbf{ORLIS}} & \rothead{\textbf{Han et al.}} & \rothead{\textbf{OSSPOLICE}} & \rothead{\textbf{LibD}} & \rothead{\textbf{LibScout}} & \rothead{\textbf{LibRadar}} & \rothead{\textbf{LibSift}} & \rothead{\textbf{PEDAL}} & \rothead{\textbf{AdDetect}} \\ 
\midrule[1pt]
\Xhline{0.3mm}
\multicolumn{1}{|c|}{\multirow{5}{*}{\textbf{\begin{tabular}[c]{@{}c@{}}Pre-processing\\ Tool\end{tabular}}}} & \multicolumn{1}{c|}{\textbf{Apktool}} & \multicolumn{1}{c|}{} & \multicolumn{1}{c|}{\color{Green}{\cmark}} & \multicolumn{1}{c|}{} & \multicolumn{1}{c|}{\color{Green}{\cmark}} & \multicolumn{1}{c|}{} & \multicolumn{1}{c|}{} & \multicolumn{1}{c|}{} & \multicolumn{1}{c|}{\color{Green}{\cmark}} & \multicolumn{1}{c|}{} & \multicolumn{1}{c|}{} & \multicolumn{1}{c|}{\color{Green}{\cmark}} & \multicolumn{1}{c|}{\color{Green}{\cmark}} & \multicolumn{1}{c|}{\color{Green}{\cmark}} \\ \cline{2-15} 
\multicolumn{1}{|c|}{} & \multicolumn{1}{c|}{\textbf{Androguard}} & \multicolumn{1}{c|}{} & \multicolumn{1}{c|}{} & \multicolumn{1}{c|}{\color{Green}{\cmark}} & \multicolumn{1}{c|}{\color{Green}{\cmark}} & \multicolumn{1}{c|}{} & \multicolumn{1}{c|}{\color{Green}{\cmark}} & \multicolumn{1}{c|}{} & \multicolumn{1}{c|}{\color{Green}{\cmark}} & \multicolumn{1}{c|}{} & \multicolumn{1}{c|}{} & \multicolumn{1}{c|}{} & \multicolumn{1}{c|}{} & \multicolumn{1}{c|}{} \\ \cline{2-15} 
\multicolumn{1}{|c|}{} & \multicolumn{1}{c|}{\textbf{Soot}} & \multicolumn{1}{c|}{} & \multicolumn{1}{c|}{} & \multicolumn{1}{c|}{} & \multicolumn{1}{c|}{} & \multicolumn{1}{c|}{\color{Green}{\cmark}} & \multicolumn{1}{c|}{} & \multicolumn{1}{c|}{\color{Green}{\cmark}} & \multicolumn{1}{c|}{} & \multicolumn{1}{c|}{\color{Green}{\cmark}} & \multicolumn{1}{c|}{} & \multicolumn{1}{c|}{} & \multicolumn{1}{c|}{} & \multicolumn{1}{c|}{} \\ \cline{2-15} 
\multicolumn{1}{|c|}{} & \multicolumn{1}{c|}{\textbf{baksmali}} & \multicolumn{1}{c|}{\color{Green}{\cmark}} & \multicolumn{1}{c|}{} & \multicolumn{1}{c|}{} & \multicolumn{1}{c|}{} & \multicolumn{1}{c|}{} & \multicolumn{1}{c|}{} & \multicolumn{1}{c|}{} & \multicolumn{1}{c|}{} & \multicolumn{1}{c|}{} & \multicolumn{1}{c|}{} & \multicolumn{1}{c|}{} & \multicolumn{1}{c|}{} & \multicolumn{1}{c|}{} \\ \cline{2-15} 
\multicolumn{1}{|c|}{} & \multicolumn{1}{c|}{\textbf{LIBDEX}} & \multicolumn{1}{c|}{} & \multicolumn{1}{c|}{} & \multicolumn{1}{c|}{} & \multicolumn{1}{c|}{} & \multicolumn{1}{c|}{} & \multicolumn{1}{c|}{} & \multicolumn{1}{c|}{} & \multicolumn{1}{c|}{} & \multicolumn{1}{c|}{} & \multicolumn{1}{c|}{\color{Green}{\cmark}} & \multicolumn{1}{c|}{} & \multicolumn{1}{c|}{} & \multicolumn{1}{c|}{} \\ \Xhline{0.3mm}
\multicolumn{1}{|c|}{\multirow{4}{*}{\textbf{\begin{tabular}[c]{@{}c@{}}Module \\ Decoupling\\ Feature\end{tabular}}}} & \multicolumn{1}{c|}{\textbf{Package Structure}} & \multicolumn{1}{c|}{} & \multicolumn{1}{c|}{\color{Green}{\cmark}} & \multicolumn{1}{c|}{\color{Green}{\cmark}} & \multicolumn{1}{c|}{\color{Green}{\cmark}} & \multicolumn{1}{c|}{} & \multicolumn{1}{c|}{} & \multicolumn{1}{c|}{\color{Green}{\cmark}} & \multicolumn{1}{c|}{} & \multicolumn{1}{c|}{\color{Green}{\cmark}} & \multicolumn{1}{c|}{\color{Green}{\cmark}} & \multicolumn{1}{c|}{} & \multicolumn{1}{c|}{\color{Green}{\cmark}} & \multicolumn{1}{c|}{} \\ \cline{2-15} 
\multicolumn{1}{|c|}{} & \multicolumn{1}{c|}{\textbf{Homogeny Graph}} & \multicolumn{1}{c|}{} & \multicolumn{1}{c|}{} & \multicolumn{1}{c|}{} & \multicolumn{1}{c|}{} & \multicolumn{1}{c|}{} & \multicolumn{1}{c|}{} & \multicolumn{1}{c|}{} & \multicolumn{1}{c|}{\color{Green}{\cmark}} & \multicolumn{1}{c|}{} & \multicolumn{1}{c|}{} & \multicolumn{1}{c|}{} & \multicolumn{1}{c|}{} & \multicolumn{1}{c|}{} \\ \cline{2-15} 
\multicolumn{1}{|c|}{} & \multicolumn{1}{c|}{\textbf{PDG}} & \multicolumn{1}{c|}{} & \multicolumn{1}{c|}{} & \multicolumn{1}{c|}{} & \multicolumn{1}{c|}{} & \multicolumn{1}{c|}{} & \multicolumn{1}{c|}{} & \multicolumn{1}{c|}{} & \multicolumn{1}{c|}{} & \multicolumn{1}{c|}{} & \multicolumn{1}{c|}{} & \multicolumn{1}{c|}{\color{Green}{\cmark}} & \multicolumn{1}{c|}{} & \multicolumn{1}{c|}{\color{Green}{\cmark}} \\ \cline{2-15} 
\multicolumn{1}{|c|}{} & \multicolumn{1}{c|}{\textbf{Class Dependency}} & \multicolumn{1}{c|}{\color{Green}{\cmark}} & \multicolumn{1}{c|}{} & \multicolumn{1}{c|}{} & \multicolumn{1}{c|}{} & \multicolumn{1}{c|}{\color{Green}{\cmark}} & \multicolumn{1}{c|}{} & \multicolumn{1}{c|}{} & \multicolumn{1}{c|}{} & \multicolumn{1}{c|}{} & \multicolumn{1}{c|}{} & \multicolumn{1}{c|}{} & \multicolumn{1}{c|}{} & \multicolumn{1}{c|}{} \\ \Xhline{0.3mm}
\multicolumn{1}{|c|}{\multirow{7}{*}{\textbf{\begin{tabular}[c]{@{}c@{}}Extracted \\ Features\end{tabular}}}} & \multicolumn{1}{c|}{\textbf{Fuzzy Method Sig.}} & \multicolumn{1}{c|}{\textbf{}} & \multicolumn{1}{c|}{\textbf{\color{Green}{\cmark}}} & \multicolumn{1}{c|}{\textbf{\color{Green}{\cmark}}} & \multicolumn{1}{c|}{\textbf{}} & \multicolumn{1}{c|}{\textbf{\color{Green}{\cmark}}} & \multicolumn{1}{c|}{\textbf{}} & \multicolumn{1}{c|}{\textbf{\color{Green}{\cmark}}} & \multicolumn{1}{c|}{\textbf{}} & \multicolumn{1}{c|}{\textbf{\color{Green}{\cmark}}} & \multicolumn{1}{c|}{\textbf{}} & \multicolumn{1}{c|}{\textbf{}} & \multicolumn{1}{c|}{\textbf{}} & \multicolumn{1}{c|}{\textbf{}} \\ \cline{2-15} 
\multicolumn{1}{|c|}{} & \multicolumn{1}{c|}{\textbf{CFG}} & \multicolumn{1}{c|}{\textbf{}} & \multicolumn{1}{c|}{\textbf{}} & \multicolumn{1}{c|}{\textbf{\color{Green}{\cmark}}} & \multicolumn{1}{c|}{\textbf{}} & \multicolumn{1}{c|}{\textbf{}} & \multicolumn{1}{c|}{\textbf{\color{Green}{\cmark}}} & \multicolumn{1}{c|}{\textbf{}} & \multicolumn{1}{c|}{\textbf{\color{Green}{\cmark}}} & \multicolumn{1}{c|}{\textbf{}} & \multicolumn{1}{c|}{\textbf{}} & \multicolumn{1}{c|}{\textbf{}} & \multicolumn{1}{c|}{\textbf{}} & \multicolumn{1}{c|}{\textbf{}} \\ \cline{2-15} 
\multicolumn{1}{|c|}{} & \multicolumn{1}{c|}{\textbf{APIs}} & \multicolumn{1}{c|}{\textbf{}} & \multicolumn{1}{c|}{\textbf{}} & \multicolumn{1}{c|}{\textbf{}} & \multicolumn{1}{c|}{\textbf{}} & \multicolumn{1}{c|}{\textbf{}} & \multicolumn{1}{c|}{\textbf{}} & \multicolumn{1}{c|}{\textbf{}} & \multicolumn{1}{c|}{\textbf{}} & \multicolumn{1}{c|}{\textbf{}} & \multicolumn{1}{c|}{\textbf{\color{Green}{\cmark}}} & \multicolumn{1}{c|}{\textbf{}} & \multicolumn{1}{c|}{\textbf{\color{Green}{\cmark}}} & \multicolumn{1}{c|}{\textbf{\color{Green}{\cmark}}} \\ \cline{2-15} 
\multicolumn{1}{|c|}{} & \multicolumn{1}{c|}{\textbf{Class Dependency}} & \multicolumn{1}{c|}{\textbf{\color{Green}{\cmark}}} & \multicolumn{1}{c|}{\textbf{}} & \multicolumn{1}{c|}{\textbf{\color{Green}{\cmark}}} & \multicolumn{1}{c|}{\textbf{\color{Green}{\cmark}}} & \multicolumn{1}{c|}{\textbf{}} & \multicolumn{1}{c|}{\textbf{}} & \multicolumn{1}{c|}{\textbf{}} & \multicolumn{1}{c|}{\textbf{}} & \multicolumn{1}{c|}{\textbf{}} & \multicolumn{1}{c|}{\textbf{}} & \multicolumn{1}{c|}{\textbf{}} & \multicolumn{1}{c|}{\textbf{}} & \multicolumn{1}{c|}{\textbf{}} \\ \cline{2-15} 
\multicolumn{1}{|c|}{} & \multicolumn{1}{c|}{\textbf{CFG Centroid}} & \multicolumn{1}{c|}{\textbf{}} & \multicolumn{1}{c|}{\textbf{}} & \multicolumn{1}{c|}{\textbf{}} & \multicolumn{1}{c|}{\textbf{}} & \multicolumn{1}{c|}{\textbf{}} & \multicolumn{1}{c|}{\textbf{}} & \multicolumn{1}{c|}{\textbf{\color{Green}{\cmark}}} & \multicolumn{1}{c|}{\textbf{}} & \multicolumn{1}{c|}{\textbf{}} & \multicolumn{1}{c|}{\textbf{}} & \multicolumn{1}{c|}{\textbf{}} & \multicolumn{1}{c|}{\textbf{}} & \multicolumn{1}{c|}{\textbf{}} \\ \cline{2-15} 
\multicolumn{1}{|c|}{} & \multicolumn{1}{c|}{\textbf{Permission,component,UI}} & \multicolumn{1}{c|}{\textbf{}} & \multicolumn{1}{c|}{\textbf{}} & \multicolumn{1}{c|}{\textbf{}} & \multicolumn{1}{c|}{\textbf{}} & \multicolumn{1}{c|}{\textbf{}} & \multicolumn{1}{c|}{\textbf{}} & \multicolumn{1}{c|}{\textbf{}} & \multicolumn{1}{c|}{\textbf{}} & \multicolumn{1}{c|}{\textbf{}} & \multicolumn{1}{c|}{\textbf{}} & \multicolumn{1}{c|}{\textbf{}} & \multicolumn{1}{c|}{\textbf{\color{Green}{\cmark}}} & \multicolumn{1}{c|}{\textbf{\color{Green}{\cmark}}} \\ \cline{2-15} 
\multicolumn{1}{|c|}{} & \multicolumn{1}{c|}{\textbf{Strings}} & \multicolumn{1}{c|}{\textbf{}} & \multicolumn{1}{c|}{\textbf{}} & \multicolumn{1}{c|}{\textbf{}} & \multicolumn{1}{c|}{\textbf{}} & \multicolumn{1}{c|}{\textbf{}} & \multicolumn{1}{c|}{\textbf{}} & \multicolumn{1}{c|}{\textbf{\color{Green}{\cmark}}} & \multicolumn{1}{c|}{\textbf{}} & \multicolumn{1}{c|}{\textbf{}} & \multicolumn{1}{c|}{\textbf{}} & \multicolumn{1}{c|}{\textbf{}} & \multicolumn{1}{c|}{\textbf{\color{Green}{\cmark}}} & \multicolumn{1}{c|}{\textbf{\color{Green}{\cmark}}} \\ \Xhline{0.3mm}
\multicolumn{1}{|c|}{\multirow{5}{*}{\textbf{\begin{tabular}[c]{@{}c@{}}Comparison\\ Method\end{tabular}}}} & \multicolumn{1}{c|}{\textbf{Similarity Comparison}} & \multicolumn{1}{c|}{\textbf{\color{Green}{\cmark}}} & \multicolumn{1}{c|}{} & \multicolumn{1}{c|}{} & \multicolumn{1}{c|}{} & \multicolumn{1}{c|}{} & \multicolumn{1}{c|}{\textbf{\color{Green}{\cmark}}} & \multicolumn{1}{c|}{} & \multicolumn{1}{c|}{\textbf{\color{Green}{\cmark}}} & \multicolumn{1}{c|}{} & \multicolumn{1}{c|}{\textbf{\color{Green}{\cmark}}} & \multicolumn{1}{c|}{-} & \multicolumn{1}{c|}{-} & \multicolumn{1}{c|}{-} \\ \cline{2-15} 
\multicolumn{1}{|c|}{} & \multicolumn{1}{c|}{\textbf{\begin{tabular}[c]{@{}c@{}}Fuzzy Class Match\\ (Adaptive Match)\end{tabular}}} & \multicolumn{1}{c|}{} & \multicolumn{1}{c|}{\textbf{\color{Green}{\cmark}}} & \multicolumn{1}{c|}{} & \multicolumn{1}{c|}{\textbf{\color{Green}{\cmark}}} & \multicolumn{1}{c|}{} & \multicolumn{1}{c|}{} & \multicolumn{1}{c|}{} & \multicolumn{1}{c|}{} & \multicolumn{1}{c|}{} & \multicolumn{1}{c|}{} & \multicolumn{1}{c|}{-} & \multicolumn{1}{c|}{-} & \multicolumn{1}{c|}{-} \\ \cline{2-15} 
\multicolumn{1}{|c|}{} & \multicolumn{1}{c|}{\textbf{Fuzzy Hash}} & \multicolumn{1}{c|}{} & \multicolumn{1}{c|}{} & \multicolumn{1}{c|}{} & \multicolumn{1}{c|}{} & \multicolumn{1}{c|}{\textbf{\color{Green}{\cmark}}} & \multicolumn{1}{c|}{} & \multicolumn{1}{c|}{} & \multicolumn{1}{c|}{} & \multicolumn{1}{c|}{} & \multicolumn{1}{c|}{} & \multicolumn{1}{c|}{-} & \multicolumn{1}{c|}{-} & \multicolumn{1}{c|}{-} \\ \cline{2-15} 
\multicolumn{1}{|c|}{} & \multicolumn{1}{c|}{\textbf{Hierarchical Indexing}} & \multicolumn{1}{c|}{} & \multicolumn{1}{c|}{} & \multicolumn{1}{c|}{} & \multicolumn{1}{c|}{} & \multicolumn{1}{c|}{} & \multicolumn{1}{c|}{} & \multicolumn{1}{c|}{\textbf{\color{Green}{\cmark}}} & \multicolumn{1}{c|}{} & \multicolumn{1}{c|}{} & \multicolumn{1}{c|}{} & \multicolumn{1}{c|}{-} & \multicolumn{1}{c|}{-} & \multicolumn{1}{c|}{-} \\ \cline{2-15} 
\multicolumn{1}{|c|}{} & \multicolumn{1}{c|}{\textbf{LSH}} & \multicolumn{1}{c|}{} & \multicolumn{1}{c|}{} & \multicolumn{1}{c|}{\textbf{\color{Green}{\cmark}}} & \multicolumn{1}{c|}{} & \multicolumn{1}{c|}{} & \multicolumn{1}{c|}{} & \multicolumn{1}{c|}{} & \multicolumn{1}{c|}{} & \multicolumn{1}{c|}{} & \multicolumn{1}{c|}{} & \multicolumn{1}{c|}{-} & \multicolumn{1}{c|}{-} & \multicolumn{1}{c|}{-} \\ \Xhline{0.3mm}
\multicolumn{1}{|c|}{\multirow{4}{*}{\textbf{\begin{tabular}[c]{@{}c@{}}Identification\\ Granularity\end{tabular}}}} & \multicolumn{1}{c|}{\textbf{Class-level}} & \multicolumn{1}{c|}{} & \multicolumn{1}{c|}{} & \multicolumn{1}{c|}{} & \multicolumn{1}{c|}{} & \multicolumn{1}{c|}{\textbf{\color{Green}{\cmark}}} & \multicolumn{1}{c|}{} & \multicolumn{1}{c|}{} & \multicolumn{1}{c|}{} & \multicolumn{1}{c|}{} & \multicolumn{1}{c|}{} & \multicolumn{1}{c|}{} & \multicolumn{1}{c|}{} & \multicolumn{1}{c|}{} \\ \cline{2-15} 
\multicolumn{1}{|c|}{} & \multicolumn{1}{c|}{\textbf{Ad Library}} & \multicolumn{1}{c|}{} & \multicolumn{1}{c|}{} & \multicolumn{1}{c|}{} & \multicolumn{1}{c|}{} & \multicolumn{1}{c|}{} & \multicolumn{1}{c|}{} & \multicolumn{1}{c|}{} & \multicolumn{1}{c|}{} & \multicolumn{1}{c|}{} & \multicolumn{1}{c|}{} & \multicolumn{1}{c|}{} & \multicolumn{1}{c|}{\textbf{\color{Green}{\cmark}}} & \multicolumn{1}{c|}{\textbf{\color{Green}{\cmark}}} \\ \cline{2-15} 
\multicolumn{1}{|c|}{} & \multicolumn{1}{c|}{\textbf{Library-level}} & \multicolumn{1}{c|}{\textbf{\color{Green}{\cmark}}} & \multicolumn{1}{c|}{\textbf{\color{Green}{\cmark}}} & \multicolumn{1}{c|}{} & \multicolumn{1}{c|}{\textbf{\color{Green}{\cmark}}} & \multicolumn{1}{c|}{} & \multicolumn{1}{c|}{} & \multicolumn{1}{c|}{} & \multicolumn{1}{c|}{\textbf{\color{Green}{\cmark}}} & \multicolumn{1}{c|}{} & \multicolumn{1}{c|}{\textbf{\color{Green}{\cmark}}} & \multicolumn{1}{c|}{} & \multicolumn{1}{c|}{} & \multicolumn{1}{c|}{} \\ \cline{2-15} 
\multicolumn{1}{|c|}{} & \multicolumn{1}{c|}{\textbf{Version-level}} & \multicolumn{1}{c|}{} & \multicolumn{1}{c|}{} & \multicolumn{1}{c|}{\textbf{\color{Green}{\cmark}}} & \multicolumn{1}{c|}{} & \multicolumn{1}{c|}{} & \multicolumn{1}{c|}{} & \multicolumn{1}{c|}{\textbf{\color{Green}{\cmark}}} & \multicolumn{1}{c|}{} & \multicolumn{1}{c|}{\textbf{\color{Green}{\cmark}}} & \multicolumn{1}{c|}{} & \multicolumn{1}{c|}{} & \multicolumn{1}{c|}{} & \multicolumn{1}{c|}{} \\ \Xhline{0.3mm}
\end{tabular}
\label{tbl:tool_cmp}
	\begin{center}
		\textit{ PDG: package dependency graph; LSH: Locality-Sensitive Hashing; CFG: Control Flow Graph, '-': not applicable; \\
		{~\color{Green}\cmark}: represents the use of a tool, a feature or a technique }
	\end{center}
\vspace{-2ex}
\end{table*}

\vspace{1mm}
\noindent$\bullet$ \textbf{Module Decoupling}

\noindent Considering the module decoupling, the features used by existing tools can be divided into four types: i.e., the package structure and package name, homogeny graph, package dependency graph (PDG), and class dependency. In fact, the homogeny graph and PDG also involve the package hierarchy structures.
Based on the TABLE~\ref{tbl:tool_cmp}, we can find most TPL detection tools adopt the package structure as the module decoupling features,  
{since Java uses packages to organize class files, and almost all existing detection tools focus on Java TPLs}. 
% Firstly, almost all existing tools focus on Java TPLs. Secondly, Java uses packages to organize class files. 
In general, an independent TPL usually corresponds to an independent package structure. However, when these TPLs are imported into apps, things will be more complicated. Some different TPLs may share the same root package when these TPLs are imported into an app. 
For instance, Google Android GMS~\cite{google_GMS} and Google Android Library~\cite{google_android} share the same root package ``com.google.android''.
Besides, TPL files can also depend on other TPLs, which also are called nested TPLs~\cite{centris2021ICSE}. %TPL clones in some literature. 
This type of TPLs usually has several parallel root packages, while these interdependent parts together constitute one TPL~\cite{ATVHunter2021ICSE}. If a tool only uses the package hierarchy structure as the module decoupling feature, it may generate incorrect TPL instances.
%divide the same TPL into different parts, leading to false negatives (the host TPL) and false positives (the invoked TPLs). \tm{bit confusing for reviewers}

LibD employs the \textit{Homogeny Graph} as a basic unit in TPL partition to construct TPL instance candidates. A homogeny graph is a directed graph, where each node indicates a package or a class file, and each edge denotes the nodes with inclusion or inheritance relations.

PDG is a weighted directed graph, which includes the package homogeny relationship (parent-child or sibling relationships among packages) and class dependency (e.g., field references, call relations and class inheritance). Existing tools (LibSift and AdDetect) adopt the hierarchical agglomerative clustering (HAC) by cutting a PDG into different modules. These separated modules will be treated as TPL candidates. Considering the accuracy of module decoupling, PDG can achieve better performance than that of homogeny graph because PDG considers the degree of correlation between different parts. This PDG-based method can effectively split modules that only exist the package inclusion dependency without code dependency.

However, homogeny graph and PDG sometimes cannot generate accurate TPL instances because different TPLs can share the same root package or even the sub-packages; different TPLs also can be nested in one package hierarchy tree within different level. If the nested package levels exceed a certain number, PDG also cannot effectively separate these modules.

Another commonly-used module decoupling feature is class dependency. Although ORLIS and LibExtractor adopt class dependency as the code feature, the class dependency relations they adopted are somewhat different. ORLIS just uses the call graph to construct the TPL candidates while LibExtractor adopts class inheritance and interface relations, function invocation relations, and field references as the decoupling features. The advantage of class dependency relations is that it does not depend on the package structures. Therefore this method is resilient to package flattening.

\vspace{1mm}
\noindent$\bullet$\textbf{ Extracted Features}

\noindent As can be seen from TABLE~\ref{tbl:tool_cmp}, although there are various TPL detection tools, we find that extracted features of existing techniques can be divided into five categories (i.e., fuzzy method signature, CFG, API call, class dependency, and function centroid). Note that some tools may use several code features to represent a TPL at the same time.

The fuzzy method signature~\cite{libscout2016ccs} means using a placeholder X to replace the developer-defined variables and types in a method signature. For example, \code{int methodA(classA, int, classB)} is a normal method signature, its fuzzy method signature is \code{int X(X, int, X)}. We need to use X to replace the developer-defined variable (i.e., methodA) and type (e.g., classA).
The purpose is to defend against renaming obfuscation. LibRoad and LibScout use fuzzy method signature as the only code features. Apart from the fuzzy method signatures, LibID and OSSPoLICE also employ other features in signature generation,{as mentioned in the following paragraphs}.

Both the tool~\cite{identifyads2018WPC} and LibD choose the opcode of the basic block of CFG (control flow graph) as the TPL code feature. They first build the CFG of each method and then extract the opcode of each basic block on the CFG. This feature is hashed as the method feature, and all of the method feature values are concatenated in a certain order as the class feature and each class feature is hashed again as the class-level feature. All class features are sorted by using a certain order according to the hash value and then these features are hashed again as the TPL features. LibID also extracts the CFG from the TPL instance candidate and uses all instructions of basic blocks as a part of a method signature. Besides, the complete method signature also includes the class access flag, superclass name, class interface, and fuzzy method signature. LibID adopts the Locality-Sensitive Hashing to calculate the compared pairs. It then uses the binary integer identification to determine whether a potential matched pairs has a corresponding class dependency.

LibRadar extracts the Android API calls, the total number of APIs and total kinds of APIs to construct the feature vector and calculate a hash value for each feature vector.

LibID uses the class dependency to judge whether the compared TPL instance has the corresponding class dependency with the potential matched TPL in the database. Each class signature of LibID does not contain the other dependency class features. %\tm{(does not contain the features of the classes it depends on), 'depended' might not be a word}
In contrast, LibPecker and LibExtractor encode the direct class dependency relations into each class signature. The signature of each class contains the dependent classes with class dependencies, method invocation, or field reference relations.
%The difference between LibExtractor and LibPecker is in the class dependency relations. LibPecker does not choose the interface relationship because some obfuscation tools can delete the interface class.(\tm{Maybe: 
The difference between LibExtractor and LibPecker is that LibPecker does not include the interface relationship in the class dependency relations because some obfuscation tools may delete the interface class.
%})

A function centroid~\cite{chen14ICSE} can be constructed via a deterministic traverse of the CFG. A function centroid is a three-dimensional vector composed of basic block index, outgoing degree, and loop depth. %OSSPoLICE uses the string constants and fuzzy method signature as the code features and hashes them to get the method signature. All method feature values are sorted and hashed again as the class signature. OSSPoLICE first employs the method of LibScout (string constants + fuzzy method signatures) to identify the potential in-app TPLs. By exploiting the function centroid to determine the specific versions. 
OSSPoLICE first uses string constants and fuzzy method signatures to identify the potential in-app TPLs, then determines the specific versions of TPLs by exploiting the function centroid.

Among these features, API calls and fuzzy method signatures only contain the syntactic information. The remaining code features contain both syntactic and semantic information at the same time, which can achieve better performance of resiliency to code obfuscation and adversarial attacks. However, such semantic features usually consume more computing resources. %The advantage of syntactic features usually can save more CPU and memory resources. \tm{('The advantage of syntactic features...' sentence seems a bit redundent) However, such semantic features usually consume more computing resources than syntactic features.}}

\vspace{1mm}
\noindent$\bullet$ \textbf{Comparison Method}

\noindent  Most existing tools just adopt a simple {similarity} comparison method to identify the in-app TPLs by comparing the features with the signatures in database. LSH and hierarchical indexing search are used to improve searching efficiency. Fuzzy hash can effectively handle the code obfuscation since it can tolerate some changes in the in-app TPLs caused by code obfuscation. Similar to the fuzzy hash, LibPecker introduces an adaptive class similarity threshold and weighted class similarity score to identify potential in-app TPLs. This method will give a higher weight value to more important classes, which can effectively distinguish different TPLs and improve the accuracy.

\subsubsection{Obfuscation-resilient Capability Comparison %\ling{why is it a subsection here? it seems it is not in the parallel position of other subsections?}
} 
\label{sec:tpldetection:codeobfuscation}

{Code obfuscation is often used to protect Android apps by hiding the actual logic of the apps as well as the used libraries. The commonly-used obfuscation strategies such as API hiding, control flow randomization, dead code removal can modify the code of in-app TPLs, which leads to the code features of in-app TPLs to be different from the original TPL files.
%Additionally, some TPL detection tools adopt the package structure as the supplementary features to construct the in-app TPL candidates. 
The capability of code obfuscation-resilience is one of the most important indexes to evaluate the performance of TPL detection tools.
Therefore, we {aim to investigate} the impact of code obfuscation on these state-of-the-art TPL detection tools.}
To achieve it, we summarize a comparative result towards common obfuscation techniques in Table~\ref{tbl:obfuscation_cmp}. Note that LibSift only implements the TPL separation without TPL identification, so we do not discuss the resiliency to code obfuscation here.

\settowidth\rotheadsize{\theadfont{corandomization}}
\begin{table}[t]
		\centering \scriptsize
		\caption{Comparison of obfuscation-resilient capability of different TPL detection systems}
		\vspace{-2mm}
		\scalebox{1.1}{
		\begin{tabular}{lccccc}
			\Xhline{1.2pt}
			& \rothead{\textbf{Dead code removal}} & \rothead{\textbf{Control-flow randomization}} & \rothead{\textbf{Identifier renaming}} & \rothead{\textbf{String encryption}} & \rothead{\textbf{Package flattening}} \\
			\midrule[1pt]
\rowcolor{gray!15}
		LibDX & \moon[scale=0.8]{7.2} & \moon[scale=0.8]{7.2} & \color{Green}\checkmark & \xmark & \xmark \\
LibExtractor & \moon[scale=0.8]{7.2} & \moon[scale=0.8]{7.2} & \color{Green}\checkmark & \color{Green}\checkmark & \moon[scale=0.8]{7.2} \\
\rowcolor{gray!15}
LibRoad & \xmark & \moon[scale=0.8]{7.2} & \color{Green}\checkmark & \color{Green}\checkmark & \moon[scale=0.8]{7.2} \\
LibID & \xmark & \xmark & \color{Green}\checkmark & \color{Green}\checkmark & \xmark \\
\rowcolor{gray!15}
LibPecker & \moon[scale=0.8]{7.2} & \moon[scale=0.8]{7.2} & \color{Green}\checkmark & \color{Green}\checkmark & \xmark \\
ORLIS & \color{Green}\checkmark & \moon[scale=0.8]{7.2} & \color{Green}\checkmark & \color{Green}\checkmark & \color{Green}\checkmark \\
\rowcolor{gray!15}
Han et al.~\cite{identifyads2018WPC} & \moon[scale=0.8]{7.2} & \xmark & \color{Green}\checkmark & \color{Green}\checkmark & \xmark \\
OSSPoLICE & \xmark & \xmark & \color{Green}\checkmark & \xmark & \xmark \\
\rowcolor{gray!15}
LibD & \moon[scale=0.8]{7.2} & \xmark & \color{Green}\checkmark & \color{Green}\checkmark & \xmark \\
LibScout & \xmark & \moon[scale=0.8]{7.2} & \color{Green}\checkmark & \color{Green}\checkmark & \xmark \\
\rowcolor{gray!15}
LibRadar & \xmark & \moon[scale=0.8]{7.2} & \color{Green}\checkmark & \color{Green}\checkmark & \xmark \\
PEDAL & \xmark & \moon[scale=0.8]{7.2} & \moon[scale=0.8]{7.2} & \xmark & \xmark \\
\rowcolor{gray!15}
AdDetect & \xmark & \moon[scale=0.8]{7.2} & \moon[scale=0.8]{7.2} & \xmark & \xmark  \\ 
			\bottomrule[1pt]
		\end{tabular}
		}
		\begin{center}
			\color{Green}{\checkmark} \color{black}: can; \xmark\ : cannot; \  \moon[scale=0.8]{7.2} : partially effective

		\end{center}
		\label{tbl:obfuscation_cmp}
		\vspace{-4mm}
\end{table}

\noindent $\bullet$ \textbf{Dead code removal} (a.k.a. code elimination), is able to delete the TPL code that is not invoked by the host app. When judging whether a tool is resilient to dead code removal, we mainly depend on two points: 1) the module decoupling features and 2) the extracted features. 
To sum up, if a tool is resilient to dead code removal, the constructed TPL instances should include at least the method invocation relations and the extracted features should include the invoked methods.

LibRoad, LibID, LibScout, and OSSPoLICE more or less adopt the fuzzy method signatures as the TPL signatures. However, fuzzy method signatures do not include the method invocation information. Some methods can be deleted in dead code removal so these tools are not resilient to dead code removal.
LibDX uses the read-only DATA in binary as the code feature and failed to consider the method call relationship. Therefore, LibDX also can be affected by dead code removal.
LibD and this tool~\cite{identifyads2018WPC} use the opcode of CFG as the code feature. CFGs include the semantic information of TPLs but some opcode can be deleted in dead code removal so LibD and \cite{identifyads2018WPC} are partially resilient to this obfuscation.

LibExtractor, LibID, LibPecker, and ORLIS all include the class dependencies in their code features but the role of class dependencies is slightly different. LibExtractor uses class dependencies to construct the TPL instances and encode the class dependencies into the TPL signature. LibID uses class dependencies in the match stage. LibPecker encodes the class dependencies in TPL signatures.
ORLIS just considers the call graph to construct the in-app TPL candidate so the code features do not contain the dead code. Besides, ORLIS just reports the class files that belong to the TPLs instead of a complete TPL. Therefore, ORLIS is resilient to dead code removal. However, in practice, the detection rate of ORLIS may decrease to some extent when it handles the obfuscated apps because the detection rate is also affected by the extracted feature granularity, comparison strategy and algorithm.

LibExtractor considers the interface class in feature extraction. However, the interface classes can be deleted by dead code removal. Therefore, dead code removal can affect code features of in-app TPLs, which may decrease the detection rate of LibExtractor.
Similar to LibExtractor, LibPecker also uses class dependencies as the code features but it does not consider the interface classes that can be deleted by some obfuscators. However, both LibPecker and LibID adopt the package structures in module decoupling stages, which leads to the TPL instances including irrelevant code (may including other TPLs or uncalled code), {hence reducing the resiliency to dead code removal}.
% \tm{hence reducing the resiliency}
%\tm{Maybe stress the impact}. 
%\tm{(Maybe change to hence reducing the ... or sth alike, and lose the newly-added sentence below.)}
%\tm{Even though encoding the class dependencies into code signature can achieve a good resiliency to code obfuscation, the consequence of using the package structure in module decoupling can lead to the incorrect TPL instances that have more impact on the final accuracy.}
Therefore, the resiliency to dead code removal of ORLIS is better than that of LibExtractor, and LibExtractor is better than LibID and LibPecker.

%TPL detection tools that use the fuzzy method signature (i.e., ORLIS and LibScout) are resilient to this kind of obfuscation technique since fuzzy methods rename the user-defined methods and variable types to generate the signature. Besides, LibPecker does not use APIs as basis of the identification process, thus it is also resilient to API hiding.
%	The remaining tools rely on extracting APIs to identify TPLs, thus will fail to identify TPLs that are hardened by API hiding.

\noindent $\bullet$\textbf{ Control flow randomization} usually involves modifying the original control flows. Therefore, such obfuscation method can directly affect TPL detection tools that rely on CFG as the code features, including LibID, OSSPoLICE, ~\cite{identifyads2018WPC} and LibD.
Both LibD and \cite{identifyads2018WPC} use the opcode of the basic block of CFG as the signature. OSSPoLICE extracts the CFG centroid as the fine-grained feature to identify specific versions. LibID employs the basic block signature as one of the TPL signatures. Besides, the modification of the CFG also could change method dependencies, therefore, it also affects tools {that extract the method invocation relations in feature generation (e.g., LibPecker, LibExtractor).} %\tm{(use two 'some' in one sentence seems vague, be specific or maybe change expressions)}

	%The general method of control flow obfuscation usually includes the inserting redundant control flow or flattening control flows.
	%	Such obfuscation method can affect TPL detection techniques that rely on CFG (e.g., LibD and  this system~\cite{identifyads2018WPC}.) Both of them depends on the CFG to extract the opcode sequence information. Therefore, they are  not resilient to control flow randomization. 

\noindent $\bullet$ \textbf{Identifier renaming} is also called renaming obfuscation, which can modify the identifiers such as the class name, method name, field name, variables into a meaningless string or hash value. PEDAL and AdDetect leverage the class name as one of the features to identify the classes that belong to ad libraries. As can be seen from TABLE~\ref{tbl:obfuscation_cmp}, apart from PEDAL and AdDetect, the remaining tools are resilient to renaming obfuscation.

\noindent $\bullet$ \textbf{String encryption} is to use the encryption algorithms to encrypt constant strings to protect sensitive information such as the URL, username, and email address. LibDX, OSSPoLICE, PEDAL, and AdDetect all adopt the constant strings as one of the features so the string encryption can discount their detection performance.

\noindent $\bullet$ \textbf{Package flattening} can change the package hierarchy structures and package names. Even worse, it can delete the whole package hierarchy structure.
We can find that most tools are not resilient to package flattening.
ORLIS is the only tool that does not depend on package hierarchy structures and package names in TPL identification so it is completely resilient to code obfuscation. LibRoad has two strategies to identify in-app TPLs. For TPLs with package flattening, LibRoad just adopts a simple method (whether the package is a single character or not) to judge whether this TPL is obfuscated. However, determining whether a TPL is obfuscated by the package flattening technique is non-trivial. The judgment method of LibRoad can lead to false negatives. LibExtractor extracts the relative path of the dependency class as the code feature and uses a placeholder A to replace each path segment. LibExtractor attempts to use this method to enhance the resiliency to code obfuscation. However, different developers may adopt different strategies to change the hierarchy structures, and even the same TPL can be obfuscated into different package patterns.%\tm{(even the same TPL can be applied to different obfuscation patterns)} 
Therefore, this method of LibExtractor can generate different code features for the same in-app TPLs.

{Our analysis finds tools that include more semantic code features and use class dependency to build TPL instances can achieve better resiliency to code obfuscation.}

%Tools such as LibPecker, Han et al.~\cite{identifyads2018WPC} and LibRadar can be affected since they all depend on the package hierarchy structure to identify libraries.

\begin{comment}

\revise{We can find that all of these systems are resilient to identifier renaming and string encryption, while none of them is resilient to visualization-based protection.
%
The system proposed by Han et al.~\cite{identifyads2018WPC} and LibD~\cite{LibD2017ICSE} depend on CFG to extract the code features. Therefore, they cannot defend against control-flow randomization. LibPecker, Han et al.~\cite{identifyads2018WPC}, LibRadar and LibScout require the package structure as the supplementary feature to identify TPLs. Therefore, they are not resilient to the package flattening technique. 
The obfuscation-resilient capability of LibSift remains unclear as it only implements TPL separation, without TPL identification. 
%
{LibRadar extracts all the Android APIs in each file as the code feature, thus is not resilient to API hiding and dead code removal. The code elimination can occur at both the class and field/method level, therefore, it can affect tools that choose the class and method/field as the code features, including LibPecker, LibD, this system~\cite{identifyads2018WPC} as well as LibScout.}}

\end{comment}

\subsubsection{Disadvantages Analysis}

{In this section, we summarize the disadvantages of existing tools, aiming to inspire future researchers to develop better tools.}

\vspace{1mm}
\noindent$\blacksquare$ \textbf{Disadvantages of Clustering-based Methods}

The intuition of clustering-based methods to detect TPLs is that TPLs are usually widely-used by many apps~\cite{Wukong2015issta}. One of the advantages of this approach is that we do not need to collect TPL files in advance; we can directly get the TPL features with clustering and identify in-app TPLs without prior knowledge. However, this method also has the following disadvantages:
1) \textit{The decision of clustering parameters.} Clustering algorithms often require developers to decide the number of clusters or set appropriate parameters to get the clustering results, while in fact deciding the parameters is very challenging. 
% and the version diversity of different TPLs are also various.
The code similarity of different versions of the same TPLs could be various. For example, 
% we employed a similarity algorithm to calculate the code similarity between different versions of the same TPL. 
we empirically find the code similarity of different versions of the TPL \textit{okio} ranges from about 80\% to nearly 100\%. While the code similarity of \textit{okhttp} 2.7.x and \textit{okhttp} 3.x only reaches less than 30\%.
% Without a doubt, the diversity of TPLs and versions adds the difficulty for clustering parameter decision. 
The version diversity makes the parameter decision of clustering-based algorithms difficult, and it is impossible to find the perfect parameters to get completely correct clustering results. 
The millions of input apps also add difficulties. Some clustering may contain a single TPL version, while others may contain several versions of the same TPL, or even contain different TPLs. 
2) \textit{The clustering result may contain impurities.} The quality of clustering-based results generally depend on the number of input apps and the usage rate of TPLs. 
Apart from the aforementioned cases, some situations also can lead to the results containing noises. For instance, if the input dataset includes several apps that are cloned many times, the clustering results may include the host apps instead of the TPLs.
3) \textit{Labor-intensive verification.} When we get the clustering-based results, we cannot directly use them before labeling the clusters.
Developers need to conduct verification to guarantee the clustering results correct. 
However, this process is labor-intensive and time-consuming. 
And the verification is also error-prone.% whether conducted manually or through an algorithm.
% Whether using manual verification or algorithm verification, both of verification cannot ensure 100\% correct.
 Based on the above-mentioned analysis, we can find that it is impossible for clustering-based methods to identify a specific in-app TPL version.
4) \textit{Identify new and niche TPLs.}  In fact, clustering-based methods only can identity commonly-used TPLs and may miss some niche and new TPLs, whose recall depends on the number of input apps and the reuse rate of TPLs. 
% Even if we can cluster the new and niche TPLs, due to lack of the GT, we still cannot ensure who they are. 
Additionally, the input apps could be out-of-date, and updating the new results is really labor-insensitive, which requires clustering and labeling again.

Apart from these common disadvantages of clustering-based methods, existing clustering tools also have their own shortcomings.

{LibExtractor generates class signatures by hashing the relative paths of their direct dependency classes. For each class, LibExtractor gets the relative path from this class to its dependency classes, and uses a placeholder A to replace the package name in this relative path link, to gain resiliency to the obfuscation of package renaming.
% The relative paths are hashed as the signatures.
% LibExtractor tries to gain resiliency to package renaming in this way. 
% Even the same TPL can generate different signatures by this method.
However, the package flattening obfuscation technique can change the hierarchy structures via customizing. Users can choose different packages to obfuscate and the hierarchy level also can be changed into different forms. Therefore, the same TPL can be transformed into different obfuscated hierarchy structures, leading to the mutation of relative paths for the same TPL. In short, the method of LibExtractor is not resilient to package flattening.}

LibRadar uses package hierarchy as the module decoupling feature. However, for in-app TPLs, an independent package may not correspond to an independent TPLs; 
% an in-app package tree could map to several independent TPLs. 
Several different TPLs may have the same root package and nested sub-packages. 
An independent TPLs also may have several parallel independent root packages due to the TPL dependency. Using the package structure as module decoupling features is error-prone, which could cut a complete TPL into different parts or cluster several TPLs as the one.

LibD also has the aforementioned issues, although it adopts homogeny graphs as the module decoupling features. Besides, LibD adopts the package-level hash values to detect the in-app TPLs. A small change could lead to the final signatures different. Many obfuscation techniques can easily change the code features, leading to false negatives.

\vspace{1mm}
\noindent$\blacksquare$ \textbf{Disadvantages of Similarity Comparison Methods}

%{Compared with clustering-based methods, similarity comparison methods have less disadvantages.
Similarity comparison methods require developers to collect the TPLs files to build a predefined feature database. Thus, the size of feature database can directly affect the recall of similarity comparison-based tools.
Apart from that, we find existing similarity comparison-based tools have other disadvantages.
%\tm{(Compared with clustering-based methods, similarity comparison methods have fewer defects in theory, only that similarity comparison methods require developers to collect the TPLs files to build a predefined feature database. Thus, the size of feature database can directly affect the recall of similarity comparison based tools. However, existing similarity-based tools all have their own defects.
%)}

{One of the obvious disadvantages for most existing similarity comparison tools is they more or less depend on the package structure to construct the TPL candidates. However, package structures are not stable. The package structure of different versions could mutate, and package flattening technique can easily change the package structures.
% Each independent package tree cannot map to a independent TPL. 
Sometimes an independent package tree could include several different TPLs. Such as the Google ads and Android TPL, they are two different TPLs but they have the same root package ``com.google.android''.}

LibDX is a cross-platform version detection tool that extracts the DATA read-only segment of binary files to identify different TPLs. However, these features may not be very effective against Java libraries, especially after obfuscation, such as string encryption.

Both LibID and LibPecker set too strong assumption on package hierarchy structures.
LibPecker assumes the package hierarchy information of a library is retained during the obfuscation~\cite{libpecker2018}. LibID assumes that the inter package hierarchy structure will not be changed during obfuscation~\cite{LibID2019issta}.
 They only can identify the in-app TPLs without inter hierarchy structure modification but package flattening can easily change the structures, which directly discounts their recalls.

{LibRoad adopts a combined strategy to identify in-app TPLs. For the non-obfuscated parts of TPLs, LibRoad uses the package name to match the potential TPLs. It assumes that each root package corresponds to a TPL. However, this assumption may not be valid in reality because some different TPLs can share the same root package and one TPL also can have different parallel root packages.
% Therefore, using this policy could lead to false negatives and false positives. 
Besides, LibRoad uses a simple strategy to judge whether the package names are obfuscated or not, which also can affect detection performance.}

{The code feature granularity of LibScout and LibRoad are too coarse. Both of them use the fuzzy hash method as the signature, however, the code features only include the syntactic information and cannot find some tiny changes of the inside method, such as the statement insertion and deletion. OSSPoLICE adopts a two-stage method to identify the in-app TPL versions. It first uses string constants and fuzzy method signature as the coarse-grained features to find the potential in-app TPLs. %\tm{(perhaps should maintain accordance with before about osspolice-libscout relationship)}. 
LibScout also uses the fuzzy method signatures as the signatures.
Therefore, OSSPoLICE inherits the shortcomings of LibScout. Additionally, OSSPoLICE adopts the CFG centroid as the fine-grained feature to ensure the specific version of potential TPLs. However, getting the circle loop of a CFG is 
time-consuming, especially for the method with rich functionalities, which is not good for large-scale analysis.}

{ORLIS just adopts the call relationship to construct the in-app TPL candidates, which could lose classes that exist in the inheritance relationship, field reference dependencies, etc., 
%lead it to miss classes that exist the inheritance relationship and field reference dependencies and so on, 
resulting in false negatives in identification. ORLIS just reports the matched classes, which is not practical for most users either.}

\vspace{1mm}
\noindent $\blacksquare$ \textbf{Summary}

Although different tools have different disadvantages, the most primary issue of existing tools is using package structures in TPL candidate construction, which directly leads to many false negatives and false positives.
{Apart from ORLIS, the remaining tools more or less depend on the package structure to conduct the module decoupling. Firstly, these tools are not resilient to package flattening obfuscation. If the TPLs are without the package structure, these tools cannot handle TPLs without packages that could cause false negatives. Secondly, using package structures to build the TPL candidates cannot handle the TPL dependency and different TPLs with nested packages.}

\subsubsection{Essential Findings}

{Based on above analyses, we summarize the essential findings in TPL identification.}

{$\bullet$ We find that most existing TPL detection tools more or less depend on the package structure to conduct the module decoupling. However, package structures are not stable, which can be easily obfuscated by code obfuscation.
Besides, an independent TPL also can include several parallel root packages. One package tree also can contain multiple TPLs.  
Therefore, package structure cannot accurately split the in-app TPL instances, leading to the low recall. TPL detection tools use the class dependencies as the module decoupling features can resiliency to package flattening because this feature does not depend on the package structure.}

$\bullet$ For TPL detection tools, the extracted features within more semantic features can achieve better resiliency to code obfuscation than tools only include the syntactic information. In the comparison stage, class-level features can achieve better resiliency to code obfuscation than package-level features.

$\bullet$ Most TPL detection tools have high precision but low recall. As for the version identification, we have a long way to go. 
We need to overcome many challenges, 
especially the diverse code differences among different versions of the same TPL. We find the code differences among some versions are huge and the code differences among some versions are very tiny. How to choose the feature granularity can effectively reflect the differences and does not affect the detection efficiency is what future researchers need to consider.

$\bullet$ Future researchers can consider handling the challenges, such as the TPL shrink and optimization.

\subsubsection{Implications and Future Work}

Even though current tools have various disadvantages, they also have many advantages in different detection steps. We think existing techniques are highly complementary to each other. By combining their advantages, we can design better tools. For instance, we can learn from the method of ORLIS and LibExtractor in module decoupling. We suggest using the class dependencies but does not include the interfaces to build TPL candidates, which can achieve better resiliency to package flattening. For feature extraction, we can refer to the method of LibPecker; each class signature includes their direct dependency classes. This method includes rich semantic information, which can achieve better resiliency to code obfuscation. For the comparison stage, we can adopt the strategy of LibRoad; regarding the non-obfuscate TPLs, we can first use the package name to narrow the search scope and use signature-based methods to locate the specific versions.

Based on our analysis, we find that all current tools adopt the static analysis method to identify in-app TPLs. Therefore, all of them cannot identify the dynamically-loaded TPLs and classes. Besides, they also are not resilient to the sophisticated code obfuscation, such as API hiding that is a kind of obfuscation by leverage the Java reflection and dynamic class load to hide a part of codes, visualization-based protection that translates the code into a  stream of pseudo-code bytes that is hard to be recognized by the machine and human and only can be interpreted during the runtime. None of these problems have a good solution. Hence, we suggest future researchers try to solve these problems.

On top of that, most research just focuses on the detection of Android Java TPLs while only two tools can detect native libraries but they are not resilient to string encryption, and just use simple syntactic features, leading to not resilient to other sophisticated code obfuscation. We highlight that future researchers pay more attention to the native library detection.

\subsection{TPL Security-related Issue Analysis}

{This section introduces research on existing TPL security-related issues.}

\subsubsection{Research Background}
TPL is a double-edged sword. On the one hand, it can facilitate app development and decrease the release time; on the other hand, TPLs could also bring various security risks. As previously mentioned, the permission-control mechanism of Android only works at the app-level; therefore, the TPLs and the main app share the same permissions. This permission model obviously violates the principle of least privilege~\cite{hammad2017ICSA,DelDroid201983HAMMAD}. Malicious TPLs could abuse the permissions of host apps and easily access privacy data and resources, resulting in data and privacy leakage. Besides, some serious vulnerabilities in TPLs are disclosed as Common Vulnerabilities and Exposures (CVEs). Researchers~\cite{libscout2016ccs} also found that developers usually do not replace the vulnerable TPLs in time, which aggravates the spread of vulnerabilities.
Besides, some attackers attempt to masquerade malicious libraries by modifying the package names as existing legal libraries. For example, DroidKungFu~\cite{droidkungfu} is a well-known malware, which uses names like ``com.google.update'' and ``com.google.ssearch'' to confuse users. These malicious libraries usually can leak users' private data, hijack the SNS account, read/send text messages, or lead to money loss.
Furthermore, some unscrupulous developers try to violate the ad library guidelines and develop ads that trick users into clicking or watching the in-app ads~\cite{MadFraud2014mobisys}. 
% Therefore, it is necessary to take a closer look at existing research on TPL security and privacy analysis.

Another security risk comes from the direct and transitive use of TPLs. 
Android apps are widely built on the top of TPLs; however, if the imported TPL includes a vulnerability and can be easily exploited, without a doubt, it can bring %non-assessable 
inestimable risks to downstream apps and app users. 
Therefore, it is necessary to take a closer look at existing research on TPL-related security issues to understand the current research status and the existing gap.
% Therefore, understanding current research on TPL-related security issues is necessary. We can understand the current research status and the existing gap.

%We have known that there are many program analysis, anti-virus software techniques using in mobile malware detection, repackaged apps detection, and these techniques are very mature and can achieve good performance~\cite{MassVet2015chen}. However, research on TPL security analysis alone appeared later these technologies.
%\revise{why?} 
%Many undiscovered issues are waiting for us to explore.
%\revise{What are the concrete issues or problems? }

\subsubsection{Existing Research}
{Considering existing threats, many researchers conduct different studies in these directions. Based on our analysis, we find current research primarily focuses on the following fields: 1) privacy leakage analysis, 2) vulnerability identification, 3) malicious TPL detection and analysis, 4) ad frauds.}
%1) violation detection, 2) vulnerability analysis, 3) Ad frauds, 4) privacy leakage,  and 5) dependency conflicts.
The research related to each field can be seen in TABLE~\ref{tbl:sec&privacy_SUM}.

%According to the different research purposes, we can find from the Table% We classify the security-related work of TPLs into the following five categories: 1) security analysis, 2) ad fraud, 3) data leakage, 4) privacy analysis, 5) vulnerabilities analysis.

\begin{table}
%\vspace{-5pt}
\centering
\caption{A summary of TPL security-related issue analysis}
\vspace{-2mm}
%\scalebox{0.9}{
\begin{tabular}{llc}
\toprule
\textbf{Function}  &\textbf{Tool/First Author} & \textbf{Year} \\
\midrule
\multirow{8}{*}{\textbf{Privacy Leakage}}  & Son et al.~\cite{son2016mobile}  & 2016 \\
& Wei et al.~\cite{price2016NDSS} & 2016 \\
                            & Pluto~\cite{pluto2016} & \multicolumn{1}{c}{2016} \\
                            & Paturi et al.~\cite{paturi2015NDSS} & 2015 \\
                            & Moonsamy et al.~\cite{privacyleak} & 2014 \\
                            & Short et al.~\cite{privacyleak2014adhoc} & 2014 \\
                         & Leontiadis et al.~\cite{Leontiadis12HotMobile} & 2012 \\
                           & Stevens et al.~\cite{Most2012}  & 2012 \\
                                   
  \midrule

\multirow{2}{*}{\textbf{Vulnerability Identification} } & Droid-V~\cite{Watanabe2017MSR} & 2017 \\
& AdRisk~\cite{AdRisk2012Wisec} & 2012 \\

\midrule                                   
\multirow{8}{*}{\textbf{Malicious TPL Analysis}} 
    &   MadDroid~\cite{liu2020maddroid}  & 2020 \\
   & MadLife~\cite{chen2019revisiting} & 2019 \\
              & Rastogi et al.~\cite{ad2016NDSS}  & 2016 \\
             & \multicolumn{1}{l}{LibFinder~\cite{LibFInder2016sp}} & \multicolumn{1}{c}{2016} \\
&  K\"{u}hnel et al.~\cite{kuhnel2015fast} & 2015  \\
            & APKLancet~\cite{Apklancet2014ASIACCS} & 2014 \\
           & \multicolumn{1}{l}{Duet~\cite{Duet2014wisec}} & \multicolumn{1}{c}{2014} \\
         &  Brahmastra~\cite{bhoraskar2014brahmastra}  & 2014   \\

\midrule
\multirow{3}{*}{\textbf{Ad Frauds}}          & Madfraud~\cite{MadFraud2014mobisys} & 2014 \\
                                       & DECAF~\cite{Decaf2014NSDI}  & 2014 \\
                                       &  ClickDroid~\cite{cho2015empirical}  & 2015 \\
                                       & \multicolumn{1}{l}{Dong et al.~\cite{Dong2018HotMibile}} & 2018 \\
                                   & \multicolumn{1}{l}{FraudDroid~\cite{FraudDroid2018FSE}} & \multicolumn{1}{c}{2018} \\

\bottomrule
\end{tabular}

\vspace{-2mm}
\label{tbl:sec&privacy_SUM}
\end{table}

\noindent \textbf{Privacy leakage detection.}
% \revise{@TM MISS .... Son et al~\cite{son2016mobile} !!! here }
Son et al.~\cite{son2016mobile} first introduced how Android ad libraries isolated the in-app ads to prevent them from sharing the privileges of the host apps.
They find the ad libraries confined the ads to a separate WebView instance, which can prevent the ads from reading the local storage.
They further proposed an attack model under this protective measure, where malicious advertisers could infer whether local files with a certain name exist within the storage.
They demonstrated how sensitive information of users, including browsing history and social graph, could be revealed by this existing information alone.
They also proposed defensive measures towards such exploits.

Wei et al.~\cite{price2016NDSS} conducted a survey on 231 real users and also verify ad revenue heavily depends on both users' privacy data such as user interests and demographic information. They point out that a user's demographics information plays an essential part in determining the targeted ads.
{Pluto~\cite{pluto2016} is an automatic modular framework for privacy risk assessment of in-app ad libraries. This work systematically explored the data collection of ad libraries from four channels: using unprotected APIs to learn other apps' information on the mobile; using protected APIs through permissions inherited from the apps to access sensitive data, gaining access to the storage of host app, and getting user inputs from the host apps. Pluto combines the static and dynamic analysis and NLP techniques to assess 2,553 real-world apps from Google Play and finds a trend in ad libraries to become more aggressive towards reachable user information.
}
{
As we all know, the permission lists only display information accessible to the host app but ad libraries can gain access to sensitive information from users' devices due to the shared permission mechanism.
Paturi et al.~\cite{paturi2015NDSS} proposed a novel icon-based privacy threat interface that displays privacy risks from app providers and TPLs separately to substitute the traditional permissions list before installation. 
They considered three privacy granules: location, identity, and query (refers to a user's search queries). Besides, they performed two online usability studies to obtain user feedback on their new permission module.
They detect related data access through both static and dynamic analysis.
Their user studies show that users are positive to the new interface, which helps them better understand privacy threats within apps.}
Moonsamy et al.~\cite{privacyleak} analyzed the device ID leakage problem from the in-app ads. They studied 123 apps by combining static analysis and dynamic analysis. They exploit DroidBox~\cite{Droidbox} to track the information leakage and find that 13 apps leak device-related information.
To detect data leakage from TPLs, Short et al.~\cite{privacyleak2014adhoc} extended DroidBox by modifying TaintDroid{~\cite{TaintDroid}}, a taint tracking tool. They use dynamic analysis to catch the data leakage.
Leontiadis et al.~\cite{Leontiadis12HotMobile} analyzed more than 250,000 Android apps and proved that the demographic information has a close link with the ad-supported revenue delivered to developers. It is conducive to more revenues if a precise targeting ad is delivered to the customers in need.
At the same time, they also point out that there are no conflicts between privacy protection and developers' revenue. Therefore, they proposed a feasible privacy control framework, which can achieve an equilibrium between the private information flow and the generated advertisement revenue. This framework established a feedback control loop mechanism that can adjust the level of privacy protection on smartphones based on the generated ad revenues.
Stevens et al.~\cite{Most2012} first compared the commonalities and differences between in-browser ads and in-app ads and found that in-app ads are more likely to leak users' privacy. They examined 13 Android ad networking and attempted to find the risk of privacy leakage from the perspective of permission. Besides, they also discover the vulnerabilities in the use of the JavaScript extension mechanism in several ad libraries. Finally, they proposed potential solutions to address the above issues.

\noindent \textbf{{Vulnerability identification}.}
Droid-V~\cite{Watanabe2017MSR} mainly studies four types of vulnerabilities of mobile libraries in both free apps and paid apps, including information disclosure, SSL/TLS and cryptography, inter-component communication, and WebView.
AdRisk~\cite{AdRisk2012Wisec} is a vulnerability detection framework for in-app advertisements on Android platform. This research reveals a set of privacy and security problems in 100 representative ad libraries. AdRisk first collected sensitive APIs and the required permissions; it then exploits the control flow graph to find a dataflow path from the dangerous API calls to an external sink, which could lead to personal data leakage. Apart from reporting potentially-feasible paths, AdRisk also analyzes five suspicious code patterns: use of reflection, dynamic code loading, permission probing, JavaScript extension functions, and reading installed list of packages. AdRisk found a number of security risks in 100 representative ad libraries such as uploading sensitive information to ad servers, executing suspicious code in the context of the host app environment, fetching malicious payloads.

\noindent \textbf{Malicious TPL Analysis.}
{
MadDroid~\cite{liu2020maddroid} is a dynamic framework for automatic detection of malicious ad contents in android apps.
% It first collects all relevant network traffic through automated app testing,
It adopts a novel method to identify ad traffic by building a mapping between ad libraries and ad hosts through HTTP hooking.
They classified devious ad contents into two categories: ad loading content (pre-click) and ad clicking content (post-click).
They adopted MadDroid to 40K adware apps and found 6\% of apps with devious ad contents.
}

{
Chen et al.~\cite{chen2019revisiting} studied the security threats introduced by in-app advertising, including click fraud, malvertising, and inappropriate ad contents.
They developed Madlife, a dynamic ad collection tool that automatically records all ad-related data for android apps.
They collected 83K ads from 5.7K apps on Google Play and discovered 37 apps with click fraud and 1.49\% ads related to malvertising.
They also found the click fraud and malvertising are strongly correlated.
}

Rastogi et al.~\cite{ad2016NDSS} held the opinion that some apps may be benign, but such in-app ads can redirect users to a certain website, which could play an essential role in propagating attacks.
Thus, they analyzed more than 600,000 Android apps from Google play and four other app markets from China in two months. They attempted to understand the web-app interface attacks. They identified several malware and scam campaigns propagating through in-app ads and web links.

LibFinder~\cite{LibFInder2016sp} is a cross-platform system that can detect potentially harmful libraries over both Android and iOS.
Based on the observation that many iOS libraries have counterparts in Android apps. The authors found that a considerable portion of third-party services to Apple devices are also provided to Android.
LibFinder first identifies suspicious libraries on the Android platform and then tries to find the corresponding iOS versions based on the common features of the same service with the Android and iOS libraries.

% \xian{MISS ...  K\"{u}hnel et al..~\cite{kuhnel2015fast} HERE!!!!}

K\"{u}hnel et al.~\cite{kuhnel2015fast} performed a fast detection of ad libraries within android malware apps by statically checking the smali code for the invocations of advertising APIs which are publicly available.
They further discovered a decrease in the usage of ad libraries for android malware.
Through a manual analysis of the samples from malware families that adopt fewer ad libraries, they discovered some of the malware would send premium SMS on the first run.

Brahmastra~\cite{bhoraskar2014brahmastra} is an app automation tool for dynamically testing the security of third-party components within apps.
Brahmastra first constructs a call graph for the app using static analysis, through which they build activity transition paths for exercising the TPL code.
Brahmastra then attempts to jump start the activities in the transition paths to trigger TPL code through Android Debug Bridge~\cite{adb}.
They further rewrite apps for self-execution by inserting the callback functions that trigger the desired transitions.
Their experiment shows Brahmastra outperforms start-of-the-art GUI testing tools by 170\% in triggering targeted TPL methods.

APKLancet~\cite{Apklancet2014ASIACCS} is an automated app diagnosis system that can identify the tumor payload in Android apps. Tumor payloads include the malicious code fragment and some unwelcome advertising/analytical libraries.
Duet~\cite{Duet2014wisec} is a library integrity verification tool for Android apps, which aims to detect 1) library modification threats, 2) masquerading threats, and 3) aggressive library threats. Duet extracts the library files from the DEX files and generates library digests for both the original library and the libraries from tested apps.

\begin{figure}[t]
	\vspace{-1ex}
  \centering
  \includegraphics[scale=0.4]{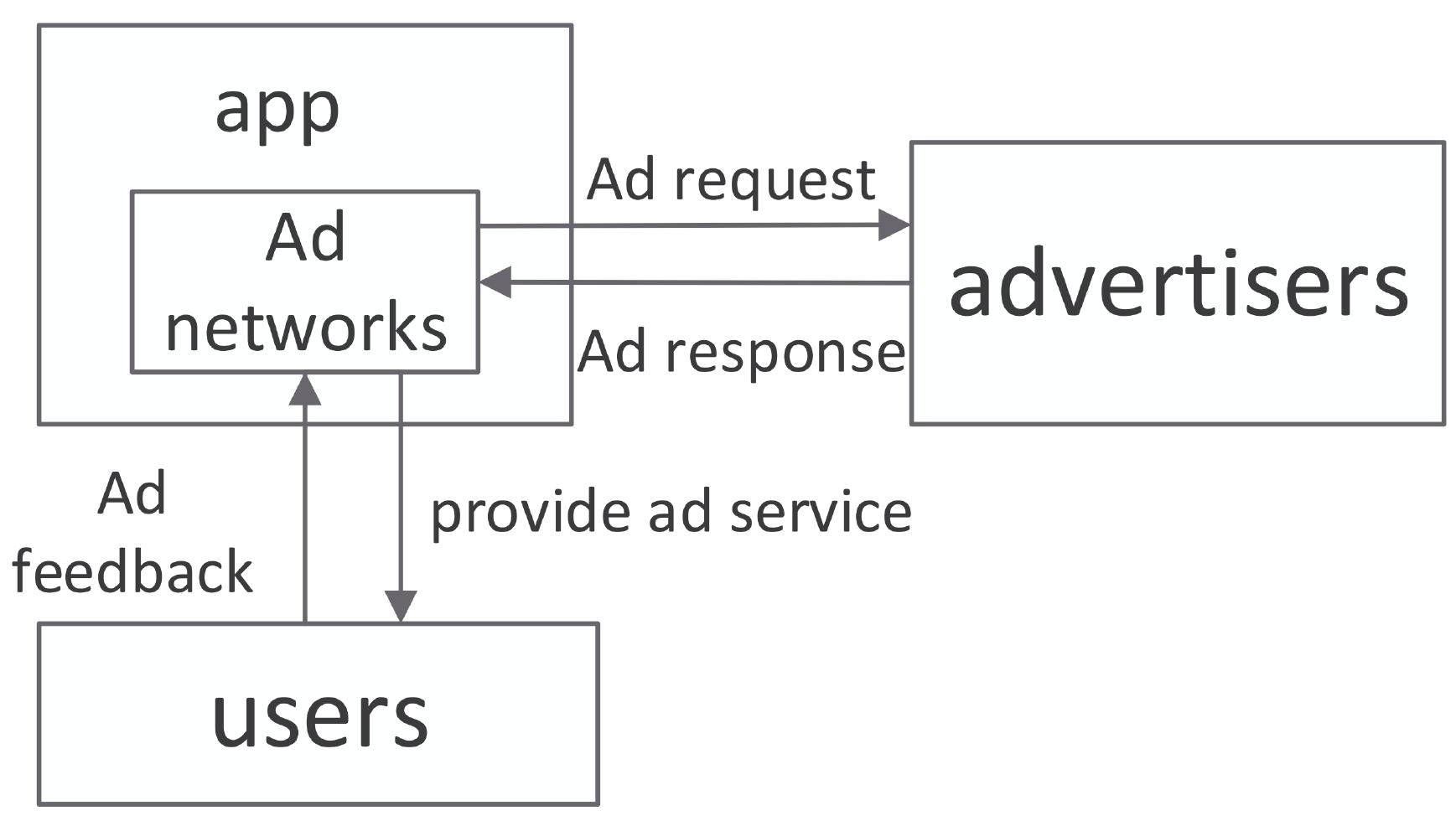}\\
  \caption{Overview of the in-app advertising ecosystem}
  \label{fig:ad-eco}
  \vspace{-2ex}
\end{figure}

\noindent \textbf{Ad fraud detection.}
%Before we discuss the concept of ad frauds, we first should understand the mobile advertising ecosystem. 
Before introducing the concept of ad frauds, we first illustrate the mobile advertising ecosystem in Fig.~\ref{fig:ad-eco}. The advertising ecosystem involves three essential roles: the advertisers, developers, and users. The ad libraries (ad networks) act as the bridge connecting the users and developers, the developers and advertisers. Advertisers usually attempt to propagate their products or service through the in-app ads.
The mobile developers who can embed the ad networks in their apps help the advertisers to disseminate their commodities. When ads are observed (called impression) or clicked by users, the developers can get revenue from the advertisers. The ad network serves as a proxy that connects the app developers and advertisers by exchanging the ad impression and revenue across the ecosystem.
%
% Advertisers pay developers based on the number of times ads are shown or clicked.
Ad networks usually require that app developers strictly follow the guidelines or documents~\cite{Dong2018HotMibile} 
that are used to instruct developers to set the in-app ads. However, some unscrupulous developers attempt to cheat either the advertisers or users by violating the guidelines. For example, unscrupulous developers can modify the code to fetch the ads but make them run in the background instead of displaying the ad impression for users. Some ads are resized by developers, making them too small to read. Some unscrupulous developers may place the ads that close to or cover the UI controllers of the main apps. These settings may affect the user experience and induce users to click the ads. We call all violations of the guidelines as \textit{ad frauds}, regardless of whether they cheat the users or the advertisers. {Dong et al.~\cite{Dong2018HotMibile} summarized a taxonomy of behavior policies based on Admob~\cite{admob}, a popular ad library, and then investigated 3,661 popular apps that all apply the Admob library to detect the advertising policy-violation apps. They first designed an automated event-driven testing tool to send random events and captured the state of the user interface (UI). After that, they removed other UI states and kept the ad-related UI states based on the UI type features, UI location features, and strings. Finally, they performed a violation detection. This research shows that behavior policy violation is a real problem in Android ecosystem. Security experts and developers need to pay more attention to these issues.
}

% \revise{MISS ClickDroid~\cite{cho2015empirical}}
{
Cho et al.~\cite{cho2015empirical} performed an empirical study of click fraud on eight popular ad networks to examine their ability to defend against such ad fraud. They developed ClickDroid, which automatically simulates click events on mobile ads repeatedly. ChickDroid updates the device identifiers after each click as an attempt to bypass the security policies of ad networks and gain more profit. Their experiment showed 6 out of 8 ad networks failed to detect their fraudulent clicks. They also discuss countermeasures against such click fraud.
}

According to our analysis, we compare existing ad fraud detection techniques in Table~\ref{tbl:highlevel_adfrauds}. All of them employ the dynamic method to detect ad frauds. 
MadFraud~\cite{MadFraud2014mobisys} attempts to find two kinds of ad frauds in mobile apps: 1) requesting ads while the app is in the background. 2) clicking ads without user interaction. MadFraud builds the HTTP request trees and extracts the features from the HTTP request pages. By using the machine learning method, it can identify the ad-impression and ad-clicking frauds. 
DECAF~\cite{Decaf2014NSDI} focuses on the placement ad frauds. It uses monkey~\cite{monkey} to get the UI state transition graph. By extracting the information from the DOM tree, DECAF can identify the placement ad frauds. 
FraudDroid~\cite{FraudDroid2018FSE} defines {two kinds of ad frauds}, i.e., the static placement fraud and dynamic interaction fraud, including nine types. It can detect both of them by extracting the UI features and traffic features. The work of FraudDroid is based on the work of Dong et al.~\cite{Dong2018HotMibile}.

\begin{table}
	\centering
	\caption{Summary of Ad fraud detection techniques}
	\vspace{-2mm}
	\scalebox{0.84}{
	\begin{tabular}{lccc}
		\toprule
		
		\textbf{Tool Name} & \textbf{Dynamic Detection} & \textbf{Network Traffic Data} & \textbf{UI Features} \\
		\midrule
				\rowcolor{gray!20}
		\textbf{MadFraud} & \color{Green}\checkmark & HTTP request pages & - \\

		\textbf{DECAF}    & \color{Green}\checkmark & -    & DOM tree  \\
				\rowcolor{gray!20}
		\textbf{FraudDroid} & \color{Green}\checkmark & HTTP request data & DOM Tree \\
		\textbf{Dong et al.} & \color{Green}\checkmark  & -   & Ad views \\
			\rowcolor{gray!20}
		\textbf{ClickDroid}  & \color{Green}\checkmark  & - & - \\
		\hline
	\end{tabular}
}
	\label{tbl:highlevel_adfrauds}
	\vspace{-1mm}
	\begin{center}
	 ``-'' means not use
	\end{center}
	\vspace{-5mm}
\end{table}

As aforementioned, advertisers pay the publishers based on the number of ad impressions and ad clicks.
Based on the previous research~\cite{MadFraud2014mobisys,Decaf2014NSDI,FraudDroid2018FSE}, we summarized ten different ways of ad frauds as follows.
%
%The specific explanation of each ad fraud type is as follows:
1) \textit{Automatic Click Fraud:} Clicking on ads without user interaction. 2) \textit{Ad Hidden Fraud:} Ads are placed under other controls or are hidden so that users cannot find them, giving users a wrong impression that it is an ``ad-free app''. 3) \textit{Ad Size Fraud:} Developers resize the ads to make them too small to read, or so large that users are forced to click or view the ads. 4) \textit{Ad Number Fraud:} The number of ads exceeds a normal quantity in one UI page. Developers want to attract users' attention and increase the probability of interacting with the ads, which obviously affects user experience. 5) \textit{Ad Overlap Fraud:} The ads are placed close to the actionable components or cover the normal functional UI components of the host app. Developers want to trigger accidental clicks in this way to earn illegal revenue. 6) \textit{Interaction Fraud:} When users interact with host apps, the advertisement pops up unexpectedly, causing the user to click. 7) \textit{Driven-by download:} Triggering the unintentional download of other apps when clicking on the ads. 8) \textit{Outside Ad Fraud:} Displaying the ad even when the app is running in the background without any interaction with users. 9) \textit{Frequency Fraud:} Ads being popped up too often upon different user operations. 10) \textit{Non-content Ad Fraud:} Placing ads at non-content-based pages such as the login or exit screen, which may cause users to mistake the ads for the real app content. 
{Following the taxonomy of FraudDroid, we divide them into two categories:}
{the static fraud involving the ad placement issues and the dynamic fraud involving the user interaction.
Among these ad frauds, Ad hidden fraud, Ad size fraud, Ad number fraud and Ad overlap fraud are static frauds; the remaining frauds are dynamic frauds. }
%\tm{Add definition of static and dynamic ad fraud}

TABLE~\ref{tbl:adfraudscmp} compares the detection capabilities of MadFraud, DECAF, FraudDroid, and ClickDroid. As can be seen from the Table~\ref{tbl:adfraudscmp}, MadFraud can detect two types of ad frauds, including the \code{automatic click} and \code{outside ad fraud}. DECAF can detect all the static ad frauds. FraudDroid has the most robust detection capability, which can detect all the ad frauds except the \code{automatic click}. In contrast, ClickDroid can only detect automatic click and cannot detect other types of ad frauds.

\begin{table}[t]
	\centering
	\caption{Comparison of detection capabilities of different ad fraud detection tools}
	\vspace{-2mm}
	\scalebox{0.88}{
	\begin{tabular}{lcccc}
		\toprule
		\textbf{Fraud Types}     & \textbf{MadFraud }& \textbf{DECAF}  & \textbf{FraudDroid}  & ClickDroid \\
		\midrule
		\rowcolor{gray!20}
		\textbf{Automatic Click} & \color{Green}\cmark & \xmark  & \xmark       &  \color{Green}\cmark \\
		\textbf{Ad Hidden }      & \xmark & \color{Green}\cmark  & \color{Green}\cmark  & \xmark\\
		\rowcolor{gray!20}
		\textbf{Ad Size}         & \xmark   &  \color{Green}\cmark & \color{Green}\cmark  & \xmark \\
		\textbf{Ad numbers }     &  \xmark  & \color{Green}\cmark  & \color{Green}\cmark  & \xmark \\
		\rowcolor{gray!20}
		\textbf{Ad Overlap}      &  \xmark & \color{Green}\cmark & \color{Green}\cmark & \xmark \\
		\textbf{Interaction }       & \xmark & \xmark & \color{Green}\cmark  & \xmark \\
		\rowcolor{gray!20}
		\textbf{Driven-by download} &  \xmark  &  \xmark   & \color{Green}\cmark  & \xmark \\
		\textbf{Outside Ad fraud}   & \color{Green}\cmark &  \xmark & \color{Green}\cmark  & \xmark  \\
		\rowcolor{gray!20}
		\textbf{Frequent}           &   \xmark  & \xmark  & \color{Green}\cmark  & \xmark  \\
		\textbf{Non-content}        &  \xmark   &  \xmark  &  \color{Green}\cmark  & \xmark  \\
		\bottomrule
	\end{tabular}
	}
	\label{tbl:adfraudscmp}
	\begin{center}
		{\xmark}  means cannot; {\color{Green}\cmark} means can
	\end{center}
	\vspace{-5mm}
\end{table}

\subsubsection{Future Work}

According to our survey, we find current research mainly focus on the security problem of Ad libraries, such as the privacy leakage of ad libraries, violation of ad fraud. Besides, some violation behaviors of TPLs are also investigated. However, we find only two papers focus on vulnerable TPL analysis. In addition, the research scope of existing work on vulnerability of TPL analysis is also very limited. For example, Droid-V just studied four types of vulnerabilities of TPLs. AdRisk detected the potential bugs in ad libraries only. {In fact, existing vulnerabilities of Android TPLs are far more than those that have been revealed by current research. Many vulnerabilities of Android TPLs can be found in National Vulnerability Database (NVD)~\cite{NVD} or other related website~\cite{sourceclear}.  
% There are many blind spots in our understanding of these vulnerabilities. 
Currently, there lacks a systematic study revealing the threats of these vulnerabilities. 
There still exist many blind spots in our understanding of these vulnerabilities,
% We know nothing about these vulnerabilities; 
such as their number and types, their severity and threats to apps and users, whether these vulnerabilities are easy to exploit, and how to exploit them. Apart from these known vulnerabilities, there may be a large number of 0-day vulnerabilities in TPLs that have not been found. How to detect these vulnerabilities and reveal their threats is also another tough work.}
%As far as we know, existing types and numbers of TPLs are not limited to this, not along many 0-day vulnerabilities of TPLs. \tm{(According to \cite{}, existing 1-day TPL vulnerabilities covers a much wider range in both type and quantity, and causes great harm to the Android ecosystem, not to mention 0-day TPL vulnerabilities.)}
We believe that revealing the vulnerability issues of TPLs would be a significant contribution to our community, which also can help improve the quality of apps.
We suggest future researchers conduct more related work in this direction.

\begin{comment}

\vspace{2mm}
\noindent\fbox{
	\parbox{0.95\linewidth}{
%Vulnerable TPLs have already become a serious threats to users' privacy and finance security. We can find that there are only two papers focus on vulnerable TPLs analysis.
%		\sen{Can we provide a summary for security/privacy analysis. Any useful info can be provide to readers?}
\revise{Compared with other security issues of TPLs, vulnerable TPLs have already become a serious threat to user's privacy and security. These TPLs would introduce more significant security implications compared with source code vulnerabilities within Android apps. Because vulnerable TPLs are more communicable in practice.} {However, we just find that only two papers focus on vulnerable TPL analysis. We believe the security issues of TPL is an essential parts to community. Future researchers may can conduct more related work in this direction. It must be benefit to the software eco-system.}
	}
}

\end{comment}

%################################################
\subsection{TPL Privilege De-escalation}
\label{Sec: TPL isolation}

\subsubsection{Research Background}
There are two security mechanisms in Android system: the sandbox mechanism and the permission framework. Regarding the sandbox mechanism for isolating apps, Android system assigns a unique user ID (UID) to each app and lets it run in a separate process. Regarding the permission framework for controlling the privileges of each app, Android system allows apps to access the system resources (e.g., telephone ID/status, location, camera) with the corresponding permissions.
This permission-based security model can restrict apps from accessing resources and private data.
% When an app attempts to access some sensitive resources (e.g., locations and Internet), Android system would request access permissions from users and let users decide whether to grant the permissions or not.
%can let users decide whether to give this permission to this app. 
\textit{However, these two security mechanisms have a flaw that they only work at the app level.} Apps and their incorporated third-party libraries share the same permissions and UID with the host app, which means TPLs could also access the same sensitive data and resources if the host apps own these permissions.

%\begin{figure}[t]
%	\centering
%	
%	\includegraphics[scale=0.4]{IPC.png}\\
%	
%	\caption{An example of an Android app security mechanism. Diagram showing how TPLs work in Android system. The permission module is working at the app level. All TPLs share the same process and UID and the same privileges with the host apps.}
%	\label{fig:IPCmodule}
%\end{figure}

We provide an example to explain this security mechanism in Fig.~\ref{fig:IPCmodule}, which illustrates the relationship between the host app, TPLs, as well as the Android permission system. When users try to install an app on a mobile device, Android system will parse the APK file and get the component information and the requested permissions during the installation process. In Android, Package Manager Service (PMS) is responsible for creating a new user and private storage for this new app. 
When the app is launched, Activity Manager Service (AMS) initiates a process to run this app. 
%This process creation is initiated by the Activity Manager Service (AMS). 
AMS first retrieves the process information (i.e., UID, GID, and storage information) from the PMS, and sends a process creation request to the Zygote process that will fork a new process (sandbox) for this new app. This security mechanism works at the app level. Therefore, the bundled TPLs and the host app share the same permissions, UID, GID, and storage space. We can see an example of this mechanism from Fig.~\ref{fig:IPCmodule}. This app has one ad library A and other two third-party libraries, B and C, running in the same process. Both the host app and the TPLs can access the Android system service via inter-process communication (IPC) if they have the corresponding permissions to control these resources.

\subsubsection{Existing Problem}
Based on previous introduction, we can find that the permission mechanism and the sandbox mechanism can bring potential risks because both of them lead to over-privileged problems~\cite{DelDroid201983HAMMAD,hammad2017ICSA}, which could pose threats to the privacy and security of users. To handle this situation, many researchers attempted to solve this problem through \textit{TPL isolation}. They tried to find methods to implement privilege de-escalation.

Besides, we also need to note that Android provides two file storage methods for apps, i.e., external storage and internal storage. The external storage is global storage that can be accessed by any application while the internal storage is a private space that can only be visited by the app itself. Since the TPLs and the host app share the same UID and process storage, the TPLs are able to get sensitive data from the host apps. Therefore, it is necessary to give separated storage to the host app and TPLs.

\subsubsection{Existing Solutions}

%\begin{figure*}[htpb]
%	%\vspace{1ex}
%	
%	\centering
%	\subfigure[split the TPLs into independent process]{%
%		
%		\includegraphics[scale=0.4]{IPCM1.png}
%		
%	}
%	\quad
%	\quad
%	\quad
%	\quad
%	\centering
%	\subfigure[cut off direct communication between the system service with libs]{%
%		\includegraphics[scale=0.4]{IPCM2.png}
%	}
%	
%	\caption{Two TPLs isolation schemes Based on the example in the Fig.~\ref{fig:IPCmodule}}
%	\label{fig:modification_scheme}
%	%\vspace{-1ex}
%\end{figure*}

\begin{figure*}
	\centering
	\begin{minipage}{0.27\linewidth}
		\centering
		\includegraphics[scale=0.32]{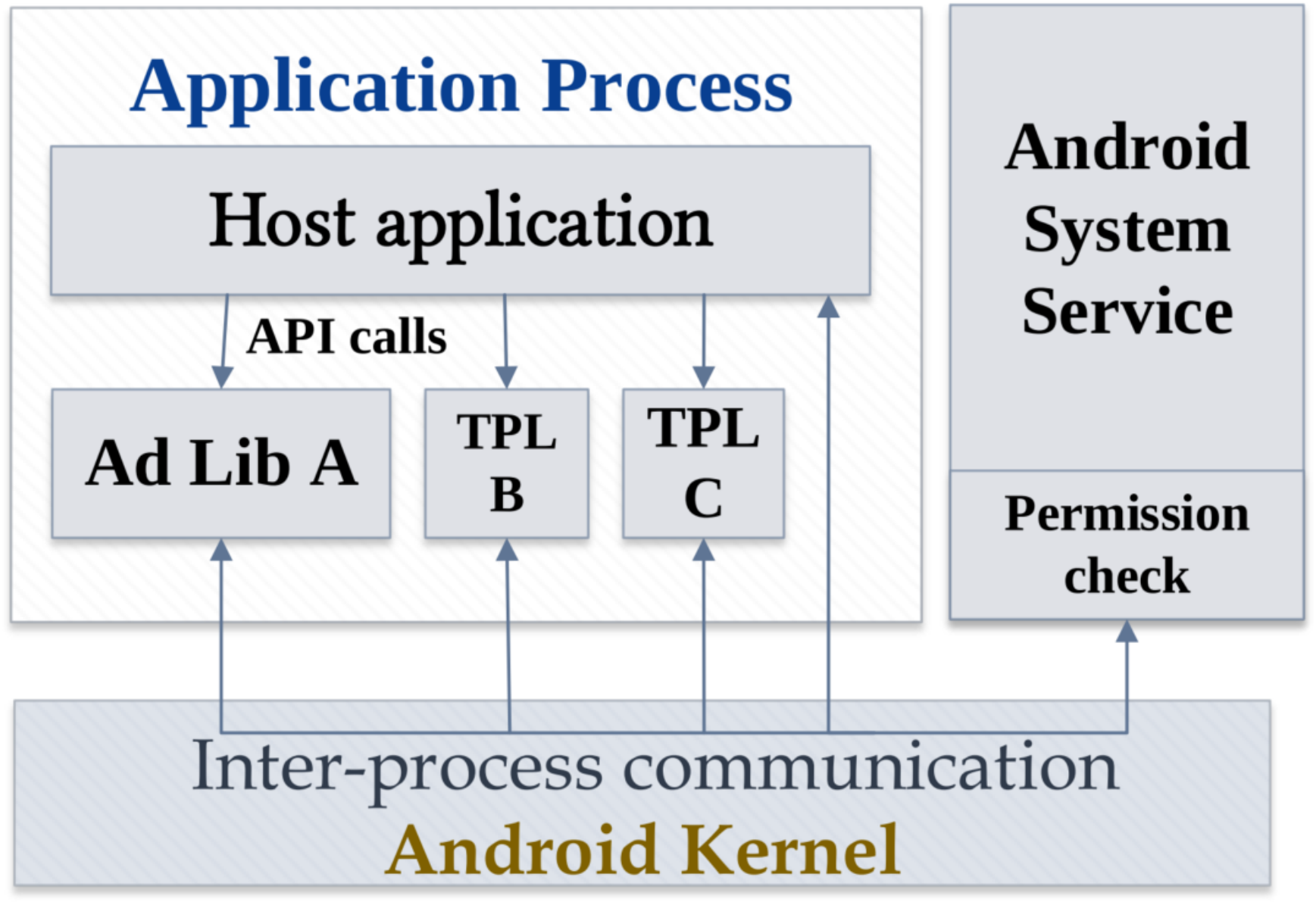}
		\caption{An example of security mechanism of Android apps 
			%Diagram showing how TPLs work in Android system. The permission module is working at the app level. All TPLs share the same process and UID and the same privileges with the host apps.
			}
		\label{fig:IPCmodule}
	\end{minipage}
	\begin{minipage}{0.72\linewidth}
		\centering
		\subfigure[Split the TPLs into independent processes]
		{\label{fig:a}\includegraphics[scale=0.325]{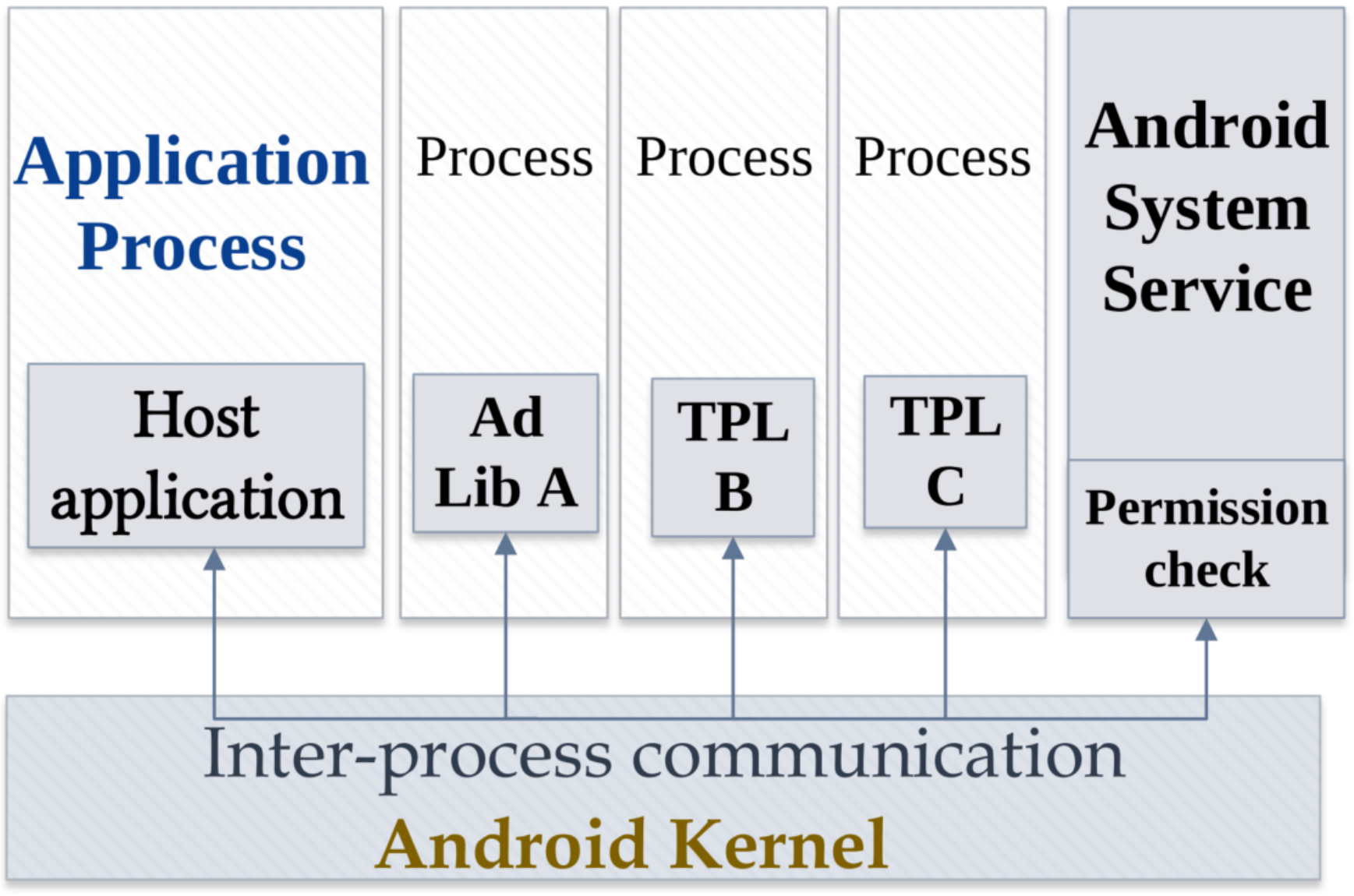}}
		\subfigure[Cut off the direct communication between the system service and the libraries]
		{\label{fig:b}\includegraphics[scale=0.3015]{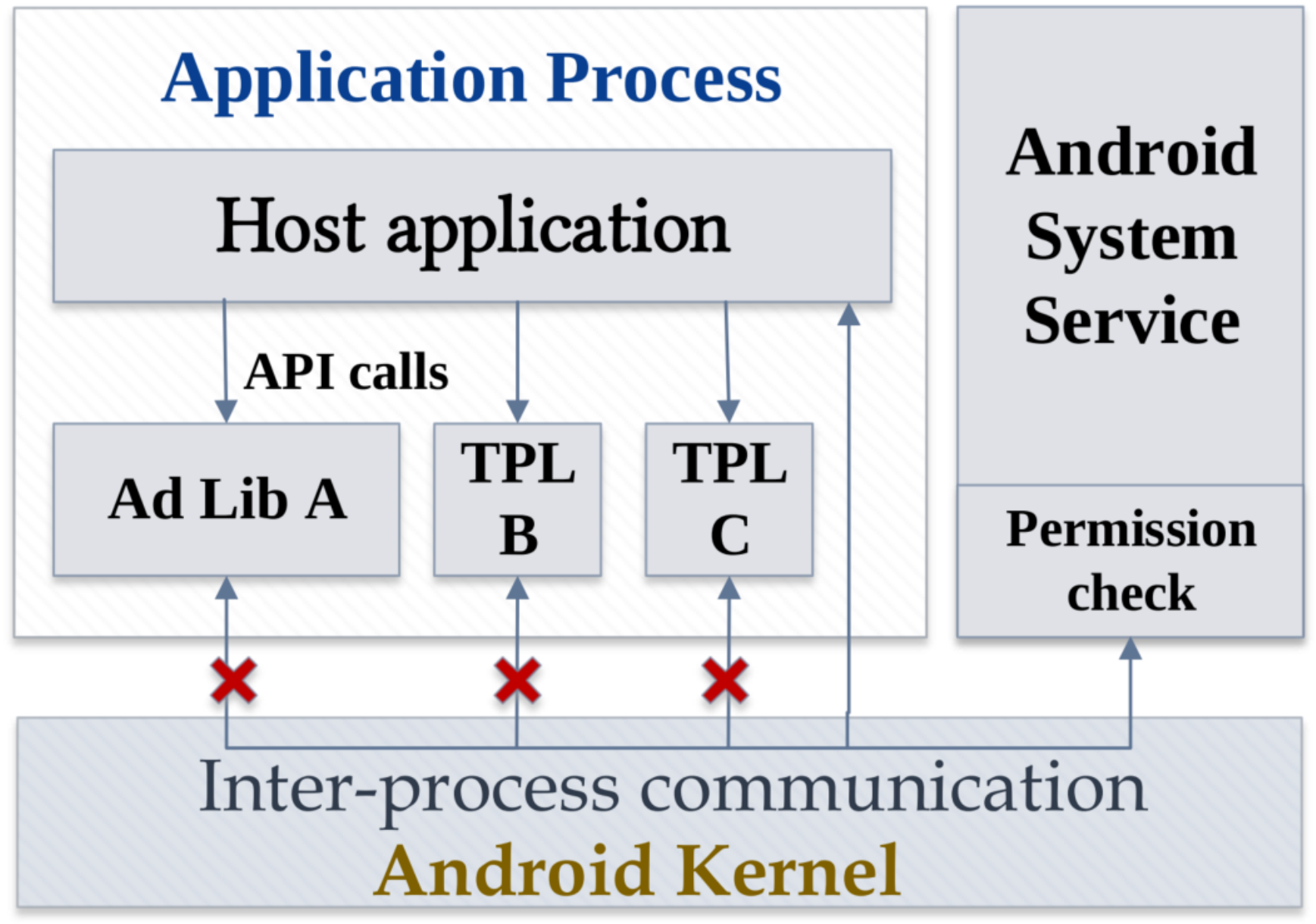}}
%		\begin{subfigure}[Split the TPLs into independent process]
%			\centering
%			\includegraphics[width=0.5\textwidth]{IPCM1.png}
%			%\caption{Split the TPLs into independent process}
%		\end{subfigure}%
%
%		\begin{subfigure}[Cut off direct communication between the system service with libs]
%			\centering
%			\includegraphics[width=0.5\textwidth]{IPCM2.png}
%			%\caption{Cut off direct communication between the system service with libs}
%		\end{subfigure}
		\vspace{-4mm}
		\caption{Two TPLs isolation schemes based on the example in the Fig.~\ref{fig:IPCmodule}}
		\label{fig:modification_scheme}	
	\end{minipage}%\hfill
\vspace{-3ex}
\end{figure*}

Existing isolation techniques can be concluded into two types, as illustrated in Fig.~\ref{fig:modification_scheme} which are based on the example in Fig.~\ref{fig:IPCmodule}. 
%We use the same example in Fig.~\ref{fig:IPCmodule} to explain these techniques. As can be seen from the Fig.~\ref{fig:modification_scheme}, we illustrate two different TPLs isolation schemes. 
The first scheme in Fig.~\ref{fig:modification_scheme} (a) tries to split these TPLs into different processes and lets Android system assign a separate storage space and permissions. Under this situation, these TPLs can only access the system service with the corresponding permissions that the TPLs required, and they cannot share the permissions with the host app. 
This scheme usually requires the modification of the Android framework.
TPLs can access the system resources via two ways: (1) directly invoke the system calls to get the system services or resources; (2) indirectly access the resource through the interfaces from the host app logic. The second scheme is shown in the Fig.~\ref{fig:modification_scheme} (b) attempts to block the direct system invocations so that TPLs can only get the resources through the host app by API calls. 
In this situation, developers can configure some fake information to the TPLs rather than share the real information with them. This scheme does not require the modification of the operating system and APIs; instead, it usually requires the developers to rewrite the resource access functions to achieve the TPLs isolation and permission separation.

\subsubsection{Existing Research}
%\ling{Merge the work according to the two aforementioned schemes.}

According to our collected papers, we summarized ten publications regarding TPL isolation techniques in Table~\ref{tbl:isolationsum}, most of which were published in security-related conferences. %We introduce these tools based on the aforementioned schemes.
{Since some tools are improved based on previous tools, we will introduce these tools in chronological order.}
\begin{table}[t]
%\vspace{-5pt}
\centering
\caption{A summary of TPL isolation literature}
\vspace{-2mm}
\begin{tabular}{lcr}
\toprule
\textbf{Tool/First Author}  & \textbf{Year}  & \textbf{Venue} \\
\midrule
\rowcolor{gray!20}
Zhan et al.~\cite{splitads2017ACISP} & 2017 & ACISP \\
FLEXDROID~\cite{FLEXDROID201NDSS} & 2016 & NDSS \\
\rowcolor{gray!20}
LibCage~\cite{Libcage2016ESORICS} & 2016 & ESORICS \\
PEDAL~\cite{PEDAL2015MobiSys} & 2015 & MobiSys \\
\rowcolor{gray!20}
NativeGuard~\cite{NativeGuard2014Wisec} & 2015 & Wisec  \\
COMPAC~\cite{COMPAC2014Wang} & 2014  & CODASPY \\
\rowcolor{gray!20}
AFrame~\cite{aframe2013ACSAC} & 2013 & ACSAC \\
SanAdBox~\cite{sanAdBox2013ICC} & 2013 & ICC \\
\rowcolor{gray!20}
AdDroid~\cite{AdDroid20120ASISCCS} & 2012 & ASIACCS \\
AdSplit~\cite{adsplit2012USENIX} & 2012 & USNIX Security \\
\bottomrule
\end{tabular}
%\vspace{-2mm}
\label{tbl:isolationsum}
\end{table}

\noindent$\bullet$ \textbf{Scheme of splitting TPLs into independent processes.}
%\noindent$\bullet$\textbf{First Solution Scheme}

{\textbf{AdSplit~\cite{adsplit2012USENIX}} is built on QUIRE~\cite{Quire2011USENIX} 
{which can separate the advertising TPL and the host app into two activities within different processes but sharing the same screen.
AdSplit achieves the permission isolation since the ad libraries and the host app run in separate processes.}
% which allows the activities of ads and activities of the host app to share the same screen but run in different processes. Since the ad libraries and the host app run in different processes, AdSplit can achieve the process and permission isolation.
{The screen sharing is achieved with a transparency technique where the advertisement activity is put under the app activity with a see-through region for the in-app ad slot.}
% In AdSplit, the advertisement activity is put under the app activity. With the transparency technique, users can see the advertisement through the app activity.
However, there are some disadvantages of this method. First, the transparency technique will cause considerable overhead in drawing, especially for a mobile phone, since it requires several layers of drawing windows to be combined.
Second, the transparency technique has been wildly used by clickjacking, which can make attack detection of apps more difficult. Third, this method also may affect the normal operation of the ads and directly block the interaction with the users and ad network. %Since the ads have different types, e.g., the interstitial ads, banner ads, and video ads. If the ad is a interstitial type, this ad will not be shown.
}
%can separate the ad libraries from the host apps. 
%\xian{AdSplit uses the transparency technique that allows users to see the advertisement through the transparent region in the application activity 

\textbf{AdDroid~\cite{AdDroid20120ASISCCS}} {achieves the ad libraries and host app isolation. To accomplish this goal, AdDroid introduces extended advertising APIs into the Android SDK. Meanwhile, it also adds an AdDroid Service like the Android system service, which provides the all advertisement required permissions. Host apps offer the APIs with configuration data to identify which advertising networks and to fetch advertising contents, contextual information about the advertisement. The extended APIs handle the user interface events.
}

\textbf{SanAdBox~\cite{sanAdBox2013ICC}} {is a privilege separation framework that can separate the advertising libraries from the host app and run the advertising libraries as different independent applications. In addition, SanAdBox is an updating sandbox where the host app and ad libraries run into and only can invoke their own permissions. 
The authors conduct an installer responsible for managing the communication between the host app and dependent advertising libraries. 
The advantage of SanAdBox does not require system modification.}

\textbf{AFrame~\cite{aframe2013ACSAC}} {achieves not only the process and permissions isolation but also 
the display and input isolation. % by modifying the Android framework code and setting new rules to developers. 
AFrame is similar to a typical View component; inside this area runs the TPL process, which is called AFrame process. The main process runs the host app. From the user perspective, the host activity and the TPL activity look like the one. From the system perspective, there are two processes. Compared with AdSplit, the advantage of AFrame is that the users can interact with the advertisement. Besides, AFrame can support all TPL isolation.
AFrame needs to modify the Android framework and bytecode of apps. %\tm{AFrame paragraph need modify}
AFrame adds a new parsing module in Package Manager Service (PMS). The modified PMS will create an extra process for AFrame. In this way, AFrame can create an independent process for TPLs. TPLs in different processes cannot share the same permissions with the host apps. Therefore, AFrame achieves the permission isolation, process isolation, input/output isolation, and display isolation. }

\textbf{COMPAC~\cite{COMPAC2014Wang}} also achieves a fine-grained access control at third-party components level by modifying the Android Kernel and Framework. It keeps the original permission checking strategy to avoid compatibility issues. COMPAC extends the Android access control architecture, the Framework Reference Monitor (FRM), and Kernel Reference Monitor (KRM) by adding two Policy Managers (PMs): Framework Policy Manager (FPM) and Kernel Policy Manager (KPM). The two reference monitors are responsible for permission checking of the app while the new PMs are responsible for the permission checking of TPLs.

\textbf{NativeGuard~\cite{NativeGuard2014Wisec}} is the first work that proposes to separate the native libraries from Android apps to achieve limiting the over privileges of native libraries. It isolates native libraries by splitting one Android app into two apps. Specifically, by leveraging the reverse-engineering techniques, NativeGuard first finds the code of all the native libraries and then moves them to an entirely-new service app, which can be started by the host app. The two apps can communicate through the interfaces defined by the Android Interface Definition Language (AIDL)~\cite{AIDL}. They create a library that acts as a proxy to call the AIDL interface. NativeGuard does not need to modify the Android framework or access to the source code of an app.

\textbf{FLEXDROID~\cite{FLEXDROID201NDSS}} modifies different access control policies such as the Kernel of Android OS, the Android Framework, Dalvik VM, Bionic, the Java core library, the Binder library, and SELinux setting to achieve a fine-grained access control for TPLs. FLEXDROID allows developers to totally control the grant permissions and sensitive behaviors to access the system resources and data. Moreover, FLEXDROID implements a new permission mechanism to conduct the inter-process stack inspection and isolate TPLs, which is not affected by JNI~\cite{JNI}, Java reflection~\cite{Javareflection} and dynamic code execution~\cite{dymcode}.

\textbf{Zhan et al.~\cite{splitads2017ACISP}} {extend
the package manage service (PMS) to enable the system to identify the permission at the library-level. Besides, it allows developers to configure the permission policy at the library-level instead of the app-level. They add a module named the system libraries to get the specific Android API that is invoked by the host app and the corresponding invoking method in the host app.
% in this way, the system is able to get the runtime information of the libraries.
When the system finds the {invoking method} that belongs to TPL, it will search the corresponding privileges and then decide whether to grant it or not.  The advantage of this approach does not require considerable modification on host apps, and the modified system can easily adapt to the DVM and ART virtual machine.}

\noindent$\bullet$ \textbf{Scheme of cutting off the communication between TPLs and the host apps.}

\textbf{PEDAL~\cite{PEDAL2015MobiSys}} resets the resources access rules for the ad libraries by rewriting the bytecode on user-specified privacy policy. It implements three different levels of accessing privileges (i.e., allow, obscure, block). Users can configure different access privileges to the ad libraries of the registered apps through the controller app.

\textbf{LibCage~\cite{Libcage2016ESORICS}} allows developers to grant different permissions to each TPL and put each TPL in a separate file, and it does not need to modify the Android Framework or the bytecode of libraries. LibCage creates a new sandbox on the system-level and lets the host app and TPLs work in this process. 
{Besides, LibCage contains a permission checker that can restrict the access of TPLs to sensitive resources.}
%Besides, there is a permission checker in LibCage, which can limit the TPLs access to sensitive resources. \tm{last sentence needs modify, try to avoid 'there is'}

% {LibCage~\cite{Libcage2016ESORICS}} allows developers to grant different permissions to each TPL and put each TPL in a separate file, and it does not need to modify the Android Framework or the bytecode of libraries. LibCage creates a new sandbox on the system-level and lets the host app and TPLs work in this process. Besides, there is a permission checker in LibCage, which can limit the TPLs access to sensitive resources.

\subsubsection{State-of-the-art Techniques}

%\settowidth\rotheadsize{\theadfont{Zhan et al.~\cite{splitads2017ACISP}}}

\begin{table}[t]
	\centering \scriptsize
	\caption{A Summary of TPL isolation techniques}
	\vspace{-2mm}
	\scalebox{0.9}{
	\begin{tabular}{lcccccccccc} 
		\toprule
	\textbf{Technique}	& \rotatebox{90}{\textbf{Zhan et al.~\cite{splitads2017ACISP}}} & \rotatebox{90}{\textbf{FLEXDROID}} & \rotatebox{90}{\textbf{LibCage}}  & \rotatebox{90}{\textbf{PEDAL}} & \rotatebox{90}{\textbf{NativeGuard}}  &
		\rotatebox{90}{\textbf{COMPAC}} & 
		\rotatebox{90}{\textbf{AFrame}} & \rotatebox{90}{\textbf{SanAdBox}} & \rotatebox{90}{\textbf{AdDroid}} & \rotatebox{90}{\textbf{Adsplit}}\\
		\midrule
		\rowcolor{gray!20}
		\textbf{Dynamic} & \color{Green}{\cmark} &  \color{Green}{\cmark}  & \color{Green}{\cmark} &  \xmark   & \xmark & \color{Green}{\cmark} & \color{Green}{\cmark} &  \xmark    & \color{Green}{\cmark}& \color{Green}{\cmark}\\
		\textbf{Bytecode modification} & \xmark & \xmark &  \xmark  & \color{Green}{\cmark} & \color{Green}{\cmark} & \xmark &\color{Green}{\cmark}& \color{Green}{\cmark} &  \xmark     & \xmark  \\
		\textbf{System modification} & \color{Green}{\cmark} & \color{Green}{\cmark} &  \xmark & \xmark & \color{Green}{\cmark} & \color{Green}{\cmark} & \color{Green}{\cmark} &  \xmark   & \color{Green}{\cmark} & \color{Green}{\cmark} \\
		\rowcolor{gray!20}
		\textbf{Java reflection} & \color{Green}{\cmark} & \color{Green}{\cmark} & \color{Green}{\cmark}&  \color{Green}{\cmark}   & \xmark& \color{Green}{\cmark} & \color{Green}{\cmark}  & \color{Green}{\cmark} & \xmark     &    \xmark    \\
		\rowcolor{gray!20}
		\textbf{Dynamic code execution}  & \color{Green}{\cmark} & \color{Green}{\cmark} & \color{Green}{\cmark} & \color{Green}{\cmark} & \xmark & - &\color{Green}{\cmark} &  \color{Green}{\cmark} &  \xmark & \xmark  \\

		\textbf{DVM}     & \color{Green}{\cmark} & \color{Green}{\cmark} & \color{Green}{\cmark} & \color{Green}{\cmark} & \color{Green}{\cmark} & \color{Green}{\cmark} & \color{Green}{\cmark}& \color{Green}{\cmark} & \color{Green}{\cmark} & \color{Green}{\cmark} \\

		\textbf{ART}     & \color{Green}{\cmark} & \xmark & \color{Green}{\cmark} & \color{Green}{\cmark}  & \color{Green}{\cmark}& \xmark & \xmark & \color{Green}{\cmark} & \xmark & \xmark\\
		\rowcolor{gray!20}
		\textbf{Ad libs} & \color{Green}{\cmark} & \color{Green}{\cmark} & \color{Green}{\cmark} & \color{Green}{\cmark}& \xmark & \color{Green}{\cmark} & \color{Green}{\cmark} & \color{Green}{\cmark} & \color{Green}{\cmark} & \color{Green}{\cmark} \\
			\rowcolor{gray!20}
		\textbf{TPLs}    & \color{Green}{\cmark} & \color{Green}{\cmark} & \color{Green}{\cmark} &    \xmark & \color{Green}{\cmark} &  \color{Green}{\cmark}     & \color{Green}{\cmark} &    \xmark       &    \xmark         &       \xmark     \\
	
		\textbf{Permission separation} &  \color{Green}{\cmark} & \color{Green}{\cmark}& \color{Green}{\cmark} & \color{Green}{\cmark} & \color{Green}{\cmark} & \color{Green}{\cmark} &  \color{Green}{\cmark} & \color{Green}{\cmark} & \color{Green}{\cmark} & \color{Green}{\cmark} \\

		\textbf{Storage separation} & \color{Green}{\cmark} & \xmark & \color{Green}{\cmark} & \color{Green}{\cmark} & \xmark & \xmark & \color{Green}{\cmark} & \color{Green}{\cmark} & \color{Green}{\cmark} & \color{Green}{\cmark}  \\
		
		\bottomrule
	\end{tabular}
     }
	\label{tbl:isolationCMP}
	\begin{center}
	  {\color{Green}{\cmark}} means can; {\xmark} means cannot; - means not mention
	\end{center}
\vspace{-4ex}
\end{table}

Generally speaking, the main purpose of existing tools is to achieve TPL isolation and privilege de-escalation.
Table~\ref{tbl:isolationCMP} provides the comparison details of ten different TPL isolation tools. We summarized existing techniques from the following aspects.

%In order to let readers understand the TPLs isolation techniques, we summarized the existing techniques from different aspects. Table~\ref{tbl:isolationCMP} provides the comparison details of ten different TPLs isolation tools.

\noindent \textbf{Modification mode.}
The modification mode usually involves three ways: 1) modify the bytecode of Android apps 2) modify the Android system. 3) both 1) and 2).
Both PEDAL and SanAdBox involve modifying the bytecode of the app, while the remaining five tools, i.e., FLEXDROID, COMPAC, AdDroid, Adsplit, and the tool of Zhan et al.~\cite{splitads2017ACISP} extend the Android framework to achieve the TPL isolation and privilege separation. AFrame and NativeGuard modify both of them.

\noindent\textbf{Dynamic feature support.}
Simply modifying the permission check mechanism cannot check certain dynamic sensitive behaviors such as the dynamic class load, JNI method, and Java reflection.
These dynamic methods can easily bypass the DVM permission checking module in a virtual machine while the dynamic class loading or dynamic code execution and reflection technologies generally cannot affect bytecode rewrite methods. But FLEXDROID, PEDAL, SanAdBox, LibCage, and \cite{splitads2017ACISP} can also handle the java reflection and dynamic code execution.

\noindent\textbf{Virtual machine support.}
As shown in Table~\ref{tbl:isolationCMP}, PEDAL, LibCage, NativeGuard, and ~\cite{splitads2017ACISP} support the Dalvik Virtual Machine (DVM) and Android Runtime (ART) virtual machines. The remaining systems AFrame, SanAdBox, AdDroid, COMPAC, FLEXDROID, and Adsplit only can adapt to the DVM. 
Generally speaking, if the modification is on Android apps, it can support both ART and DVM. If the changes just happen on DVM-specific features, it cannot support the ART mechanism.

\noindent\textbf{Library type support.}
PEDAL, SanAdBox, AdDroid, and AdSpilt only achieve the ad TPL isolation; NativeGuard isolates the native libraries; the remaining tools all implement the Java TPL isolation.

\noindent\textbf{Separation items.}
TPL isolation usually includes two parts, i.e., \textit{permission separation} and \textit{storage separation}. Both of them involve limiting the privileges to TPLs and improving the security performance of apps. All of these systems have already achieved permission separation, while FLEXDROID, NativeGuard, and COMPAC do not support storage separation. Besides, AFrame achieves not only  the process and permission isolation but the display and input isolation.

\subsubsection{Summary}

According to our observation, TPL isolation techniques are directly related to the Android security mechanism, and the isolation process usually leverages dynamic analysis techniques. Based on our analysis, we know that the PMS involves process allocation and permission management. Hence,
existing tools usually involve PMS modification or sandbox mechanism updating.
Besides, we also find that most existing tools only can handle Java TPLs; only NativeGuard can handle Native libraries. 
We expect future work can conduct more research on Native libraries, even though the task is challenging.
%One limitation is that most of them only focus on TPLs that are written in Java except for NativeGuard. However,
%
Also, the state-of-the-art isolation methods are not very practical because they need to modify either the Android apps or the Android system.
Providing the new process for each TPLs will add resource consumption, especially for mobile devices. Unlike the communication with other apps, TPLs have more interaction with the host app. We hope future researchers can propose more practical and valuable approaches.
Furthermore, none of them are publicly available. Therefore, it is impossible to compare their advantages and disadvantages from practical perspectives.
%It is impossible for general users to apply them.

\begin{comment}

%%%%%%%%%%%%%%%%%%%%%%%%%%%%%%%
\vspace{2mm}
\noindent\fbox{
	\parbox{0.95\linewidth}{
%We can find most of TPL isolation techniques just focus on TPLs are written in Java, and only {NativeGuard} supports the native code library isolation.
	%	\sen{Can we provide a summary for TPL isolation. Any useful info can be provide to readers?}
	{According to our observation, we find that TPL isolation techniques are directly related to Android security mechanism, and the isolation process usually leverages dynamic analysis techniques. One limitation is that most of them only focus on TPLs that are written in Java except for NativeGuard.} However,
	{The state-of-the-art isolation methods are not very practical. They need to modify the Android apps or Android system. Furthermore, none of them are publicly available. It is impossible for general users to apply them.}
	}
}
\end{comment}

%################

\subsection{TPL Maintenance}
\label{Sec:TPL maintenance}

{For TPL maintenance, we mainly introduce research on TPL dependency conflicts and in-app TPL updating. The related work can be seen from TABLE~\ref{tbl:TPL_maintenance}.}

\begin{table}
%\vspace{-5pt}
\centering
\caption{A summary of related work on TPL maintenance}
\vspace{-2mm}
\scalebox{0.9}{
\begin{tabular}{llc}
\toprule
\textbf{Function}  &\textbf{Tool/First Author} & \textbf{Year} \\
\midrule

\multirow{3}{*}{\textbf{Dependency Conflicts}} &        \multicolumn{1}{l}{LibHarmo~\cite{huang2020interactive}} & \multicolumn{1}{l}{2020}\\
    &  \multicolumn{1}{l}{DECCA~\cite{DC2018FSE}}   & \multicolumn{1}{l}{2018}   \\    
    &  RIDDLE~\cite{RIDDLE2019ICSE}  & 2019 \\
\midrule

\multirow{8}{*}{\textbf{TPL Updating}}   &          Wang et al.~\cite{Wang2020ICSME}  & 2020 \\
     &  Yasumatsu et al.~\cite{Yasumatsu2019codaspy} & 2019    \\

   & APPCOMMUNE~\cite{APPCOMMUNE2019SANER}  & 2019   \\
& Salza et al~\cite{TPLsurvey2019} & 2019  \\

&Salza et al.~\cite{Salza2018ICPC}  & 2018 \\
& Ogawa et al.~\cite{CANDARW2018}    & 2018  \\

& {Derr et al.~\cite{Derr2017ccs}}  &  {2017}  \\
& Ruiz et al.~\cite{Israel2016software} & 2016 \\
\bottomrule
\end{tabular}
}
\vspace{-2mm}
\label{tbl:TPL_maintenance}
\end{table}

\subsubsection{Dependency Conflicts}

%\revise{在这里写一个总分的结构，现有的DC的工具有三个，分别是。。。。，然后我们依次介绍一下 these tools}
\noindent $\bullet$ \textbf{Research Background.}

\noindent{Dependency conflict (DC) is another essential issue in Android TPLs. Nowadays, many apps and TPLs directly or transitively depend on other TPLs. 
% The imported TPLs may depend on the same TPL but different versions. 
% Dependency conflicts occur when JVM loads one version and shadows the others but the loaded version cannot cover the required features by the project, leading to runtime exceptions. 
{As a result of the intensive dependencies on TPLs, an app may depend on multiple versions of the same TPL or class, among which only one version will be loaded.
Dependency conflict occurs when the loaded version cannot cover the features required by the app, leading to runtime exceptions or system crashes~\cite{DC2018FSE}.
%Apart from this, extensive uses of TPLs in Android apps can also take potential security risks of dependency conflicts. The dependency conflicts can also lead to serious consequences such as system crashes.
}

\noindent $\bullet$ \textbf{Existing Research.}

\noindent In our research scope, we find three related work. 
{Even though all of them focus on Java projects, it still provides insights to the research on the DC issues of Android platform, which involves extensive use of Java libraries.}
% Even though all of them focus on Java libraries, Android apps also extensively use various Java libraries.
Therefore, we include the three research in our paper repository.}

\textbf{\textbf{DECCA}~\cite{DC2018FSE}} is a DC detection tool based on static analysis, which supports the assessment of DC issues' severity.
% They performed an empirical study on 135 real-world DC issues from 71 open-source Java projects to characterize the common manifestation patterns and fixing patterns of the DC issues.
Through an empirical study on real-world DC issues, they categorized DC issues in three patterns: conflicts in library versions, conflicts in classes among libraries, and conflicts in classes between projects and libraries.
% They argue that DC issues are overlooked by developers because the existing build tools fail to analyze the real impact of DC in depth.
DECCA first constructs the TPL dependency tree from the dependency management script (e.g,pom.xml) and identifies duplicate classes with the same fully-qualified name.
DECCA further deduces whether the duplicate classes are loaded or shadowed based on the class loading mechanisms of the build tools.
DECCA then extract methods referenced by the host project from the duplicate classes and deduce a DC severity level according to the subset relationship among the referenced, loaded, and shadowed method set.

% \tm{
% \textbf{\textbf{Decca}~\cite{DC2018FSE}} is a DC detection tool based on static analysis.
% They performed an empirical study on 135 real-world DC issues collected from 71 open-source Java projects to characterize the common manifestation patterns and fixing patterns of the DC issues.
% They argue that DC issues are overlooked by developers because the existing build tools fail to analyze the real impact of DC in depth, and developed their own DC detection tool Decca which supports the assessment of the DC severity levels.
% Decca first constructs the TPL dependency tree from the dependency management script (e.g,pom.xml) and identifies duplicate classes with the same fully-qualified name.
% Based on the class loading mechanisms of the build tools, Decca further deduces whether the duplicate class are loaded or shadowed, and extracted the method within.
% Decca then constructs a referenced method set containing the method from the duplicate classes that is directly or indirectly referenced by the host project, and deduce a DC severity level according to the subset relationship among the referenced, loaded and shadowed method set.}
% % They categorized DC issues in three patterns: conflicts in library versions; conflicts in classes among libraries; and conflicts in classes between project and libraries.

{\textbf{RIDDLE}~\cite{RIDDLE2019ICSE}} is an automated test generation approach for JAVA projects with DC issues. It further collects crashing stack traces for reproducing and debugging the relevant issues.
% \textbf{RIDDLE}~\cite{RIDDLE2019ICSE} is a dependency conflicts (DC) detection system, which automatically generates test cases and collects crashing stack traces for the project, and is validated to detect 20 DCs in 19 real-world Java projects. 
%
%For test generation, RIDDLE firstly takes the project with DC issues as input and uses Soot~\cite{soot} to generate Control Flow Graphs (CFG) for static analysis to identify risky method sets. 
%Then, RIDDLE forces the condition branches of CFG to be TRUE or False to call those risky methods. 
%Based on EVOSUITE, the system applies a genetic algorithm to develop test case sets.
%Besides, RIDDLE checks whether the generated test case sets covers all risky methods by restoring mutated branches in CFG.  
To assist developers in facilitating diagnosing and fixing DC issues, RIDDLE can report a) the root cause of dependency conflicts in the project, b) risky method set, and c) test cases with the fewest unrestored branch conditions and corresponding program variant.
{RIDDLE is a follow-up work of DECCA.}
DECCA base on the static analysis to identify dependency conflicts and can provide the severity levels of dependency conflicts. 
RIDDLE uses DECCA to identify the risky methods causing the dependency conflict issues, and can also offer stack trace information and failure-introducing conditions with the help of dynamic analysis.
RIDDLE provides more practical value to developers, which can help them reproduce and debug the dependency conflict issues.

{\textbf{LibHarmo~\cite{huang2020interactive}} tackles the TPL version inconsistency problem in JAVA projects, which is one of the major cause of the DC issues according to DECCA.
LibHarmo is an interactive and effort-aware library version harmonization technique.
It first detects TPL version inconsistencies and false consistencies (a separate declaration of the same versions of TPL) through statically analyzing the inheritance relationships and the declared TPL versions among POM files.
LibHarmo then recommends harmonized TPL versions for developers to choose, with detailed harmonization efforts manifested by the number of called libraries APIs deleted or changed in the harmonized version and the number of calls to those libraries APIs in the project. 
% Developer can choose a version according to the harmonization effort report.
% , and LibHarmo will automatically refactor the POM files based on developer's decision.
LibHarmo significantly outperforms Maven's \code{enforcer} Plugin with 4X inconsistencies detected in their evaluation.
}

{
Compared to DECCA and RIDDLE which mainly focus on the examination of DC issues, LibHarmo focuses more on the maintenance and harmonization of TPLs.
Specifically, LibHarmo differs from DECCA and RIDDLE in that,
1) When detecting version inconsistencies, DECCA and RIDDLE work on the class level, a finer granularity, while LibHarmo works only on the library level.
2) DECCA and RIDDLE only focus on inconsistencies that cause DC issues, while LibHarmo also detects library version inconsistencies that do not cause DC issues for maintenance purposes.
3) LibHarmo does not support transitive library dependencies, since the inconsistencies in transitive dependencies are often out of developers' control hence cannot be harmonized.
}

\subsubsection{TPL Updating}

$\bullet$\textbf{ Research Background}

\noindent As mentioned before, software reuse has become a very common practice in mobile app development. Many developers use third-party libraries to facilitate the development progress. However, in-app third-party libraries could amplify their host apps' attack surface at the same time, if mobile apps contain vulnerable TPLs. Derr et al.~\cite{libscout2016ccs} reported that about 70\% apps with TPLs have the library outdated problem, which means even though the vulnerabilities of TPLs have been fixed, developers may still use the vulnerable versions in apps. Fast response to these vulnerable TPLs can decrease these threats in the mobile app ecosystem.

From developers' perspectives, they need to consider many factors when updating apps. For example, whether the new TPLs will affect users' experience or introduce bugs, or whether it is worth the extra effort to update a TPL, the cost/benefit ratio. Current research help reveal many unknown parts in TPL updating. We introduce their research one by one in the following subsections.

\noindent$\bullet$\textbf{Existing Research}
                                     
%The five papers~\cite{Yasumatsu2019codaspy, Salza2018ICPC,CANDARW2018,Derr2017ccs,Israel2016software} mainly focus on the library updating.
\noindent Table~\ref{tbl:TPL_maintenance} provides a summary of {eight} papers focus on the library updating. 
{Except two papers~\cite{CANDARW2018, APPCOMMUNE2019SANER} propose prototypes to help developers update TPL automatically,} the remaining papers mainly investigate the factors about library updating.
% Except this paper~\cite{CANDARW2018} propose a prototype to help developers update TPL automatically, the remaining papers investigate the factors about library updating.  
%One of them~\cite{Derr2017ccs} comes from the security field, and the other three are from software engineering. 
We can see that compared to other studies, the research on library updating appeared later, where the earliest began in 2016.

\begin{comment}

%
\begin{table}[h]
%\vspace{-5pt}
\centering
\caption{A summary of TPL updating literature}
\vspace{-2mm}
\scalebox{0.9}{
\begin{tabular}{lcr}
\toprule
\textbf{Tool/First Author}  & \textbf{Year}  & \textbf{Venue} \\
\midrule
\rowcolor{gray!20}
Wang et al.~\cite{Wang2020ICSME}  & 2020 & ICSME \\
Yasumatsu et al.~\cite{Yasumatsu2019codaspy} & 2019   & CODASPY \\
\rowcolor{gray!20}
APPCOMMUNE~\cite{APPCOMMUNE2019SANER}  & 2019 & SANER  \\
Salza et al~\cite{TPLsurvey2019} & 2019 & ESE(J) \\
\rowcolor{gray!20}
Salza et al.~\cite{Salza2018ICPC}  & 2018 & ICPC \\
Ogawa et al.~\cite{CANDARW2018}    & 2018 & CANDARW \\
\rowcolor{gray!20}
{Derr et al.~\cite{Derr2017ccs}}  &  {2017} & CCS \\
Ruiz et al.~\cite{Israel2016software} & 2016 & IEEE Software \\
\bottomrule
\end{tabular}
}
% \begin{center}
%     ESE: Empirical Software Engineering
% \end{center}
\vspace{-2mm}
\label{tbl:libupdating}
\end{table}
\end{comment}

\textbf{Wang et al.~\cite{Wang2020ICSME}} performed an empirical study of TPL updates on 806 well-maintained JAVA projects from Github.
They find that only 3.5\% of these projects were using the latest versions of TPLs. 
14.1\% projects have never updated any declared libraries and 40.8\% updated at most 50\% of the declared libraries.
% while 38\% adopt TPLs over 10 versions away from the latest.
As for the update delays, only 23.1\% of projects update their dependencies within a month, while over half TPLs updated at a lag of over 60 days.
They also reveal that 56\% of projects adopt buggy library versions.
They further propose a bug-driven prototype to alert users of risky library API calls in their projects based on whether buggy library methods are invoked.
They define buggy library methods as the changed methods in the bug-fixing security patch.
The prototype system also attempts to evaluate the effort of updating a TPL version based on the changes (deleted or modified) of called library APIs in the project.
Though they mainly focused on the TPL updates of JAVA projects, this study still provides insights into Android TPL updates.

\textbf{Yasumatsu et al.~\cite{Yasumatsu2019codaspy}} attempt to find how long it takes for developers to update a TPL and what are the essential factors that determine developers into update a TPL. 
% They collected a dataset including apps and the TPLs with versions. 
{They find that 50\% of apps update the new version libraries more than 3 months after the release of new libraries. About 50\% of apps still use outdated libraries for more than 10 months. Besides, they also found that popular apps get faster library updating responses from developers. 
% Developers respond to library updating based on libraries' type, they usually give a quick response to ad libraries.
Also, developers tend to update the advertising libraries faster.}
Furthermore, this paper not only discusses the updating of TPLs from time dimension, apps and TPLs related attributes, but also studies the reasons that can promote developers to fix vulnerable TPLs. They found that if a vulnerability has been targeted by Google's App Security Improvement Program (ASI), developers tend to fix the vulnerability. 

\textbf{APPCOMMUNE~\cite{APPCOMMUNE2019SANER}} {proposes a novel TPL sharing system in Android platform which separates TPLs from app codes and centrally managed all TPLs in a new app.
Apps can still access the separated TPLs (Java \& Native libraries) through a dynamic loading mechanism.
The new manager app updates the TPLs with a conservative strategy to ensure stability and each TPL will be updated to the latest version.% possible provided that the signatures of the library APIs used in app codes remain unchanged.
In addition to providing in-time TPL updates, APPCOMMUNE also saves storage and bandwidth by sharing TPLs.
%[An evaluation covering 212 popular TPLs and 502 real-world apps on different Android devices and API levels demonstrates the feasibility of the system.(Not sure if this sentense is needed.)]
}

\textbf{Salza et al.~\cite{TPLsurvey2019}} conducted an empirical study on 2,752 mobile apps by interviewing 73 mobile developers to reveal the updating problems of TPLs in mobile apps. They find that developers seldom update TPLs and they usually prioritize the updating of GUI-related TPLs. The main reason why they update the TPLs is to try to avoid the propagation of vulnerable libraries. %Nevertheless, they also just focus on one side of the third-party library (updating).

\textbf{Salza et al.~\cite{Salza2018ICPC}} analyzed TPL updating on the evolution history of 291 open-source Android projects on F-Droid. 
% They conducted empirical studies on five research questions to understand TPL updating. 
They find that developers merely update the old version of TPLs in their apps: only 15\% are updated constantly, and about 63\% are never updated. Besides, the TPL updating includes not only the upgrades but also the downgrades of TPLs when issues are induced in the previous upgrade.
Also, TPLs related to UI and support tools are more likely to be updated; developers prefer to update the TPLs in high-rating apps. These findings are consistent with that of Yasumatsu et al.~\cite{Yasumatsu2019codaspy}.

\textbf{Ogawa et al.~\cite{CANDARW2018}} proposed a prototype to automatically update TPLs with the help of an external server. When users send an app to the remote server, this server first identifies the TPLs and their versions and updates TPLs if necessary, and then generates a new apk file with the updated TPLs and sends it back to the users.

% If the server finds that this app employs an old version, it will update the TPLs and generate a new apk file and then send it back to the user side. 
%The mobile device receives the new app and replaces the original apk file.
%

\textbf{Derr et al.~\cite{Derr2017ccs}} investigated 203 app developers and attempted to find the reason why app developers do not update TPLs in Android apps. 
% They also tried to give some implications to library developers with the poor TPL updating rate. 
Based on their survey, we can find that developers do not choose to update TPLs because of incompatible problems, difficulties at debugging, cost/benefit ratio, and unawareness of library updating. Besides, the main reason that developers choose to update TPLs is bug fixing.
%

%They found that many TPLs in apps are outdated. Since updating TPLs require developers' extra efforts and cost, developers need to trade off the pros and cons of updates. Therefore, these papers ~\cite{Salza2018ICPC,CANDARW2018,Israel2016software} try to find which TPLs should be updated by the developers.
%

\textbf{Ruiz et al.~\cite{Israel2016software}} conducted an empirical study regarding the ad library updates in Android apps. They conclude three reasons why developers try to update the ad libraries: 1) fix bugs, 2) add new functionalities, 3) improve personal information management.

\noindent$\bullet$\textbf{Summary of TPL Updating}

%As for the research on TPL updating, we can find most papers are empirical study and there are only one paper about how to update TPL automatically. 
% (the relations between the vulnerable in-app TPLs and apps)
\noindent Regarding current empirical study on TPL updating, these studies have analyzed TPL updating from different perspectives (e.g., the app developers, security problems).
Overall, existing empirical studies have revealed most mysteries on TPL updating, while a few studies focus on how to solve in-app TPL updating automatically. We believe that if future researchers can invent more effective methods to help developers automatically update vulnerable TPLs in time, it will be a significant contribution to the community. 
% If some data about the vulnerable TPLs can be released, that can help decreasing the security threats for mobile apps.

Based on the aforementioned studies, we can find many of the research insights share intersections.
We provide the significant conclusions of existing library updating research in the following:
\begin{itemize}
	\item Most library upgrades (85.6\%) do not require modification of the host app code~\cite{Derr2017ccs,Israel2016software}.
	\item Most commonly-used library versions (97.8\%) with a known vulnerability could be easily fixed by replacing a fixed library version~\cite{Derr2017ccs}.
	\item Most Apps have problems with delaying the update of TPLs~\cite{Salza2018ICPC,Yasumatsu2019codaspy,Derr2017ccs}.
	\item Developers are more willing to update GUI-related TPLs and advertisement libraries~\cite{Salza2018ICPC,Yasumatsu2019codaspy}.% and developers seldom update TPLs that provide support tools~\cite{Salza2018ICPC,Yasumatsu2019codaspy}.
	\item Developers seldom update the TPLs if the update parts of the TPL are not invoked by the host apps~\cite{Salza2018ICPC}.
	%TPLs performing tasks that go beyond the development of functional requirements of apps are seldom updated~\cite{Salza2018ICPC}.
	\item TPL update frequency of apps with high rating score is higher than that of apps with low rating~\cite{Salza2018ICPC,Yasumatsu2019codaspy}.
	
\end{itemize}

{
\noindent The common delay response of library updating is due to the lack of timely information and incentives. We suggest that TPL vendors and app markets should help decrease the spread of vulnerable and outdated TPLs with an online platform for developers to tell if their apps may include vulnerable or outdated TPLs.
% We suggest that all app markets take part in security fix campaign. They should offer an online platform to developers and security researcher to tell if their apps may include vulnerable or outdated TPLs. 
At the same time, the app markets should set up a penalty system:
TPL vulnerabilities will be reported to app developers should update the fixed version within a certain period; otherwise, they will be fined or their app will be deleted from the market. We believe this strategy can effectively decrease the risks of TPL vulnerabilities and outdatedness to users.}

\subsection{TPL Attribute Understanding}
\label{sec: tpl attribute}

\subsubsection{Research Background}
{TPL attributes understanding is also a critical part of TPL analysis. These papers help to understand the in-app TPLs from various perspectives.
We discuss this direction by splitting current research into two parts: ad library related topics and general library related topics.
Because in-app ads have a significant position in TPLs and mobile app ecosystem. For one thing, mobile advertising is essential for monetizing and ensuring that developers can earn revenue. For another, ad libraries have abundant UIs, which can directly affect users' experiences. Therefore, there are many interesting points we can explore and study.}

{The scope of TPL attribute-related studies covers a wide range, such as the relationship between TPLs and app maintenance~\cite{structure2012ICSM}, analyzing the relationship between apps quality/rating and TPLs {\cite{Gui2017WhatAO, Mojia2014software}}, and the impact of TPLs on repackaging detection or malware detection {\cite{Lili2016SANER}}.
For the ad-related studies, previous studies revealed the UIs, permission characteristics of ad libraries {\cite{Madscope2015Mobisys}}, the symbiotic relationship understanding between different types of advertising and apps, and the collected targeting information understanding.
Gui et al.~\cite{ICSE2015jiapping} conducted an empirical study to analyze the extra cost for host apps due to the inserting ad libraries {\cite{ICSE2015jiapping}}. Some researchers figured out what specific information that ad collects from users and what targeting information they published to users {\cite{adempirical2015compscience,wkshps2014}}. Some research focused on APIs and permissions of in-app ads {\cite{book2013spsm,MOST2013}}.}

%the specific classification can be seen from the Table~\ref{tbl:empirical}. Based on the Table~\ref{tbl:empirical}, we also can find that the importance of ad analysis in the analysis of TPLs, it also accounts for the largest rate.
%
%Strictly speaking, the empirical studies of TPLs analysis do not have clear boundaries because some papers may cover several research questions. 

\subsubsection{Existing Research}

According to our paper repository, we classify existing literature into four categories as shown in Table~\ref{tbl:empirical}.
These studies mainly discuss the impacts of TPLs on app rating~\cite{Gui2017WhatAO,jin2019madlens,ICSE2015jiapping,Mojia2014software}, the relations between the apps' quality and TPLs~\cite{jin2019madlens,Gui2017WhatAO,ICSE2015jiapping,Mojia2014software}, what information the ad libraries will collect from the target users and how this information is exploited, library recommendation~\cite{AppLibRec2017Internetware}, the impact of TPL usage on the app maintenance~\cite{structure2012ICSM} and how in-app TPLs impact the downstream detection~\cite{Lili2016SANER}. More detailed analysis is as follows.

\begin{comment}

In these categories, many articles discuss the impacts of ad libraries on app rating. 

These papers~\cite{jin2019madlens,Gui2017WhatAO,ICSE2015jiapping,Mojia2014software,Lili2016SANER} can also be considered as libraries exterior analysis. Apart from the study conducted by Li et al.~\cite{Lili2016SANER} that they provide the specific reasons why TPLs can affect the detection result of apps analysis, the remaining papers~\cite{jin2019madlens,Gui2017WhatAO,ICSE2015jiapping,Mojia2014software} study the relations between the apps' quality and TPLs.

Considering the library rating analysis, most of the research on library rating {\cite{Gui2017WhatAO,ICSE2015jiapping, jin2019madlens,Mojia2014software}} usually involves the review analysis, as negative comments could affect other users' choices, and scores and complaints can affect the rating directly. By analyzing users' comments, we can summarize the corresponding issues and give some implications to developers.
%From Table~\ref{tbl:HighLevel_empirical_cmp}, we can also find that most TPL analysis needs first to identify TPLs, from which we can see that the identification of TPLs plays a vital role in the whole study.
\end{comment}

\begin{table}[t]
%\vspace{-5pt}
\centering
\caption{A summary of TPLs attribution understanding}
\vspace{-2mm}
\begin{tabular}{llc}
\toprule
\textbf{Function}  &\textbf{Tool/First Author} & \textbf{Year} \\
\midrule
\multirow{11}{*}{\textbf{Ad Analysis}}  &
 Ahasanuzzaman et al.~\cite{md2020studying} &  2020 \\
 & Ahasanuzzaman et al.~\cite{ahasanuzzaman2020longitudinal} &  2020 \\
 & MAdLens~\cite{jin2019madlens} & 2019 \\
& \multicolumn{1}{l}{Gui et al.~\cite{Gui2017WhatAO}} & \multicolumn{1}{c}{2017} \\
                              & Book et al.~\cite{adempirical2015compscience} & 2015 \\
                              & \multicolumn{1}{l}{Madscope~\cite{Madscope2015Mobisys}} & \multicolumn{1}{c}{2015} \\
                              & Ullah et al.~\cite{wkshps2014} & 2014 \\
                              & \multicolumn{1}{l}{Book et al.~\cite{book2013spsm}} & \multicolumn{1}{c}{2013} \\
                              & Tongaonkar et al.~\cite{PAM2013} & 2013 \\
                              & \multicolumn{1}{l}{Book et al.~\cite{MOST2013}} & \multicolumn{1}{c}{2013} \\
 & Vallina-Rodriguez et al.~\cite{vallina2012breaking} & 2012 \\

\midrule
\multirow{2}{*}{\textbf{Rating Analysis}}  
                                  & \multicolumn{1}{l}{Gui et al.~\cite{ICSE2015jiapping}} & \multicolumn{1}{c}{2015} \\
                                  & Ruiz et al.~\cite{Mojia2014software} & 2014 \\
\midrule
\textbf{Lib Recommendation}   & {AppLibRec~\cite{AppLibRec2017Internetware}}  & {2017}  \\
\midrule
\textbf{Miscellaneous Analysis}    & Li et al.~\cite{Lili2016SANER} & 2016 \\
                                
                                  &  Bauer et al.~\cite{structure2012ICSM}  & 2012 \\

\bottomrule
\end{tabular}
\vspace{-2mm}
\label{tbl:empirical}
\end{table}

\noindent$\bullet$\textbf{ In-app ad attributes.}

{
Ahasanuzzaman et al.~\cite{md2020studying} studied the integration strategies of ad libraries in 1,837 top free apps on Google Play. 
They classified the apps into ad-displaying apps and non-ad-displaying apps through statically analysing whether ad-displaying methods were invoked in-app activities. 
They discovered 22.5\% of the non-ad-displaying apps integrated Google AdMob ad library for analytical purposes instead of ad displaying.
They also find 57.9\% of the ad-displaying apps integrate more than one ad library, which is a common practice for more popular apps.
They manually analyzed 10\% of the apps with multiple ad libraries, and identified four integration strategies: external-mediation, self-mediation, scattered, and mixed.
}

{
Ahasanuzzaman et al.~\cite{ahasanuzzaman2020longitudinal} studied the evolution of 8 most popular ad libraries from Apr. 2016 to Dec. 2018. 
They find ad libraries are evolving continuously with a median release interval of 34 days.
While the size of all but two studied ad libraries are increasing over time, three approaches were used by ad library developers trying to reduce the library size.
They investigated the motivations for a new release of the ad library through manual analysis of the release notes, which include supporting new Android platform and the video ad functionality.
They also proposed a reference architecture of the ad libraries which could be useful for ad library developers.
}

Jin et al.~\cite{jin2019madlens} proposed a taxonomy of mobile ads that classified them into five types: embedded, popup, notification, offerwall, and floating. 
They further developed MAdLens, a static analysis system for Android apps to identify ad networks and relevant APIs.
Then they performed a large-scale study using MAdLens and discovered that developers tend to be conservative when embedding advertising TPLs, with 71\% apps containing at most one ad network.
They also find that using too many advertising TPLs in one app will annoy users and lead to a low rating.
% He et al.~\cite{MAdLens2018infocom} developed a prototype named MAdLens. Based on the ad impression models and UI size, they classify the ads into five types: Embedded, popup, notification, offerwall, and floating. They also investigated the popular ad types and whether different ad types can co-exist in the same app and how many ad libraries are acceptable in an app for users.
%

{Gui et al.~\cite{Gui2017WhatAO} investigated diverse topics of ad-related complaints from users. They found most complaints about ads are UI-related topics, including the display frequency, the timing of when ads are displayed and the ads display location. They found that app developers usually attempt to give more exposure of ads to users help add chances of ad impressions and clicks; in this way, they hope earn more ad revenues. Based on this work, they pointed out that improper exposure may lead to the bad user experience and even has negative impact on app's rating.}
%
%mainly analyzed the complaint reviews on ads and summarized the common topics and finally gave some implications for users and developers. \tm{what implications?}}
% Their team also conducted the other work to find the hidden cost of in-app ads for developers~\cite{ICSE2015jiapping}.
%

{Suman ~\cite{Madscope2015Mobisys} tried to understand what targeting information the mobile apps sent to the ad networks (ad libraries) and how ad networks employ this information for targeting users. Thus, the author proposed a tool named MAdScope which can probe ad networks to characterize their targeting mechanism and learn the targeting behaviors of users.
The author found the ad libraries usually tend to collect the users' location, device information, demographics and long-term behaviors. Besides, they found that the targeting information has a statistically significant impact on how an ad library selects ads.}

Ullah et al.~\cite{wkshps2014} conducted a comprehensive analysis of the ads serving mechanism of AdMob~\cite{admob}. They pointed out that the level of targeting service in the Mobile Ads market is still quite low. We still have a long way to go in personalization and targeting advertising services. They suggested that future researchers and developers can make more efforts on efficient usage of collected users' date, which can help make better service to users based on the targeting service.

Besides, Book et al.~\cite{adempirical2015compscience} also investigated the ad targeting of Google AdMob library. They showed that AdMob was targeted on the application, user location, time, and real profile of users. They found that the targeting of mobile ads has some relations with the users' profile. The in-app ads were associated with the device IDs, which may bring the hidden dangers of privacy leakage. 

Tongaonkar et al.~\cite{PAM2013} investigated the mobile apps' behavior patterns from the in-app ad flow. They mainly analyzed the different traffic patterns in different TPLs. They believed that is a new direction for analyzing the usage behavior of mobile apps based on ad flows.

Based on our previous analysis and Fig.~\ref{fig:IPCmodule}, we can know that the ad libraries can obtain sensitive information through the system calls or the library API calls. Book et al.~\cite{book2013spsm} collected 103 ad-related APIs and surveyed the relationship between these APIs and the privacy leakage of mobile apps. They found that these APIs can have access to users' personal information and device profile information. Besides, these privacy-related APIs are widely used by the top popular libraries. They also found that the system calls and library API calls to get sensitive information are two independent processes.
Book et al.~\cite{MOST2013} researched the ad permission ecosystem and showed the use of the permission of ad libraries. Book et al. adopted the static analysis by extract the APIs within the ad libraries and corresponding permissions to investigate the particular risks to user privacy and security.
They investigated the use of the eight permissions of the ad libraries. They reported that ad libraries were increasingly making use of permissions which were requested by the host apps. They witnessed a growth in the usage of various dangerous permissions that could pose potential privacy risks.

{
Vallina-Rodriguez et al.~\cite{vallina2012breaking} performed a large-scale measurement study of mobile ad traffic on an anonymized data set from a major European mobile network containing 1.7 billion traffic connections. They proposed a rule-based approach to identify and classify HTTP-based ad traffic. 
They found that ad traffic is a significant component of all mobile traffic, and objects in ad traffic are constantly re-downloaded.
To alleviate the redundant energy overhead introduced by mobile ad traffic, they further proposed AdCache, a cached-based ad delivery system.
% They found that ad traffic is a significant component of all mobile traffic, taking up 5\% of the total traffic for 50\% of Android users. They discovered many objects in ad traffic are constantly re-downloaded, mobile ad traffic 
}

\begin{comment}

\begin{table}[h]
\centering
\caption{Permissions used in Ad libraries}
\vspace{-2mm}
\scalebox{0.9}{
\begin{tabular}{cc}
\toprule
\multicolumn{2}{c}{\textbf{Permissions}}  \\
\midrule
\multicolumn{1}{l}{INTERNET} & \multicolumn{1}{r}{ACCESS\_NETWORK\_STATE} \\

\multicolumn{1}{l}{READ\_PHONE\_STATE} & \multicolumn{1}{r}{ACCESS\_FINE\_LOCATION} \\

\multicolumn{1}{l}{WAKE\_LOCK} & \multicolumn{1}{r}{ACCESS\_COARSE\_LOCATION}  \\

\multicolumn{1}{l}{VIBRATE}  & \multicolumn{1}{r}{ACCESS\_WIFI\_STATE} \\
\bottomrule
\end{tabular}
}
\vspace{-2mm}
\label{tbl:permissions}
\end{table}
\end{comment}

\noindent $\bullet$ \textbf{ Rating Analysis.}
Gui et al.~\cite{ICSE2015jiapping} studied the hidden cost of in-app ads. They selected 21 {top popular} apps from Google play and analyzed the extra costs of ad libraries from five aspects: app performance, energy consumption, network usage, maintenance effort for ad-related code, and app reviews. The results show that the apps with ads consume: 48\% more CPU time, 16\% more energy and 79\% more network data. Besides, developers need to spend extra energy to maintain the apps due to the updating of ad libraries. They also found that the complaints on ads can affect the app rating.
Ruiz et al.~\cite{Mojia2014software} conducted an empirical study on the relationship between the ad libraries and the rating of apps. They found that specific ad libraries (i.e., Wooboo, Leadbolt, Airpush) have negative impacts on apps' ratings.

{
Considering the library rating analysis, most of the research on library rating {\cite{Gui2017WhatAO,ICSE2015jiapping, jin2019madlens,Mojia2014software}} usually involves the review analysis, as negative comments could affect other users' choices, and scores and complaints can affect the rating directly. By analyzing users' comments, we can summarize the corresponding issues and give some implications to developers.
}

\noindent$\bullet$\textbf{ Library Recommendation.}
Given a new app, AppLibRec~\cite{AppLibRec2017Internetware} could recommend third-party libraries based on the app's similar apps. AppLibRec combines the topic modeling techniques and collaborative filter component~\cite{recommendaerSystwms2002Bueke} to perform libraries' recommendation. It implements two steps analysis: README file (textual description) based analysis (RM-based) and Libraries based analysis (Lib-based). In the RM-based analysis, AppLibRec employs the topic model algorithm Latent Dirichlet Allocation~\cite{LDA2003} to extract topics from the README files (textual description). The libraries will be recommended based on the similar topic distribution.
In Lib-based module, collaborative filtering is used to a recommendation based on the apps' similarity.

\noindent $\bullet$ \textbf{ Lib Exterior Analysis.}
Li et al.~\cite{Lili2016SANER} extract 1,113 common libraries from Android apps on the Google store scale. At the same time, they clarified the impact of common libraries on the results of malware analysis, repackaged apps detection and app analysis. They also conduct an empirical investigation and evaluation of the use of common libraries in apps.

{Bauer et al.~\cite{structure2012ICSM} proposed a systematic approach to assess the impact of TPL usage on a project in terms of maintainability. 
They further provide a guidance for pre-selecting significant TPL candidates based on their entangledness with the project (manifested mainly by the number and scatteredness of method calls to the TPL) to reduce the assessment effort.
% They further provide a ranking of the TPLs based on their entangledness with the project, which is manifested mainly by the number and scatteredness of method calls to the TPL.
% Through this ranking TPL candidates with the most significance to the project can be pre-selected to reduce the assessment effort.
An industrial case study indicates the effectiveness of the approach.}
% They further provide a guidance and tool support for pre-selecting TPL candidates with the most significance to the project to reduce the assessment effort.

\subsubsection{Summary}

Based on our study, we think we still can do some research on targeting advertising service.
On the one hand, to offer better services to customers, ad networking need to collect some profile user information. On the other hand, whether the collected user information can be used reasonably and effectively is a problem we need to think about. Previous study have pointed out that ad networks has the problem of over collecting the user information. Besides, a lot of targeting information has not been fully utilized to provided better customized services for users. Even now some fin-grained monitoring technique has been proposed to limit the over privileged issues, we still believe more empirical studies on this area should be done.

\section{Discussion}
\label{sec:implications}

\subsection{Threats to validity}

\noindent\textbf{Paper collection.}
%\xian{
We do not consider the following types of papers, including the books, Master or Ph.D. dissertations on TPL-related research.
Instead, we search these authors' articles related to the topics we are interested in and finally include those not in our paper repository as a supplement.
%and we finally find the content of our missed articles is presented in the peer-reviewed conferences and journals. }
Even though we have tried our best to collect TPL-related papers as many as possible by following the state-of-the-art SLR methodology, our search results may still miss some relevant papers. One possible reason may be that some existing repository search engines are not very accurate; they may provide some irrelevant papers or omit some papers. The other reason may be that our keywords cannot find all the relevant papers. Considering the first reason, we reviewed all the references of our collected papers and tried to find the papers that were not in our paper repository. To mitigate the second reason, we iteratively optimized our searching keywords and tried our best to find the synonyms in order to extend the search scope and cover as many relevant papers as possible. 
%We tried our best to find the synonyms as many as possible to extend the search scope. 
However, it is still possible to miss some synonyms in this work. We consider using the Natural Language Processing (NLP) technique as our future work to enhance the keywords for SLR. 

%\noindent \textbf{Venues for paper collection.}
%{Besides, we implement the ``major venue search'' by choosing the top 10 venues from 3 fields (i.e., Software engineering, Security, and Program language). \revise{The venues ranking depends on the CCF. This ranking has a specific bias for some communities and may not include some top venues for some communities, such as MSR. Besides, the ranking could be some minor changes for a few years.} 
%\ling{Not sure we should mention it or not.}
Besides, we implement the ``major venue search'' by choosing the top-tier venues from 3 fields (i.e., Software engineering, Security, and Program language).
Some related research may belong to a certain conference which is not included in our search scope. Besides, some papers may not be from top venues. To mitigate this threat, we also include some venues that are not from the top venues but they are pretty representatives, such as CODASPY and WiSec.

%\noindent\textbf{Data sources and metrics.}
%\xian{Considering some analysis, such as the TPL detection, there lack a unified dataset to compare the performance of each system. Therefore, we may miss some significant findings. We just summarize some conclusions from the literature; we may lack some profound insights. Because the obfuscating tools are various, even the same code obfuscation can be implemented in different ways and present different effects. Thus, we do not compare the capability of obfuscation-resilient for each tool.To alleviate this situation, we have already collected some dataset including the apps and TPL files, to conduct a series of evaluations to give more informative results.}

\subsection{Threat Reasons and Arm Race}
%\sen{WILL FOCUS ON SEC 6.2 and 6.3}
Based on the aforementioned research, we provide some useful insights and implications.

%\textbf{RQ1: What are the primary reasons for data and users' privacy exposure of third-party libraries?} 
\noindent \textbf{Reasons for data exposure in third-party libraries.}
If we want to solve the data and privacy leakage problems, we should first understand the potential channels of data exposure. Therefore, we summarize the reasons of data exposure based on the existing literature we collected. The main reasons are as bellows:
(1) The sandbox and permission mechanisms are working at the app level. Therefore, TPLs and host apps share the same storage space, permissions. TPLs can use protected APIs via permissions inherited from the host app to access sensitive information, visit the storage of the host app and get users' input from the host app.
(2) To increase developers' revenue and improve the targeted advertising, some developers and advertisers will deliberately collect more user-related information (e.g., users' interests, profiles, and demographic information). The adversaries may make use of this and steal users' information by adding some malicious code into the TPLs.
(3) Some third-party libs could dynamically update. It is impossible to find any security risks of dynamically update code based on the current version detection~\cite{pluto2016}.
(4) Due to the \code{ClassLoader} functionality and reflection mechanism. Some ad libs can download the suspicious payload at the runtime from the remote servers and execute it in the context of the host apps.
(5) Android allows developers to use some public/unprotected APIs without requiring any permission. These unprotected APIs can get access to platform-wide information. For example, collect the list of all apps installed on mobile phones~\cite{AdRisk2012Wisec}.
(6) Android provides a cross-platform compatibility mechanism that allows \code{JavaScript} code to run in a WebView object to invoke a set of callback functions through an interface. Through this interface, JavaScript ads can dynamically invoke other functions at the runtime, similar to Java reflection. For instance, Mobclix ad library can get access to location information by registering the \code{gpsStart(...)} function.

%\textbf{RQ2. What are the existing solutions for these security risks?}
\noindent \textbf{Existing solutions for security risks.}
(1) Considering the over privilege problem of the permission mechanism, there are two main solutions to solve this problem. One is rewriting the bytecode of TPLs. By rewriting the resource access strategies and sharing functions, it can achieve de-escalation. The other solution is rewriting the Android framework. A set of new permissions are developed for TPLs alone and let each TPL run in an independent space.
(2) To achieve the balance between the targeted information sending to the users by advertiser and privacy information fetching from users, researchers have proposed a framework that builds a dynamic feedback control loop which can adjust the level of privacy protection on mobile phones based on the advertising revenue. The prototype first decouples the host apps and ad libraries. It implements a real-time monitor of the control flow of private information and can control the exposed data to advertisers. The prototype can generate real information and fuzzy information and send it to the ad servers based on the value of privacy information.
(3) For the Java reflection and dynamic class loading, we usually adopt dynamic analysis to catch the potential risks.

%\textbf{RQ3: What suggestions can we give to future researchers? Are there limitations in current work? what can we do in the future?}
\subsection{Open Challenges and Research Directions}

Based on our investigation of state-of-the-art research work on Android TPLs, we summarize some limitations in the existing work and point out some topics that are worth further investigation. % ork and a new direction in this direction.

\noindent$\bullet$\textbf{ TPL detection.}
(1) Even though there are many TPL detection tools, most of them can reach a low recall~\cite{libdetect2020ASE}. A previous study~\cite{libdetect2020ASE} conducts an empirical study on these publicly available tools by using a well-designed dataset, which finds that these publicly available tools can only find out about half of in-app TPLs. Besides, only a few of them claim to be able to find the specific library version, but the results usually include many false positives. Based on our study, we find that the code differences of different versions are various. Some code differences among some versions are very tiny while some code differences may be very large. Library version identification has many practical usages, 
which can be used to find the license violation, vulnerable TPL versions, and some outdated libraries. Via identifying these vulnerable TPLs, we can inform the developers in time to replace these problematic TPLs.
However, we still have a long way to go in this direction at present, some challenges such as code optimization, large-scale analysis, high precision version identification, partial import, and customize imported TPL identification. We believe that if a well-designed and accurate version identification tool is implemented, it will be meaningful and essential to industry and academia.
% Library version detection is meaningful, which can help developers find vulnerable TPL and apps, decreasing security threats to users.
%
{(2) Current detection tools still cannot be effectively resilient to some sophisticated obfuscation techniques, such as class encryption and virtualization-based protection, even if a few apps use these technologies. Besides, many detection systems cannot handle the API hiding, control-flow randomization, and package flattening technique very well, and the future tool should try to find a more effective way. We suggest researchers can include richer semantic information in TPL identification, which can achieve better resiliency to code obfuscation~\cite{ATVHunter2021ICSE,libdetect2020ASE}. Besides, we also find the feature granularity can affect the resiliency, existing features usually include two granularities: package-level features and the class-level features. We find the class-level features can achieve better resiliency to code obfuscation.}
%We suggest researchers use class-level or more finer-grained features in the comparison stage, and we find finer-grained features are more effective in version-level identification.
%
(3) Existing TPL detection tools are not good at finding emerging TPLs; we suggest that future researchers pay attention to this limitation. On the one hand, the speed of the TPL update is rapid; on the other hand, current TPL detection methods all have hysteresis. There will be more third-party libraries in the future, and it is meaningful to detect these emerging TPLs in real-time.
(4) Based on our analysis, most previous tools only focus on Java library identification;  researchers can try to focus on native library identification and multiple-language TPL identification.

\noindent$\bullet$\textbf{Security and Privacy Issues.}
(1) Current vulnerable TPL study is very limited, existing research only focuses on several typical TPLs. We think that future research directions can be studied from two branches: 1) the known vulnerabilities 2) the unknown vulnerabilities. For the known vulnerabilities, we find many work can be done here. We lack a comprehensive understanding of these vulnerabilities, their impact scope and threats and so on. It is necessary to collect these third-part libraries with vulnerabilities and conduct in-depth research on them. For the unknown vulnerabilities of TPLs, we can try to how to find these vulnerabilities and detect these vulnerabilities in both TPLs and apps.
{(3) Future researchers also can analyse whether the display contents of ad libraries is appropriate for specific groups. For example, the app is designed for children, the ad contents including the violence, sex, gambling and so on should be considered inappropriate.}

\noindent$\bullet$\textbf{TPL Isolation.}
(1) More and more Android apps use some TPLs which are written in C/C++. We hope future work should pay more attention to the native code of TPLs isolation. The sophisticated tool should work on both ART or DVM mode and can effectively detect some dynamic behaves (such as Java reflection, dynamic classload, etc.) and limit the privileges and sensitive storage access.
(2) {Based on our observation, the existing solutions are of little practical value because the performance overheads are usually very large. It is impossible to locate a independent process for each TPL or several TPLs on the mobile phone, because TPLs usually need to interact with host app. }

\noindent$\bullet$\textbf{TPL Attribution Analysis.}
{Based on the Section~\ref{sec: tpl attribute}, we can find that about 70\% (11/16) of existing research focused on ad analysis. 
}
{(1)  Without a doubt, the ad library is an essential part of TPLs. Nevertheless, other TPLs still have some unique features; we should understand them in-depth.
{We only collected papers from 2012 to 2020, but we have found a few papers have begun to focus on other type of TPLs, such as the analytic libraries~\cite{tangtse:21a,firebase2021mobilesoft}.
Harty et al.~\cite{firebase2021mobilesoft} conducted an empirical study on Google Firebase that is an analytic TPL. They found its logs are less pervasive and less maintained than traditional logging code. Tang et al.~\cite{tangtse:21a} analyzed 25 special TPLs, named Application Performance Management (APM) that is also a kind of analytic library. They explored the usage patterns of APMs and discovered the potential misuses of APMs. The other TPLs' attributes have not been widely analyzed, we think this may be a chance for future researchers.
}
A large-scale TPL analysis on their features and the connection between the apps used these TPLs can be done. These can help researchers understand more about the relations between TPLs and Android apps. 
(2) Understanding new TPLs. We know that many TPLs are developed by Java. However, more and more TPLs are also developed by Kotlin~\cite{Kotlin} nowadays. It is also necessary to investigate these new types of TPLs.
(3) The compatibility of TPLs analysis also deserves to analyze thoroughly and deeply. We find that all existing studies on TPL updating are mainly focus on the reason understanding and the effects of delay updating. In addition to the propagation of vulnerabilities, the delayed updating of TPLs may also cause some compatibility issues (cf. Section~\ref{Sec:TPL maintenance}).}
\section{Related Work}
\label{sec:relatedwork}

To the best of our knowledge, this is the first literature review in the research area of Android third-party library analysis. 
The most previous surveys usually focus on other aspects of Android apps such as Android repackaging app detection, %~\cite{lili2019TSE,repackaged_survey2018ISDFS,xian2019saner}, 
Android malware detection,%~\cite{JoWUA2015,Sufatrio2015CSUR,faruki2015survey_security,surveymalware2020Qiu}
program analysis techniques used in app analysis, %~\cite{faruki2015survey_security,lili2017static}, 
and Android app testing. %~\cite{kong2019Testing,choudhary2015automated}.
{Based on our search, most literature review only focus on Android apps instead of Android TPLs. Till now, there is no taxonomy and comprehensive survey of third-party libraries in Android apps.}

%\revise{need to collect more malware detection survey and repackaged app detection survey}
%research on Android third-party library is extremely scarce and very one-sided. 
% and app privacy leakage~\cite{}), testing (e.g., fragmentation and updating), and used techniques in these research fields~\cite{}.}
%The most previous surveys~\cite{lili2019TSE,repackaged_survey2018ISDFS,kong2019Testing,lili2017static,Li2018RebootingRO,JoWUA2015,Sufatrio2015CSUR} usually focus on Android security problems, such as Android malware detection, repackaging detection, or app privacy leakage.

%\revise{rewrite the following part}
% For example, 
{Sadeghi et al.~\cite{assessAndroidsecurity2017TSE} conducted a literature review on the assessment of Android security. They proposed a comprehensive taxonomy to classify and characterize research on 336 papers on Android security published from 2008 to the beginning of 2016. Moreover, they also highlighted the key challenges and future research direction. Based on the research, they found the gap in existing research regarding special vulnerable features of Android, such as the native and dynamically loaded code. They encouraged future researchers to pay more attention to hybrid analysis techniques instead of pure static or dynamic analysis. The survey showed that future research could consider combinations of multiple apps and Android framework.}

Li et al.~\cite{lili2019TSE} conducted a survey on 59 state-of-the-art approaches of repackaged app detection. They compared different repackaging detection techniques and elaborated on current challenges in this research direction. They found that current research on repackaging detection was slowing down. %They also presented current open challenges in this direction and compared existing detection solutions. 
Besides, they also provided a dataset of repackaged apps, which can help researchers reboot this research or replicate current approaches.
%The researchers attempted to trigger innovative thinking and provide useful implications in this area.
Zhan et al.~\cite{xian2019saner} also compared existing repackaged app detection tools from the implementation perspective. They evaluated the state-of-the-art tools on a uniform dataset~\cite{Androzoo} and pointed out the advantages and disadvantages of each tool.
Baykara et al.~\cite{repackaged_survey2018ISDFS} investigated malicious clone Android apps. They revealed potential threats that can affect users' experience. Finally, they provided some potential solutions for these risks.

Qiu et al.~\cite{surveymalware2020Qiu} systematically investigated the challenges of the latest deep learning-based Android malware detection and taxonomy.
Rashidi et al.~\cite{JoWUA2015} discussed the existing Android security problems and existing security detection solutions from 2010 to 2015. They also gave a taxonomy of these systems and investigated their functionalities. At last, they also provided a review on the advantages and disadvantages of existing systems.
Sufatrio et al.~\cite{Sufatrio2015CSUR} also provided a survey and tried to classify existing security detection tools. They inspected their similarities and showed their differences. This paper also sheds light on the limitations and existing challenges.
Faruki et al.~\cite{faruki2015survey_security} first discussed Android security mechanisms and existing problems for these security mechanisms, malware penetration, and stealth techniques. Then they analyzed the static and dynamic analysis for malware detection techniques. They compared the advantages and disadvantages of these two analysis methods. Finally, they summarized existing systems based on their research purpose, methodology, and deployment.

Li et al.~\cite{lili2017static} concluded the state-of-the-art static analysis techniques of Android apps. They followed a well-defined systematic literature review methodology and collected 124 research papers. 
They mainly investigated the fundamental methods leveraged in related papers, the implementation methods, and relevant evaluation comparisons. 
%The most important contribution of this work is that they build a comprehensive dataset of repackaged app pairs, which can help other researchers replicate these tools and provide new implications for this area.

Kong et al.~\cite{kong2019Testing} reviewed 103 papers related to automated testing of Android apps. They summarized the research trends in this direction, highlighted the state-of-the-art methodologies employed, and presented current challenges in Android app testing. They pointed out that new testing approaches should pay attention to app updates, continuous increasing app size, and the fragmentation problem in the Android ecosystem. 
% \tm{maybe add ~\cite{choudhary2015automated}, both dynamic testing} \xian{add the ASE 2015[31] already}
%
Choudhary et al.~\cite{choudhary2015ASE} conducted a comprehensive comparison of the primary existing test input generation tools for Android. 
They evaluated their advantages and disadvantages, effectiveness, and the corresponding methodology based on four criteria: ease of use, Android framework compatibility, code coverage, and fault detection capability. Their study gives a landscape of the state-of-the-art Android input test tools and provides implications for future research direction.

%\tm{(As a TPL-related empirical study instead of literature review, this should be in Section 5.)}

As we can see, there are various surveys on Android, but still lack a systematic review for third-party libraries on the Android platform. Actually, third-party libraries have become an essential part of the Android ecosystem. Furthermore, we also find many studies on Android third-party library and they focus on different perspectives. Therefore, it is necessary to conduct systematic research in this field. Thus, in this paper, we concluded the significant research achievements in third-party library analysis and conducted a detailed investigation from the following aspects: research purpose, background, application, methodologies, etc.
%We believe our research has profound meaning because it fills the gap in this research direction. 
Upon filling the gap in this research direction, we believe our work can provide a clear overview of Android TPL-related studies and inspire fellow researchers to take a step further in this direction.

\section{Conclusion}
\label{sec:conclusion}

In this paper, we conducted a systematic literature review regarding the third-party library analysis on the Android platform. We employed a well-defined SLR method to get a comprehensive paper repository that includes 74 publications for Android TPL-related analysis. 
%We introduce some basic concepts of TPL analysis.
%Moreover, we give a brief summary of existing research. 
We first summarized a taxonomy of existing Android TPL-related studies from four dimensions. For each category, we provided a thorough review of the existing work, compared the state-of-the-art research from different perspectives, and summarized the key insights which could shed light on the follow-up research of the corresponding research line.
%We also give our summaries and insight into each research topic.
Finally, We discussed the open challenges and proposed new research ideas of Android TPL-related research.
We believe our work can give researchers a clear overview of this direction, and inspire them to come up with more creative ideas in this area, and develop more effective approaches to solve current challenges.

\IEEEpeerreviewmaketitle

\ifCLASSOPTIONcaptionsoff
  \newpage
\fi

\bibliographystyle{IEEEtran}
%\balance
\footnotesize
\bibliography{LibsurveyBiB}

\begin{thebibliography}{100}
\providecommand{\url}[1]{#1}
\csname url@samestyle\endcsname
\providecommand{\newblock}{\relax}
\providecommand{\bibinfo}[2]{#2}
\providecommand{\BIBentrySTDinterwordspacing}{\spaceskip=0pt\relax}
\providecommand{\BIBentryALTinterwordstretchfactor}{4}
\providecommand{\BIBentryALTinterwordspacing}{\spaceskip=\fontdimen2\font plus
\BIBentryALTinterwordstretchfactor\fontdimen3\font minus
  \fontdimen4\font\relax}
\providecommand{\BIBforeignlanguage}[2]{{%
\expandafter\ifx\csname l@#1\endcsname\relax
\typeout{** WARNING: IEEEtran.bst: No hyphenation pattern has been}%
\typeout{** loaded for the language `#1'. Using the pattern for}%
\typeout{** the default language instead.}%
\else
\language=\csname l@#1\endcsname
\fi
#2}}
\providecommand{\BIBdecl}{\relax}
\BIBdecl

\bibitem{MOBSCANNER2017ICSE-C}
S.~A. Baset, S.-W. Li, P.~Suter, and O.~Tripp, ``Identifying android library
  dependencies in the presence of code obfuscation and minimization,'' in
  \emph{ICSE-Companion}, 2017.

\bibitem{statista}
``statista,''
  \url{https://www.statista.com/statistics/266210/number-of-available-applications-in-the-google-play-store/}.

\bibitem{exodus}
``Exodus privacy,''
  \url{https://reports.exodus-privacy.eu.org/en/trackers/stats/}.

\bibitem{Wang2017ICSE-C}
H.~Wang and Y.~Guo, ``Understanding third-party libraries in mobile app
  analysis,'' in \emph{Proc. ICSE-C}, 2017.

\bibitem{li2017simidroid}
L.~Li, T.~F. Bissyand{\'e}, and J.~Klein, ``Simidroid: Identifying and
  explaining similarities in android apps,'' in \emph{The 16th IEEE
  International Conference On Trust, Security And Privacy In Computing And
  Communications (TrustCom 2017)}, 2017.

\bibitem{li2017understanding}
L.~Li, D.~Li, T.~F. Bissyand{\'e}, J.~Klein, Y.~Le~Traon, D.~Lo, and
  L.~Cavallaro, ``Understanding android app piggybacking: A systematic study of
  malicious code grafting,'' \emph{IEEE Transactions on Information Forensics
  \& Security (TIFS)}, 2017.

\bibitem{privacyleak}
V.~{Moonsamy} and L.~{Batten}, ``Android applications: Data leaks via
  advertising libraries,'' in \emph{International Symposium on Information
  Theory and its Applications}, Oct 2014, pp. 314--317.

\bibitem{privacyleak2014adhoc}
A.~{Short} and F.~{Li}, ``Android smartphone third party advertising library
  data leak analysis,'' in \emph{2014 IEEE 11th International Conference on
  Mobile Ad Hoc and Sensor Systems}, Oct 2014, pp. 749--754.

\bibitem{pluto2016}
S.~Demetriou, W.~Merrill, W.~Yang, A.~Zhang, and C.~A. Gunter, ``Free for all!
  assessing user data exposure to advertising libraries on android,'' in
  \emph{NDSS}, 2016.

\bibitem{YLDSN16}
L.~Yu, X.~Luo, X.~Liu, and T.~Zhang, ``Can we trust the privacy policies of
  android apps?'' in \emph{Proc. DSN}, 2016.

\bibitem{YLTSE19}
L.~Yu, X.~Luo, J.~Chen, H.~Zhou, T.~Zhang, H.~Chang, and H.~K. Leung,
  ``Ppchecker: Towards accessing the trustworthiness of android apps' privacy
  policies,'' \emph{IEEE Transactions on Software Engineering}, 2019.

\bibitem{DelDroid201983HAMMAD}
\BIBentryALTinterwordspacing
M.~Hammad, H.~Bagheri, and S.~Malek, ``Deldroid: An automated approach for
  determination and enforcement of least-privilege architecture in android,''
  \emph{Journal of Systems and Software}, vol. 149, pp. 83--100, 2019.
  [Online]. Available:
  \url{https://www.sciencedirect.com/science/article/pii/S0164121218302589}
\BIBentrySTDinterwordspacing

\bibitem{hammad2017ICSA}
------, ``Determination and enforcement of least-privilege architecture in
  android,'' in \emph{2017 IEEE International Conference on Software
  Architecture (ICSA)}, 2017, pp. 59--68.

\bibitem{sanAdBox2013ICC}
H.~{Kawabata}, T.~{Isohara}, K.~{Takemori}, A.~{Kubota}, J.~{Kani},
  H.~{Agematsu}, and M.~{Nishigaki}, ``Sanadbox: Sandboxing third party
  advertising libraries in a mobile application,'' in \emph{Proc. ICC}, June
  2013.

\bibitem{aframe2013ACSAC}
X.~Zhang, A.~Ahlawat, and W.~Du, ``Aframe: Isolating advertisements from mobile
  applications in android,'' in \emph{ACSAC}, 2013.

\bibitem{PEDAL2015MobiSys}
B.~Liu, B.~Liu, H.~Jin, and R.~Govindan, ``Efficient privilege de-escalation
  for ad libraries in mobile apps,'' in \emph{MobiSys}, 2015.

\bibitem{adsplit2012USENIX}
S.~Shekhar, M.~Dietz, and D.~S. Wallach, ``Adsplit: Separating smartphone
  advertising from applications,'' in \emph{Proceedings of the 21st USENIX
  Conference on Security Symposium}, 2012.

\bibitem{AdDroid20120ASISCCS}
P.~Pearce, A.~P. Felt, G.~Nunez, and D.~Wagner, ``Addroid: Privilege separation
  for applications and advertisers in android,'' in \emph{ASIACCS}, 2012.

\bibitem{FraudDroid2018FSE}
F.~Dong, H.~Wang, L.~Li, Y.~Guo, T.~F. Bissyand{\'e}, T.~Liu, G.~Xu, and
  J.~Klein, ``Frauddroid: Automated ad fraud detection for android apps,'' in
  \emph{Proceedings of the 2018 26th ACM Joint Meeting on European Software
  Engineering Conference and Symposium on the Foundations of Software
  Engineering}, 2018.

\bibitem{Decaf2014NSDI}
B.~Liu, S.~Nath, R.~Govindan, and J.~Liu, ``Decaf: Detecting and characterizing
  ad fraud in mobile app,'' in \emph{NSDI}, 2014.

\bibitem{MadFraud2014mobisys}
J.~Crussell, R.~Stevens, and H.~Chen, ``Madfraud: Investigating ad fraud in
  android applications,'' in \emph{Proc. MobiSys}, 2014.

\bibitem{liu2020maddroid}
\BIBentryALTinterwordspacing
T.~Liu, H.~Wang, L.~Li, X.~Luo, F.~Dong, Y.~Guo, L.~Wang, T.~Bissyand\'{e}, and
  J.~Klein, ``Maddroid: Characterizing and detecting devious ad contents for
  android apps,'' in \emph{WWW}, New York, NY, USA, 2020, p. 1715–1726.
  [Online]. Available: \url{https://doi.org/10.1145/3366423.3380242}
\BIBentrySTDinterwordspacing

\bibitem{libscout2016ccs}
M.~Backes, S.~Bugiel, and E.~Derr, ``Reliable third-party library detection in
  android and its security applications,'' in \emph{CCS}, 2016.

\bibitem{Yasumatsu2019codaspy}
T.~Yasumatsu, T.~Watanabe, F.~Kanei, E.~Shioji, M.~Akiyama, and T.~Mori,
  ``Understanding the responsiveness of mobile app developers to software
  library updates,'' in \emph{Proc. CODASPY}, 2019.

\bibitem{OSSPOLICE2017CCS}
R.~Duan, A.~Bijlani, M.~Xu, T.~Kim, and W.~Lee, ``Identifying open-source
  license violation and 1-day security risk at large scale,'' in
  \emph{Proc.CCS}, 2017.

\bibitem{zhang2020empirical}
Z.~Zhang, W.~Diao, C.~Hu, S.~Guo, C.~Zuo, and L.~Li, ``An empirical study of
  potentially malicious third-party libraries in android apps,'' in \emph{The
  13th ACM Conference on Security and Privacy in Wireless and Mobile Networks
  (WiSec 2020)}, 2020.

\bibitem{ATVHunter2021ICSE}
X.~Zhan, L.~Fan, S.~Chen, F.~Wu, T.~Liu, X.~Luo, and Y.~Liu, ``Atvhunter:
  Reliable version detection of third-party libraries for vulnerability
  identification in android applications,'' in \emph{2021 IEEE/ACM 43rd
  International Conference on Software Engineering (ICSE)}, 2021, pp.
  1695--1707.

\bibitem{AdRisk2012Wisec}
M.~C. Grace, W.~Zhou, X.~Jiang, and A.-R. Sadeghi, ``Unsafe exposure analysis
  of mobile in-app advertisements,'' in \emph{Proc. WiSec}, 2012.

\bibitem{DC2018FSE}
Y.~Wang, M.~Wen, Z.~Liu, R.~Wu, R.~Wang, B.~Yang, H.~Yu, Z.~Zhu, and S.-C.
  Cheung, ``Do the dependency conflicts in my project matter?'' in \emph{Proc.
  ESEC/FSE}, 2018, p. 319–330.

\bibitem{RIDDLE2019ICSE}
Y.~{Wang}, M.~{Wen}, R.~{Wu}, Z.~{Liu}, S.~H. {Tan}, Z.~{Zhu}, H.~{Yu}, and
  S.~{Cheung}, ``Could i have a stack trace to examine the dependency conflict
  issue?'' in \emph{ICSE}, 2019.

\bibitem{huang2020interactive}
K.~Huang, B.~Chen, B.~Shi, Y.~Wang, C.~Xu, and X.~Peng, ``Interactive,
  effort-aware library version harmonization,'' in \emph{ESEC/FSE}, 2020, p.
  518–529.

\bibitem{Wang2020ICSME}
Y.~Wang, B.~Chen, K.~Huang, B.~Shi, C.~Xu, X.~Peng, Y.~Liu, and Y.~Wu, ``An
  empirical study of usages, updates and risks of third-party libraries in java
  projects,'' in \emph{ICSME}, 2020.

\bibitem{xian2019saner}
X.~Zhan, T.~Zhang, and Y.~Tang, ``A comparative study of android repackaged
  apps detection techniques,'' in \emph{Proc. SANER}, 2019.

\bibitem{Lili2016SANER}
L.~Li, T.~Bissyand{\'e}, J.~Klein, and Y.~L. Traon, ``An investigation into the
  use of common libraries in android apps,'' in \emph{SANER}, 2016.

\bibitem{MassVet2015chen}
C.~Kai, W.~Peng, L.~Yeonjoon, W.~XiaoFeng, Z.~Nan, H.~Heqing, Z.~Wei, and
  L.~Peng, ``Finding unknown malice in 10 seconds: Mass vetting for new threats
  at the google-play scale,'' in \emph{Proc. USENIX Security}, 2015.

\bibitem{DroidMOSS12CODASPY}
W.~Zhou, Y.~Zhou, X.~Jiang, and P.~Ning, ``Detecting repackaged smartphone
  applications in third-party android marketplaces,'' in \emph{Proc. CODASPY},
  2012.

\bibitem{YuruACSAC14}
Y.~Shao, X.~Luo, C.~Qian, P.~Zhu, and L.~Zhang, ``Towards a scalable
  resource-driven approach for detecting repackaged android applications,'' in
  \emph{Proc. ACSAC}, 2014.

\bibitem{YLSANER16}
L.~Yu, X.~Luo, C.~Qian, and S.~Wang, ``Revisiting the description-to-behavior
  fidelity in android applications,'' in \emph{Proc. SANER}, 2016.

\bibitem{YLTSE18}
L.~Yu, X.~Luo, C.~Qian, S.~Wang, and H.~K.~N. Leung, ``Enhancing the
  description-to-behavior fidelity in android apps with privacy policy,''
  \emph{IEEE Transactions on Software Engineering (TSE)}, 2018.

\bibitem{LibD2017ICSE}
M.~Li, W.~Wang, P.~Wang, S.~Wang, D.~Wu, J.~Liu, R.~Xue, and W.~Huo, ``Libd:
  Scalable and precise third-party library detection in android markets,'' in
  \emph{Proc. ICSE}, 2017.

\bibitem{LibRadar2016ICSE}
Z.~Ma, H.~Wang, Y.~Guo, and X.~Chen, ``Libradar: Fast and accurate detection of
  third-party libraries in android apps,'' in \emph{Proc. ICSE-C}, 2016.

\bibitem{LibID2019issta}
J.~Zhang, A.~R. Beresford, and S.~A. Kollmann, ``Libid: Reliable identification
  of obfuscated third-party android libraries,'' in \emph{Proc. ISSTA}, 2019.

\bibitem{ORLIS2018MOBILESoft}
Y.~Wang, H.~Wu, H.~Zhang, and A.~Rountev, ``Orlis: Obfuscation-resilient
  library detection for android,'' in \emph{Proc. MOBILESoft}, 2018.

\bibitem{snowballing2014}
C.~Wohlin, ``Guidelines for snowballing in systematic literature studies and a
  replication in software engineering,'' in \emph{Proc. 18thInt. Conf. Eval.
  Assessment Softw. Eng}, 2014.

\bibitem{SLR2007}
``survey,'' Guidelines for performing systematic literature reviews in software
  engineering, 2007.

\bibitem{acmdigital}
\BIBentryALTinterwordspacing
``{ACM Digital Library}.'' [Online]. Available: \url{https://dl.acm.org/}
\BIBentrySTDinterwordspacing

\bibitem{ieee}
\BIBentryALTinterwordspacing
``{IEEE Xplore Digital Library}.'' [Online]. Available:
  \url{https://ieeexplore.ieee.org/Xplore/}
\BIBentrySTDinterwordspacing

\bibitem{springer}
\BIBentryALTinterwordspacing
``{SpringerLink}.'' [Online]. Available: \url{https://link.springer.com/}
\BIBentrySTDinterwordspacing

\bibitem{sciencedirect}
\BIBentryALTinterwordspacing
``{ScienceDirect}.'' [Online]. Available: \url{https://www.sciencedirect.com/}
\BIBentrySTDinterwordspacing

\bibitem{NDSS}
``{NDSS},'' The Network and Distributed System Security Symposium.

\bibitem{lili2017static}
L.~Li, T.~Bissyand{\'e}, P.~M., R.~S., B.~A., D.~Octeau, J.~Klein, and
  L.~Traon, ``Static analysis of android apps: A systematic literature
  review,'' \emph{Information and Software Technology Vol. 88}, 2017.

\bibitem{Li2018RebootingRO}
L.~Li, T.~F. Bissyand{\'e}, and J.~Klein, ``Rebooting research on detecting
  repackaged android apps: Literature review and benchmark,'' \emph{CoRR}, vol.
  abs/1811.08520, 2018.

\bibitem{kong2019Testing}
P.~{Kong}, L.~{Li}, J.~{Gao}, K.~{Liu}, T.~F. {Bissyandé}, and J.~{Klein},
  ``Automated testing of android apps: A systematic literature review,''
  \emph{IEEE Transactions on Reliability}, vol.~68, no.~1, pp. 45--66, March
  2019.

\bibitem{MAdLens2018infocom}
B.~He, H.~Xu, L.~Jin, G.~Guo, Y.~Chen, and G.~Weng, ``An investigation into
  android in-app ad practice: Implications for app developers,'' in
  \emph{INFOCOM}, 2018.

\bibitem{haoyu2017automated}
Z.~M. Haoyu~Wang, Yao~Guo and X.~Chen, ``Automated detection and classification
  of third-party libraries in large scale android apps,'' \emph{Journal of
  Software (in Chinese)}, 2017.

\bibitem{jin2019madlens}
L.~Jin, B.~He, G.~Weng, H.~Xu, Y.~Chen, and G.~Guo, ``Madlens: Investigating
  into android in-app ad practice at api granularity,'' \emph{IEEE Transactions
  on Mobile Computing}, vol.~20, no.~3, pp. 1138--1155, 2021.

\bibitem{LibD22018TSE}
M.~{Li}, P.~{Wang}, W.~{Wang}, S.~{Wang}, D.~{Wu}, J.~{Liu}, R.~{Xue},
  W.~{Huo}, and W.~{Zou}, ``Large-scale third-party library detection in
  android markets,'' \emph{IEEE Transactions on Software Engineering}, 2018.

\bibitem{FLEXDROID201NDSS}
J.~Seo, D.~Kim, D.~Cho, T.~Kim, and I.~Shin, ``Flexdroid: enforcing in-app
  privilege separation in android,'' in \emph{Proc. NDSS}, 2017.

\bibitem{libdetect2020ASE}
X.~Zhan, L.~Fan, T.~Liu, S.~Chen, L.~Li, H.~Wang, Y.~Xu, X.~Luo, and Y.~Liu,
  ``Automated third-party library detection for android applications: Are we
  there yet?'' in \emph{ASE}, 2020.

\bibitem{LibDX2020SANER}
W.~{Tang}, P.~{Luo}, J.~{Fu}, and D.~{Zhang}, ``Libdx: A cross-platform and
  accurate system to detect third-party libraries in binary code,'' in
  \emph{SANER}, 2020, pp. 104--115.

\bibitem{LibExtractor2020wisec}
Z.~Zhang, W.~Diao, C.~Hu, S.~Guo, C.~Zuo, and L.~Li, ``An empirical study of
  potentially malicious third-party libraries in android apps,'' in \emph{Proc.
  WiSec}, 2020.

\bibitem{LibRoad2020TMC}
J.~{Xu} and Q.~{Yuan}, ``Libroad: Rapid, online, and accurate detection of tpls
  on android,'' \emph{IEEE Transactions on Mobile Computing}, 2020.

\bibitem{md2020studying}
M.~Ahasanuzzaman, S.~Hassan, and A.~E. Hassan, ``Studying ad library
  integration strategies of top free-to-download apps,'' \emph{IEEE
  Transactions on Software Engineering}, 2020.

\bibitem{ahasanuzzaman2020longitudinal}
M.~Ahasanuzzaman, S.~Hassan, C.-P. Bezemer, and A.~E. Hassan, ``A longitudinal
  study of popular ad libraries in the google play store,'' \emph{ESEM},
  no.~25, pp. 824--858, 2020.

\bibitem{chen2019revisiting}
G.~Chen, W.~Meng, and J.~Copeland, ``Revisiting mobile advertising threats with
  madlife,'' in \emph{The World Wide Web Conference}.\hskip 1em plus 0.5em
  minus 0.4em\relax ACM, 2019, pp. 207--217.

\bibitem{TPLsurvey2019}
P.~Salza, F.~Palomba, D.~D. Nucci, A.~D. Lucia, and F.~Ferrucci, ``Third-party
  libraries in mobile apps when, how, and why developers update them,''
  \emph{Springer Science}, 24 Aug. 2019.

\bibitem{APPCOMMUNE2019SANER}
B.~{Li}, Y.~{Zhang}, J.~{Li}, R.~{Feng}, and D.~{Gu}, ``Appcommune: Automated
  third-party libraries de-duplicating and updating for android apps,'' in
  \emph{SANER}, 2019, pp. 344--354.

\bibitem{libpecker2018}
Y.~Zhang, J.~Dai, X.~Zhang, S.~Huang, Z.~Yang, M.~Yang, and H.~Chen,
  ``Detecting third-party libraries in android applications with high precision
  and recall,'' in \emph{SANER}, 2018.

\bibitem{Salza2018ICPC}
P.~Salza, F.~Palomba, D.~Di~Nucci, C.~D'Uva, A.~De~Lucia, and F.~Ferrucci, ``Do
  developers update third-party libraries in mobile apps?'' in \emph{Proc.
  ICPC}, 2018.

\bibitem{Dong2018HotMibile}
F.~Dong, H.~Wang, L.~Li, Y.~Guo, G.~Xu, and S.~Zhang, ``How do mobile apps
  violate the behavioral policy of advertisement libraries?'' in \emph{Proc.
  HotMobile}, 2018.

\bibitem{identifyads2018WPC}
H.~Han, R.~Li, and J.~Tang, ``Identify and inspect libraries in android
  applications,'' \emph{Wireless Personal Communications vol 103, pp491-503},
  2018.

\bibitem{CANDARW2018}
H.~{Ogawa}, E.~{Takimoto}, K.~{Mouri}, and S.~{Saito}, ``User-side updating of
  third-party libraries for android applications,'' in \emph{Proc. CANDARW},
  Nov 2018.

\bibitem{Watanabe2017MSR}
T.~Watanabe, M.~Akiyama, F.~Kanei, E.~Shioji, Y.~Takata, B.~Sun, Y.~Ishi,
  T.~Shibahara, T.~Yagi, and T.~Mori, ``Understanding the origins of mobile app
  vulnerabilities: A large-scale measurement study of free and paid apps,'' in
  \emph{Proc. MSR}, 2017.

\bibitem{AppLibRec2017Internetware}
H.~Yu, X.~Xia, X.~Zhao, and W.~Qiu, ``Combining collaborative filtering and
  topic modeling for more accurate android mobile app library recommendation,''
  in \emph{Proceedings of the 9th Asia-Pacific Symposium on Internetware}, ser.
  Internetware'17, 2017.

\bibitem{Derr2017ccs}
E.~Derr, S.~Bugiel, S.~Fahl, Y.~Acar, and M.~Backes, ``Keep me updated: An
  empirical study of third-party library updatability on android,'' in
  \emph{Proc. CCS}, 2017.

\bibitem{splitads2017ACISP}
J.~Zhan, Q.~Zhou, X.~Gu, Y.~Wang, and Y.~Niu, ``Splitting third-party
  libraries' privileges from android apps,'' in \emph{ACISP}, 2017.

\bibitem{Gui2017WhatAO}
J.~Gui, M.~Nagappan, and W.~G.~J. Halfond, ``What aspects of mobile ads do
  users care about? an empirical study of mobile in-app ad reviews,''
  \emph{CoRR}, vol. abs/1702.07681, 2017.

\bibitem{son2016mobile}
S.~Son, D.~Kim, and V.~Shmatikov, ``What mobile ads know about mobile users,''
  in \emph{NDSS 2016}, 2016.

\bibitem{Libcage2016ESORICS}
F.~Wang, Y.~Zhang, K.~Wang, P.~Liu, and W.~Wang, ``Stay in your cage! a sound
  sandbox for third-party libraries on android,'' in \emph{ESORICS}, 2016.

\bibitem{LibFInder2016sp}
K.~Chen, X.~Wang, Y.~Chen, P.~Wang, Y.~Lee, X.~Wang, and B.~Ma, ``Following
  devil’s footprints: Cross-platform analysis of potentially harmful
  libraries on android and ios,'' in \emph{S \& P}, 2016.

\bibitem{LibSift2016soh}
C.~{Soh}, H.~B.~K. {Tan}, Y.~L. {Arnatovich}, A.~{Narayanan}, and L.~{Wang},
  ``Libsift: Automated detection of third-party libraries in android
  applications,'' in \emph{APSEC}, 2016.

\bibitem{Israel2016software}
I.~J.~M. Ruiz, M.~Nagappan, B.~Adams, T.~Berger, and S.~Dienst, ``Analyzing ad
  library updates in android apps,'' \emph{IEEE Software VOl. 33}, 2016.

\bibitem{ad2016NDSS}
V.~Rastogi, R.~Shao, Y.~Chen, X.~Pan, S.~Zou, and R.~Riley, ``Are these ads
  safe: Detecting hidden attacks through the mobile app-web interfaces,'' in
  \emph{NDSS}, 2016.

\bibitem{price2016NDSS}
M.~Wei, D.~Ren, S.~P. Chung, S.~Han, and W.~Lee, ``The price of free: Privacy
  leakage in personalized mobile in-app ads,'' in \emph{2016}, NDSS.

\bibitem{Madscope2015Mobisys}
S.~Nath, ``Madscope: Characterizing mobile in-app targeted ads,'' in
  \emph{Proc. MobiSys}, 2015.

\bibitem{adempirical2015compscience}
T.~Book and S.~W. Dan, ``An empirical study of mobile ad targeting,''
  \emph{computer science}, 2015.

\bibitem{paturi2015NDSS}
A.~Paturi, P.~G. Kelley, and S.~Mazumdar, ``Introducing privacy threats from ad
  libraries to android users through privacy granules,'' in \emph{Proc. NDSS},
  2015.

\bibitem{ICSE2015jiapping}
J.~Gui, S.~Mcilroy, M.~Nagappan, and W.~G.~J. Halfond, ``Truth in advertising:
  The hidden cost of mobile ads for software developers,'' in \emph{ICSE},
  2015.

\bibitem{cho2015empirical}
G.~Cho, J.~Cho, Y.~Song, and H.~Kim, ``An empirical study of click fraud in
  mobile advertising networks,'' in \emph{2015 10th International Conference on
  Availability, Reliability and Security}, 2015, pp. 382--388.

\bibitem{kuhnel2015fast}
M.~Kühnel, M.~Smieschek, and U.~Meyer, ``Fast identification of obfuscation
  and mobile advertising in mobile malware,'' in \emph{2015 IEEE
  Trustcom/BigDataSE/ISPA}, vol.~1, 2015, pp. 214--221.

\bibitem{AdDetect2014ISSNIP}
A.~Narayanan, L.~Chen, and C.~K. Chan, ``Addetect: Automated detection of
  android ad libraries using semantic analysis,'' in \emph{Proc. ISSNIP}, 2014.

\bibitem{Apklancet2014ASIACCS}
W.~Yang, J.~Li, Y.~Zhang, J.~S. Y.~Li, and D.~Gu, ``Apklancet: Tumor payload
  diagnosis and purification for android applications,'' in \emph{ASIACCS},
  2014.

\bibitem{COMPAC2014Wang}
Y.~Wang, S.~Hariharan, C.~Zhao, J.~Liu, and W.~Du, ``Compac: Enforce
  component-level access control in android,'' in \emph{Proc.CODASPY}, 2014.

\bibitem{Duet2014wisec}
W.~Hu, D.~Octeau, P.~D. McDaniel, and P.~Liu, ``Duet: Library integrity
  verification for android applications,'' in \emph{Proc. WiSec}, 2014.

\bibitem{NativeGuard2014Wisec}
M.~Sun and G.~Tan, ``Nativeguard: Protecting android applications from
  third-party native libraries,'' in \emph{Proc. WiSec}, 2014.

\bibitem{wkshps2014}
I.~{Ullah}, R.~{Boreli}, M.~A. {Kaafar}, and S.~S. {Kanhere}, ``Characterising
  user targeting for in-app mobile ads,'' in \emph{INFOCOM WKSHPS}, 2014.

\bibitem{Mojia2014software}
I.~J.~M. Ruiz, M.~Nagappan, B.~Adams, T.~Berger, S.~Dienst, and A.~E. Hassan,
  ``Impact of ad libraries on ratings of android mobile apps,'' \emph{IEEE
  Software}, 07 May 2014.

\bibitem{bhoraskar2014brahmastra}
R.~Bhoraskar, S.~Han, J.~Jeon, T.~Azim, S.~Chen, J.~Jung, S.~Nath, R.~Wang, and
  D.~Wetherall, ``Brahmastra: Driving apps to test the security of third-party
  components,'' in \emph{23rd $\{$USENIX$\}$ Security Symposium ($\{$USENIX$\}$
  Security 14)}, 2014, pp. 1021--1036.

\bibitem{book2013spsm}
T.~Book and D.~S. Wallach, ``A case of collusion: A study of the interface
  between ad libraries and their apps,'' in \emph{SPSM}, 2013.

\bibitem{MOST2013}
T.~Book, A.~Pridgen, and D.~S. Wallach, ``Longitudinal analysis of android ad
  library permissions,'' in \emph{MoST}, 2013.

\bibitem{PAM2013}
A.~Tongaonkar, S.~Dai, and D.~S. A.~Nucci, ``Understanding mobile app usage
  patterns using in-app advertisements,'' in \emph{PAM}, 2013.

\bibitem{structure2012ICSM}
V.~{Bauer}, L.~{Heinemann}, and F.~{Deissenboeck}, ``A structured approach to
  assess third-party library usage,'' in \emph{2012 28th IEEE International
  Conference on Software Maintenance (ICSM)}, 2012, pp. 483--492.

\bibitem{Leontiadis12HotMobile}
I.~Leontiadis, C.~Efstratiou, M.~Picone, and C.~Mascolo, ``Don't kill my ads!:
  Balancing privacy in an ad-supported mobile application market,'' in
  \emph{HotMobile}, ser. HotMobile '12, 2012.

\bibitem{Most2012}
R.~Stevens, C.~Gibler, J.~Crussell, J.~Erickson, and H.~Chen, ``Investigating
  user privacy in android ad libraries,'' in \emph{MoST}, 2012.

\bibitem{vallina2012breaking}
\BIBentryALTinterwordspacing
N.~Vallina-Rodriguez, J.~Shah, A.~Finamore, Y.~Grunenberger, K.~Papagiannaki,
  H.~Haddadi, and J.~Crowcroft, ``Breaking for commercials: Characterizing
  mobile advertising,'' in \emph{IMC}.\hskip 1em plus 0.5em minus 0.4em\relax
  New York, NY, USA: Association for Computing Machinery, 2012, p. 343–356.
  [Online]. Available: \url{https://doi.org/10.1145/2398776.2398812}
\BIBentrySTDinterwordspacing

\bibitem{assessAndroidsecurity2017TSE}
A.~Sadeghi, H.~Bagheri, J.~Garcia, and S.~Malek, ``A taxonomy and qualitative
  comparison of program analysis techniques for security assessment of android
  software,'' \emph{IEEE Transactions on Software Engineering}, vol.~43, no.~6,
  pp. 492--530, 2017.

\bibitem{tang:ase19}
Y.~{Tang}, X.~{Zhan}, H.~{Zhou}, X.~{Luo}, Z.~{Xu}, Y.~{Zhou}, and Q.~{Yan},
  ``Demystifying application performance management libraries for android,'' in
  \emph{2019 34th IEEE/ACM International Conference on Automated Software
  Engineering (ASE)}, 2019, pp. 682--685.

\bibitem{tangtse:21a}
Y.~Tang, H.~Wang, X.~Zhan, X.~Luo, Y.~Zhou, H.~Zhou, Q.~Yan, Y.~Sui, and J.~W.
  Keung, ``A systematical study on application performance management libraries
  for apps,'' \emph{IEEE Transactions on Software Engineering}, pp. 1--20,
  2021.

\bibitem{Kotlin}
``Kotlin,'' \url{https://en.wikipedia.org/wiki/Kotlin_(programming_language)}.

\bibitem{CLANDroid2016ICPC}
L.~Mario, H.~Andrew, and P.~Denys, ``On automatically detecting similar android
  apps,'' in \emph{Proc. ICPC}, 2016.

\bibitem{DroidEagle2015}
M.~Sun, M.~Li, and J.~C.~S. Lui, ``Droideagle: Seamless detection of visually
  similar android apps,'' in \emph{Wisec}, 2015.

\bibitem{Juxtapp13DIMVA}
S.~Hanna, L.~Huang, E.~Wu, S.~Li, C.~Chen, and D.~Song, ``Juxtapp: a scalable
  system for detecting code reuse among android applications,'' in \emph{Proc.
  DIMVA}, 2012.

\bibitem{ViewDroid14}
F.~Zhang, H.~Huang, S.~Zhu, D.~Wu, and P.~Liu, ``Viewdroid: Towards
  obfuscation-resilient mobile application repackaging detection,'' in
  \emph{Proc. ACM WiSec}, 2014.

\bibitem{chen14ICSE}
K.~Chen, P.~Liu, and Y.~Zhang, ``Achieving accuracy and scalability
  simultaneously in detecting application clones on android markets,'' in
  \emph{Proc. ICSE}, 2014.

\bibitem{FSquaDRA2014IFIP}
Y.~Zhauniarovich, O.~Gadyatskaya, B.~Crispo, F.~La~Spina, and E.~Moser,
  ``Fsquadra: fast detection of repackaged applications,'' in \emph{IFIP
  DBSec}, 2014.

\bibitem{Andarwin2013ESORICS}
J.~Crussell, C.~Gibler, and H.Chen, ``Andarwin: Scalable detection of
  semantically similar android applications,'' in \emph{Proc. ESORICS}, 2013.

\bibitem{Wukong2015issta}
H.~Wang, Y.~Guo, Z.~Ma, and X.~Chen, ``Wukong: A scalable and accurate
  two-phase approach to android app clone detection,'' in \emph{Proc. ISSTA},
  2015.

\bibitem{PiggyApp13CODASPY}
W.~Zhou, Y.~Zhou, M.~Grace, X.~Jiang, and S.~Zou, ``Fast, scalable detection of
  piggybacked mobile applications,'' in \emph{Proc. CODASPY}, 2013.

\bibitem{apktool}
``Apktool,'' \url{https://ibotpeaches.github.io/Apktool/}.

\bibitem{baksmali}
``Baksmali,'' \url{https://github.com/JesusFreke/smali}.

\bibitem{Androguard}
``Androguard,'' \url{https://github.com/androguard/androguard}.

\bibitem{soot}
``Soot,'' \url{https://github.com/Sable/soot}.

\bibitem{google_GMS}
``Google android gms,''
  \url{https://mvnrepository.com/artifact/com.google.android.gms}.

\bibitem{google_android}
``Google android library,''
  \url{https://mvnrepository.com/artifact/com.google.android/android}.

\bibitem{centris2021ICSE}
S.~Woo, S.~Park, S.~Kim, H.~Lee, and H.~Oh, ``Centris: A precise and scalable
  approach for identifying modified open-source software reuse,'' in
  \emph{ICSE}, 2021.

\bibitem{droidkungfu}
``Droidkungfu,''
  \url{https://www.f-secure.com/v-descs/trojan_android_droidkungfu_c.shtml}.

\bibitem{Droidbox}
``{DroidBox},'' \url{https://github.com/pjlantz/droidbox}.

\bibitem{TaintDroid}
``{TaintDroid},'' \url{http://www.appanalysis.org/}.

\bibitem{adb}
``Android debug bridge,''
  \url{https://developer.android.com/studio/command-line/adb}.

\bibitem{admob}
\url{https://admob.google.com/home/}.

\bibitem{monkey}
``Monkey,'' \url{https://developer.android.com/studio/test/monkey}.

\bibitem{NVD}
``Nvd,'' \url{https://nvd.nist.gov/}.

\bibitem{sourceclear}
\BIBentryALTinterwordspacing
``{Source Clear}.'' [Online]. Available:
  \url{https://sca.analysiscenter.veracode.com/vulnerability-database/search#query=language:java}
\BIBentrySTDinterwordspacing

\bibitem{Quire2011USENIX}
M.~Dietz, S.~Shekhar, Y.~Pisetsky, A.~Shu, and D.~S. Wallach, ``Quire:
  Lightweight provenance for smart phone operating systems,'' in \emph{Proc.
  USENIX security}, 2011.

\bibitem{AIDL}
``{AIDL},'' \url{https://developer.android.com/guide/components/aidl}.

\bibitem{JNI}
``{JNI},'' \url{https://developer.android.com/training/articles/perf-jni}.

\bibitem{Javareflection}
``{java reflection},''
  \url{http://tutorials.jenkov.com/java-reflection/index.html}.

\bibitem{dymcode}
``{Dynamic Code Execution},''
  \url{https://gerardnico.com/lang/java/dynamic\#load\_and\_run\_the\_class}.

\bibitem{recommendaerSystwms2002Bueke}
\BIBentryALTinterwordspacing
R.~Burke, ``Hybrid recommender systems: Survey and experiments,'' \emph{User
  Modeling and User-Adapted Interaction}, vol.~12, no.~4, pp. 331--370, Nov.
  2002. [Online]. Available: \url{http://dx.doi.org/10.1023/A:1021240730564}
\BIBentrySTDinterwordspacing

\bibitem{LDA2003}
D.~M. Blei, A.~Y. Ng, and M.~I. Jordan, ``Latent dirichlet allocation,''
  \emph{J. Mach. Learn. Res.}, vol.~3, pp. 993--1022, Mar. 2003.

\bibitem{firebase2021mobilesoft}
\BIBentryALTinterwordspacing
J.~Harty, H.~Zhang, L.~Wei, L.~Pascarella, M.~Aniche, and W.~Shang, ``Logging
  practices with mobile analytics: An empirical study on firebase,'' in
  \emph{2021 2021 IEEE/ACM 8th International Conference on Mobile Software
  Engineering and Systems (MobileSoft) (MobileSoft)}.\hskip 1em plus 0.5em
  minus 0.4em\relax Los Alamitos, CA, USA: IEEE Computer Society, may 2021, pp.
  56--60. [Online]. Available:
  \url{https://doi.ieeecomputersociety.org/10.1109/MobileSoft52590.2021.00013}
\BIBentrySTDinterwordspacing

\bibitem{lili2019TSE}
L.~{Li}, T.~F. {Bissyande}, and J.~{Klein}, ``Rebooting research on detecting
  repackaged android apps: Literature review and benchmark,'' \emph{IEEE
  Transactions on Software Engineering}, pp. 1--1, 2019.

\bibitem{Androzoo}
``Androzoo,'' \url{https://androzoo.uni.lu/}.

\bibitem{repackaged_survey2018ISDFS}
M.~{Baykara} and E.~{Çolak}, ``A review of cloned mobile malware applications
  for android devices,'' in \emph{Proc. ISDFS}, March 2018, pp. 1--5.

\bibitem{surveymalware2020Qiu}
J.~Qiu, J.~Zhang, W.~Luo, L.~Pan, S.~Nepal, and Y.~Xiang, ``A survey of android
  malware detection with deep neural models,'' \emph{ACM Comput. Surv.},
  vol.~53, no.~6, 2020.

\bibitem{JoWUA2015}
B.~Rashidi and C.~Fung, ``A survey of android security threats and defenses,''
  in \emph{Proc. JoWUA}, 2015.

\bibitem{Sufatrio2015CSUR}
Sufatrio, J.~Tan, C.~Wei, and L.~Thing, ``Securing android: A survey, taxonomy,
  and challenges,'' \emph{ACM Comput. Surv.}, 2015.

\bibitem{faruki2015survey_security}
P.~{Faruki}, A.~{Bharmal}, V.~{Laxmi}, V.~{Ganmoor}, M.~S. {Gaur}, M.~{Conti},
  and M.~{Rajarajan}, ``Android security: A survey of issues, malware
  penetration, and defenses,'' \emph{IEEE Communications Surveys Tutorials},
  vol.~17, no.~2, pp. 998--1022, 2015.

\bibitem{choudhary2015ASE}
S.~R. {Choudhary}, A.~{Gorla}, and A.~{Orso}, ``Automated test input generation
  for android: Are we there yet? (e),'' in \emph{ASE}, 2015, pp. 429--440.

\end{thebibliography}

%\end{IEEEbiography}
%% insert where needed to balance the two columns on the last page with
%%\newpage


%\vfill


%\vspace{-10ex}

\begin{IEEEbiography}[{\includegraphics[width=1in,height=1.25in,clip,keepaspectratio]{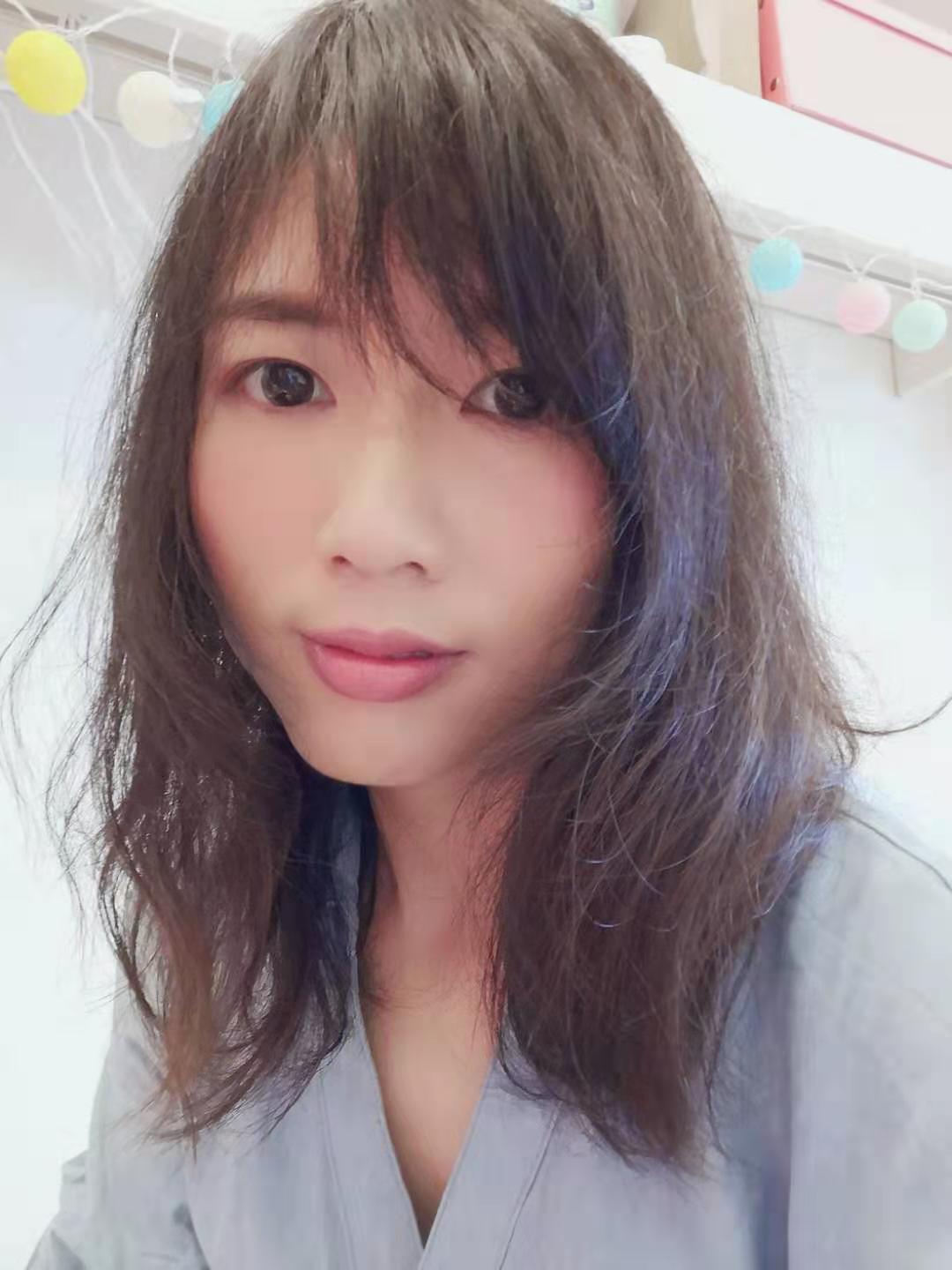}}]
	{Xian Zhan} received her BEng degree in computer science and technology from Wuhan University, Hubei, China. 
	Currently, she is a Ph.D candidate in the Department of Computing, the Hong Kong Polytechnic University. Her research interests include  program analysis, mobile privacy and security, Android analysis, third-party library and open-source software analysis and machine learning. 
	
\end{IEEEbiography}

\begin{IEEEbiography}[{\includegraphics[width=1in,height=1.3in,clip,keepaspectratio]{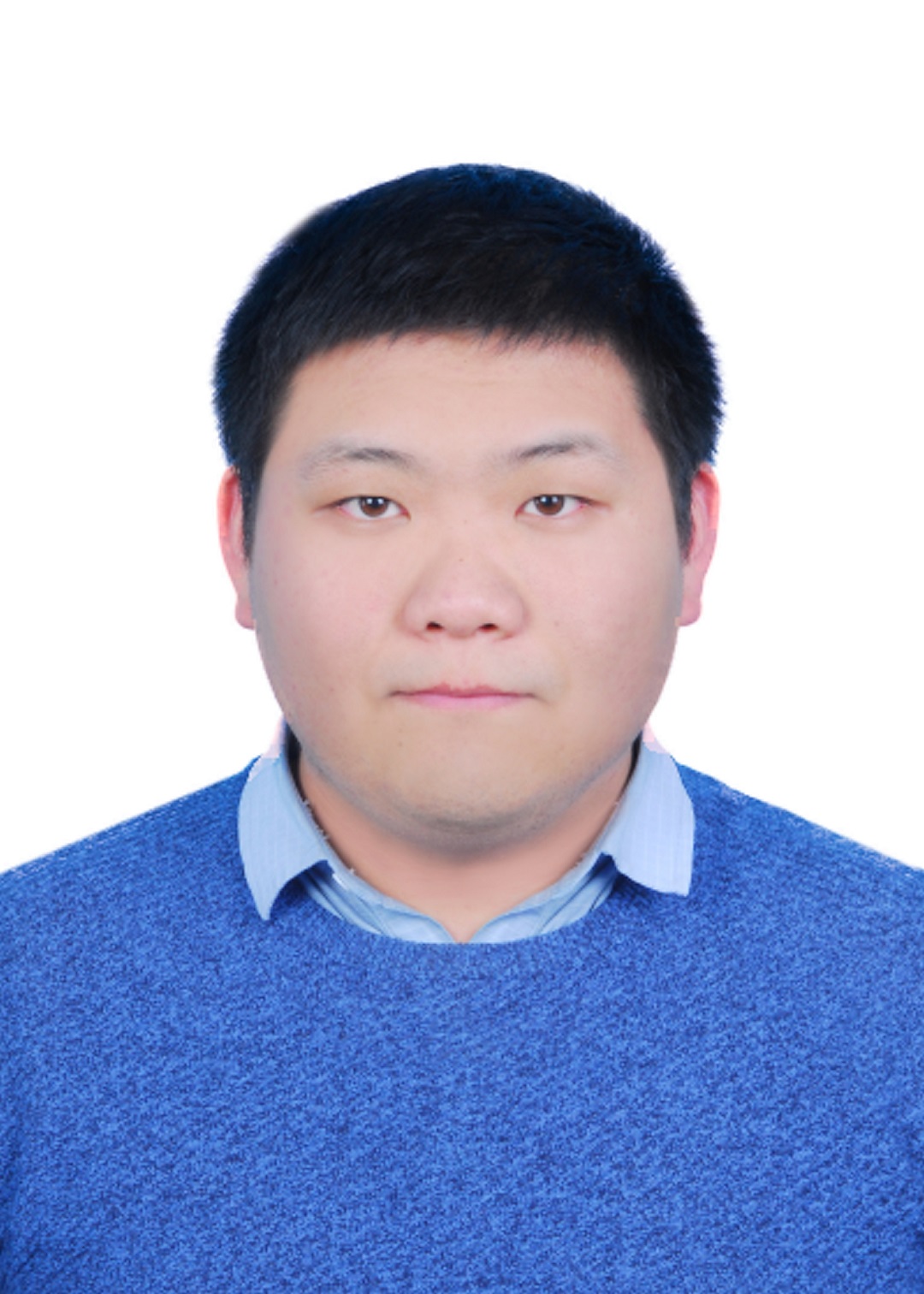}}]
    {Tianming Liu} received his M.S. degree in cyberspace security from Beijing University of Posts and Telecommunications, China, in 2020. He is currently working towards the PhD degree at Faculty of Information Technology, Monash University, Australia. His research interests include mobile security and Android application analysis.
	
\end{IEEEbiography}

\vspace{-10pt}

\begin{IEEEbiography}[{\includegraphics[width=1in,height=1.25in,clip,keepaspectratio]{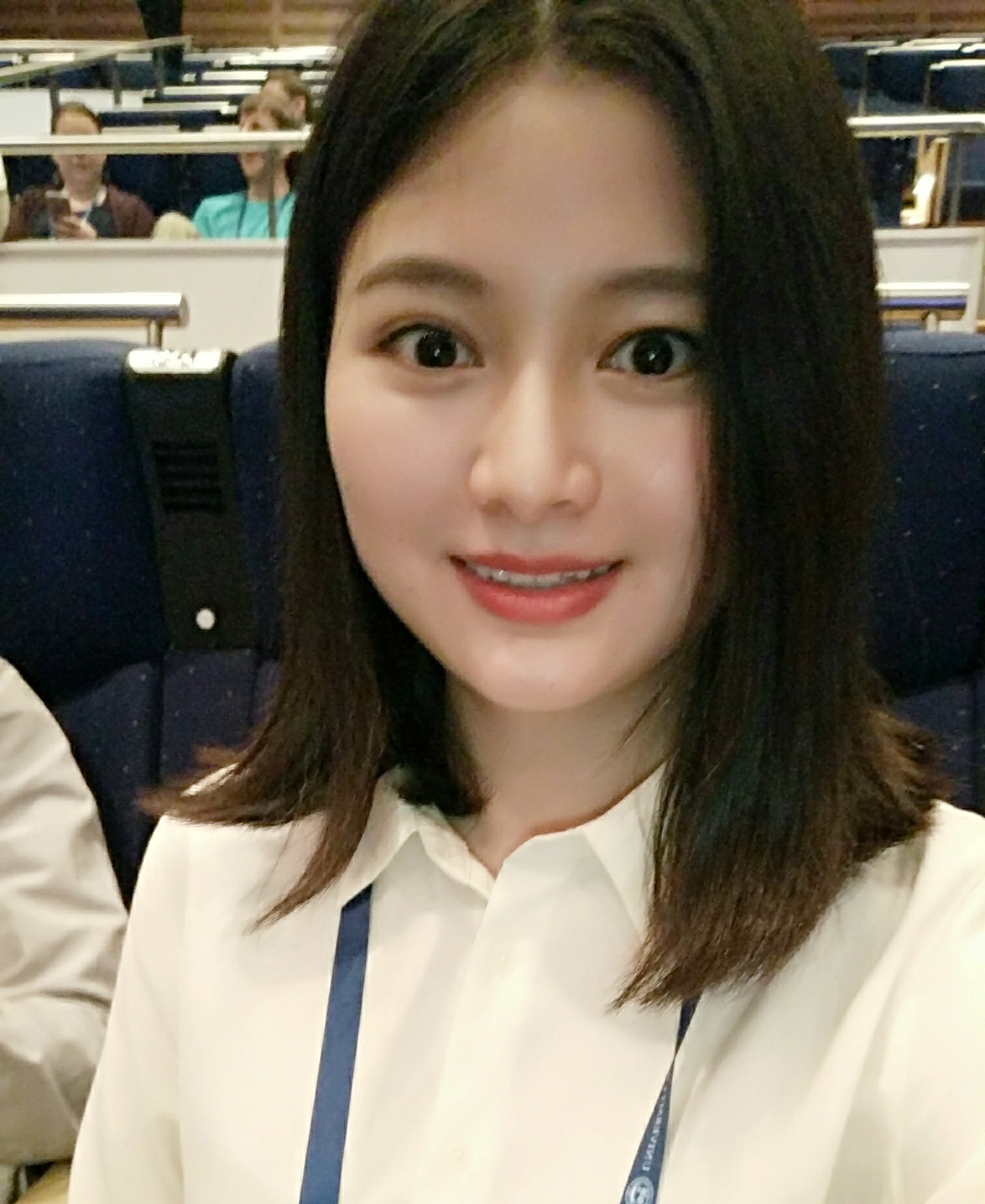}}]
	{Lingling Fan} is an Associate Professor in College of Cyber Science, Nankai University, China. She received her Ph.D and BEng degrees in computer science from East China Normal University, Shanghai, China in June 2019 and June 2014, respectively. In 2017, she joined Nanyang Technological University (NTU), Singapore as a Research Assistant and then had been as a Research Fellow of NTU since 2019. Her research focuses on program analysis and testing, software security, and Android and application analysis and testing. She got an ACM SIGSOFT Distinguished Paper Award at ICSE 2018. More information is available on ~\url{https://lingling-fan.github.io/}
\end{IEEEbiography}

\vspace{-10pt}

\begin{IEEEbiography}[{\includegraphics[width=1in,height=1.25in,clip,keepaspectratio]{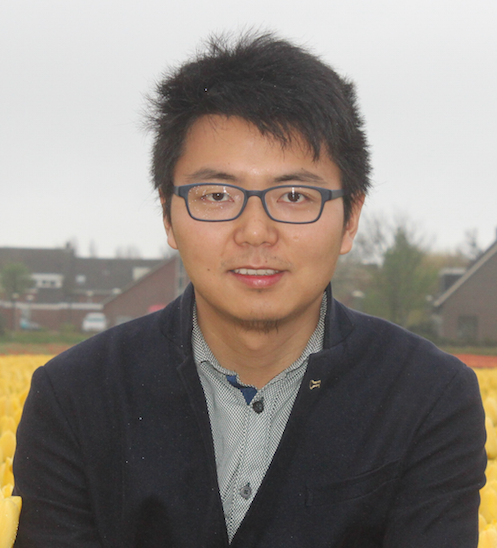}}]
{Li Li} is a lecturer (a.k.a., Assistant Professor)
and a PhD supervisor at Monash University,
Australia. Prior to joining Monash University, he
spent 1.5 years as a Research Associate at the
Serval group, SnT, University of Luxembourg. He
received his PhD degree in computer science
from the University of Luxembourg in 2016. His
research interests are in the fields of Android
security and Reliability, Static Code Analysis,
Machine Learning and Deep Learning. Dr. Li
received an ACM Distinguished Paper Award at
ASE 2018, a FOSS Impact Paper Award at MSR 2018 and a Best
Paper Award at the ERA track of IEEE SANER 2016. He is an active
member of the software engineering and security community serving
as reviewers or co-reviewers for many top-tier conferences and journals
such as ASE, ICSME, SANER, TSE, TOSEM, TIFS, TDSC, TOPS, EMSE, JSS, IST,
etc. His personal website is http://lilicoding.github.io.
\end{IEEEbiography}

\vspace{-10pt}

\begin{IEEEbiography}[{\includegraphics[width=1in,height=1.25in,clip,keepaspectratio]{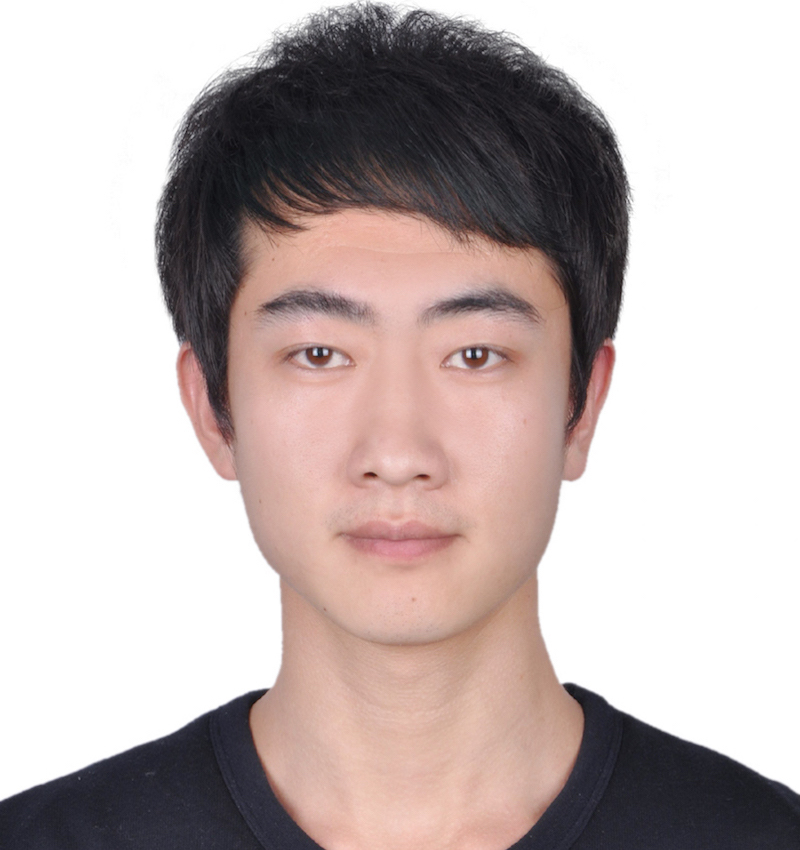}}]
{Sen Chen} received his Ph.D. degree in Computer Science from School of Computer Science and Software Engineering, East China Normal University, China, in June 2019.
Currently, he is an Associate Professor in the College of Intelligence and Computing (School of Cyber Security), Tianjin University, China. Before that, he was a Research Assistant Professor in the School of Computer Science and Engineering, Nanyang Technological University, Singapore.
Previously, he was a Research Assistant of NTU from 2016 to 2019 and a Research Fellow from 2019-2020. 
His research focuses on Security and Software Engineering such as mobile security, AI security, open-source security, and intelligent development and testing.
He has published broadly in top-tier security (S\&P, USENIX Security, CCS, IEEE TIFS, and IEEE TDSC) and software engineering venues including ICSE, FSE, ASE, ACM TOSEM, and IEEE TSE. More information is available on {\url{https://sen-chen.github.io/}}

\end{IEEEbiography}

\vspace{-10pt}

\begin{IEEEbiography}[{\includegraphics[width=1in,height=1.25in,clip,keepaspectratio]{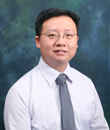}}]{Xiapu Luo}
 is an associate professor with the
Department of Computing and an Associate Researcher with the Shenzhen Research Institute,
The Hong Kong Polytechnic University. He received the Ph.D. degree in Computer Science
from The Hong Kong Polytechnic University, and
was a Post-Doctoral Research Fellow with the
Georgia Institute of Technology. His research focuses on smartphone security and privacy, network security and privacy, and Internet measurement.
More information is available at ~\url{https://www4.comp.polyu.edu.hk/~csxluo/.}

\end{IEEEbiography}

\vspace{-10pt}

\begin{IEEEbiography}[{\includegraphics[width=1.1in,height=1.65in,clip,keepaspectratio]{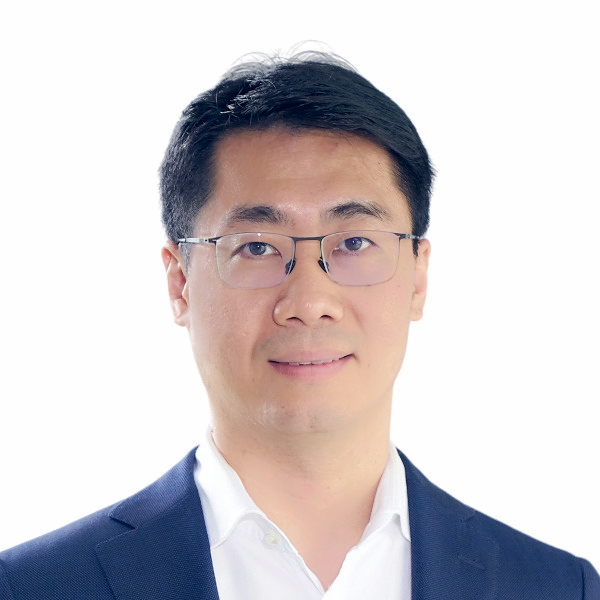}}]{Yang Liu}
Dr. Yang Liu obtained his bachelor and ph.d degree in the National University of Singapore in 2005 and 2010, respectively. In 2012, he joined Nanyang Technological University as a Nanyang Assistant Professor. He is currently a full professor, director of the cybersecurity lab, Program Director of HP-NTU Corporate Lab and Deputy Director of the National Satellite of Excellence of Singapore. In 2019, he received the University Leadership Forum Chair professorship at NTU. 
Dr. Liu specializes in software engineering, cybersecurity and artificial intelligence. His research has bridged the gap between the theory and practical usage of program analysis, data analysis and AI  to evaluate the design and implementation of software for high assurance and security. By now, he has more than 400 publications in top tier conferences and journals. He has received a number of prestigious awards including MSRA Fellowship, TRF Fellowship, Tan Chin Tuan Fellowship, Nanyang Research Award 2019, ACM Distinguished Speaker, NRF Investigatorship, and 15 best paper awards and one most influence system award in top software engineering conferences like ASE, FSE and ICSE.
	More information is available at ~\url{http://www.ntu.edu.sg/home/yangliu/}.
\end{IEEEbiography}

\clearpage
\appendices

\end{document}